\documentclass[english]{aa}
\usepackage[T1]{fontenc}
\usepackage[latin9]{inputenc}
\usepackage{txfonts}
\usepackage{natbib}
\bibpunct{(}{)}{;}{a}{}{,}
\usepackage{longtable}
\usepackage{lscape}
\usepackage{graphicx}
\usepackage{amssymb}
\usepackage{babel}
\makeatother
\providecommand{\tabularnewline}{\\}

\begin{document}

%
%%%%%%%%%%%%%%%%%%%%%%%%%%%%%% Title page.
\title{Taxonomy of asteroid families among the Jupiter Trojans: Comparison
between spectroscopic data and the Sloan Digital Sky Survey colors}

\titlerunning{Taxonomy of Trojan asteroid families} 
\authorrunning{F. Roig et al.} 

\author{F. Roig\inst{1} \and A. O. Ribeiro\inst{1} \and R. Gil-Hutton\inst{2}}

\offprints{F. Roig, \email{froig@on.br}}

\institute{Observatório Nacional, Rua Gal. Jos\'e Cristino 77, Rio de Janeiro, 20921-400, Brazil 
\and 
Complejo Astronómico El Leoncito (CASLEO) and Univ. Nacional de San Juan, Av. Espa\~na 1512 sur, San Juan, J5402DSP, Argentina}

\date{Received / Accepted }

\abstract
{}
{We present a comparative analysis of the spectral slope and color distributions of
Jupiter Trojans, with particular attention to asteroid families. We use a sample of data from
the Moving Object Catalogue of the Sloan Digital Sky Survey, together with spectra
obtained from several surveys.}
{A first sample of 349 observations, corresponding to 250 Trojan asteroids, 
were extracted from the Sloan Digital Sky Survey, and we also extracted
from the literature a second sample of 91 spectra, corresponding to 71 Trojans. 
The spectral slopes were computed by means of a least-squares fit to a 
straight line of the fluxes obtained from
the Sloan observations in the first sample, and of the rebinned spectra in
the second sample. In both cases the reflectance fluxes/spectra were renormalized 
to 1 at 6230 $\textrm{\AA}$.}
{We found that the distribution of spectral slopes among Trojan asteroids
shows a bimodality. About 2/3 of the objects have reddish slopes
compatible with D-type asteroids, while the remaining bodies show less
reddish colors compatible with the P-type and C-type classifications. The
members of asteroid families also show a bimodal distribution with a
very slight predominance of D-type asteroids, but the background is
clearly dominated by the D-types. The L4 and L5 swarms show different
distributions of spectral slopes, and bimodality is only observed in
L4. These differences can be attributed to the asteroid families since the
backgraound asteroids show the same slope distribtuions in both swarms.
The analysis of individual families indicates that the families in L5 are taxonomically
homogeneous, but in L4 they show a mixture of taxonomic types.
We discuss a few scenarios that might help to interpret these results.}
{}

\keywords{Minor planets, asteroids} 

\maketitle

%%%%%%%%%%%%%%%%%%%%%%%%%%%%%% Manuscript.
\section{Introduction}

Trojan asteroids are a very interesting population of minor bodies
due to their dynamical characteristics and physical properties. The
main hypotheses about the origin of the Jupiter Trojans assumed that
they formed either during the final stages of the planetary formation
(\citealp{1998A&A...339..278M}), or during the epoch of planetary
migration (\citealp{2005Natur.435..462M}), in any case more than
3.8 Gy. ago. The dynamical configuration kept the Trojans isolated
from the asteroid Main Belt throughout the Solar System history. In
spite of eventual interactions with other populations of minor bodies
like the Hildas, the Jupiter family comets, and the Centaurs, their
collisional evolution has been dictated mostly by the intrapopulation
collisions (\citealp{1996Icar..119..192M}, \citeyear{1997Icar..125...39M}).
Therefore, the Jupiter Trojans may be considered primordial bodies,
whose dynamical and physical properties can provide important clues
about the environment of planetary formation.

Several studies have addressed the dynamical properties of the Trojan
population. Of particular interest for the present work are the papers
by \citet{1993CeMDA..57...59M} and \citet{2001Icar..153..391B},
who computed proper elements for a large number of Jupiter Trojans
and realized the existence of several dynamical families. These authors
found that the families are mostly concentrated at the L4 swarm, and
they are much less conspicuous at the L5 swarm.

On the other hand, spectrophotometry has been used by different authors
to provide information about the surface physical properties of the
Jupiter Trojans. \citet{1985Icar...61..355Z} provided the first multiband
photometric observations of 21 of these objects. This allowed to classified
them within the D and P taxonomic classes (\citealp{1989aste.conf.1139T}),
with a significant predominance of the D class (about 90\% of bodies).
\citet{1990AJ....100..933J} obtained spectra in the visible range
of 32 Trojans and concluded that they show significant analogies with
the spectra of cometary nuclei. Eighteen Trojan asteroids have been
observed in the framework of the two major spectroscopic surveys:
the SMASS (\citealp{1995Icar..115....1X}; \citealp{2002Icar..158..146B})
and the S3OS2 (\citealp{2004Icar..172..179L}). Analyzing data form
this latter survey, \citet{2003Icar..161..356C} concluded that Trojan
spectra show differences in spectral slope with respect to the population
of D type asteroids in the main belt. Specific surveys have also contributed
to increase the spectral data of Jupiter Trojans in the visible. In
particular, \citet{2004Icar..168..374B} observed 34 Jupiter Trojans,
while \citet{2004Icar..172..221F} studied 26 asteroids, most of them
members of asteroid families detected in the L5 swarm. Spectra in
the near infrared (NIR) have been obtained by \citet{1994Icar..109..133L},
and more recently by \citet{2006Icar..183..420D} and by \citet{2007Icar..190..622F}
who provide photometric colors and spectra both in the visible and
NIR of 54 asteroids that are members of the main asteroid families
in L4 and L5. Surface mineralogy base on NIR spectra has been analyzed
by \citet{2003Icar..164..104E}, and recently by \citet{2007AJ.....134.223Y}
who addressed the presence of water ice on the Trojan surfaces. A
very complete analysis of the properties of Jupiter Trojans observed
by the Sloan Digital Sky Survey (SDSS) has been developed by \citet{2007MNRAS.377.1393S},
who addressed an interesting correlation between colors and orbital
inclinations. All these studies indicate that Jupiter Trojans seems
to be a quite homogeneous population in terms of taxonomy and surface
mineralogy.

In spite of these works, the amount of spectroscopic data of Jupiter
Trojans presently available is still small to allow a statistical
analysis of the taxonomic properties of these bodies. Moreover, since
spectra come from different sources they do not constitute a homogeneous
data sample. In this paper, we analyze the taxonomy of Jupiter Trojans
using data contained in the 3rd release of the SDSS Moving Objects
Catalog (MOC3), and compare the results to the available spectroscopic
data, \textit{with particular emphasis on Trojan asteroid families}.
The SDSS-MOC3 colors have proved to be a very useful tool to characterize
the taxonomy of Main Belt asteroids, as recently addressed by \citet{2006Icar..183..411R},
Binzel et al. (\citeyear{2006DPS....38.7106B}, \citeyear{2007LPI....38.1851B}),
\citet{2007arXiv0704.0230D}, \citet{2007arXiv0707.1012R}, \citet{2007Icar...in.press},
and \citet{2007AAS...210.8907H}. The paper is organized as follows:
Section \ref{sec:Selection-of-the} introduces the two data samples
used in this study and compares their internal accuracy. Section \ref{sec:Global-distribution-of}
is devoted to the global analysis of the color and taxonomy distributions
of the data samples. Section \ref{sec:Distribution-of-spectral} concentrates
on the particular analysis of selected asteroid families. Finally,
Section \ref{sec:Conclusions} contains the conclusions.

\section{Selection of the data samples\label{sec:Selection-of-the}}

In this work, we will analyze two different data sets containing information
on Trojan asteroids taxonomy. They are described in the following.

\subsection{The Sloan sample}

The first data set is constituted by observations from the SDSS-MOC3
and their selection required some care. The SDSS-MOC3 includes photometric
measurements of more than 204\,000 moving objects, of which only 67\,637
observations have been effectively linked to 43\,424 unique known asteroids.
The observations consist of calibrated magnitudes in the $u,g,r,i,z$
system of filters, centered at 3540, 4770, 6230, 7630 and 9130 $\textrm{\AA}$,
respectively, and with bandwidths $\sim100$ $\textrm{\AA}$ \citep{1996AJ....111.1748F}.
We adopted here a procedure similar to that of \citet{2006Icar..183..411R}.
First, we used the solar colors provided by \citet{2001AJ....122.2749I}
to compute the reflectance fluxes $F_{\nu}$ in the five bands, normalized
to 1 at the $r$ band. Then, we discarded the observations with error
$>10$\% in any of the $F_{g},F_{r,}F_{i}$ and $F_{z}$ fluxes. Observations
showing anomalous values of the fluxes, like $F_{u}>1.0$, $F_{g}>1.3$,
$F_{i}>1.5$, $F_{z}>1.7$, and $F_{g}<0.6$ were also discarded.
Note that the error in $F_{u}$ has not been constrained, which allows
to get a final data set with more than twice the amount of observations
than if we restrict this error to be less than 10\%. As we will explain
later, this error is not critical for our study.

\begin{figure}[t]
\centering 
\includegraphics[clip,width=1\columnwidth]{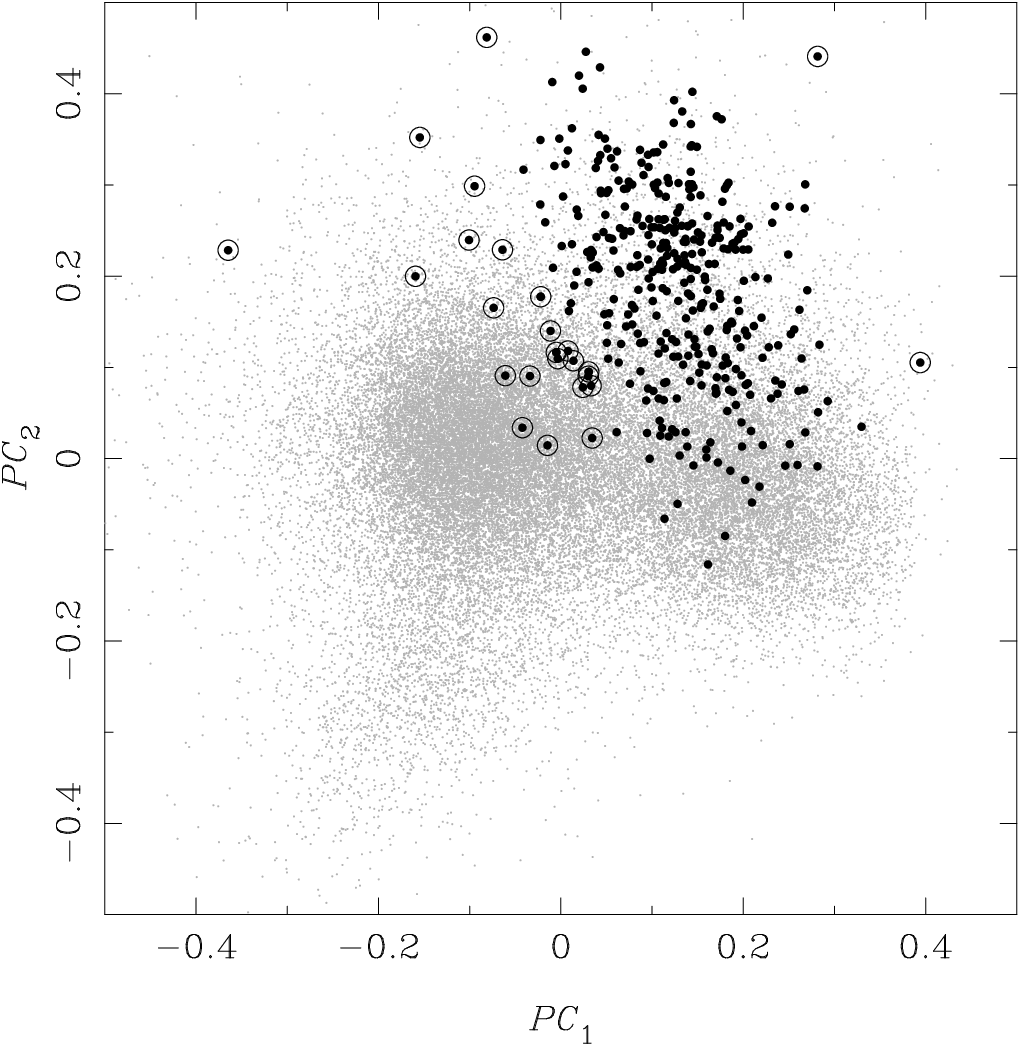} 
\caption{Distribution of 40\,863 observations selected from the SDSS-MOC3 (gray
dots) in the space of first and second principal components. The black
dots correspond to 371 observations of known Jupiter Trojans, but
those surrounded by a circle have been discarded (see text).}
\label{fig:1} 
\end{figure}

We ended up with a sample of 40\,863 observations corresponding to
28\,910 unique known asteroids. The distribution of these observations
in the space of principal components is shown in Fig. \ref{fig:1}
(gray dots), where the first and second principal components, $PC_{1}$
and $PC_{2}$, have been computed from the reflectance fluxes as:\begin{eqnarray*}
PC_{1} & = & 0.886F_{u}+0.416F_{g}-0.175F_{i}+0.099F_{z}-0.849\\
PC_{2} & = & -0.049F_{u}-0.003F_{g}+0.284F_{i}+0.957F_{z}-1.261\end{eqnarray*}
The use of principal components allows an easy interpretation of the
observations in a bidimensional space. Observations with $PC_{1}\gtrsim0$
correspond to featureless spectra (e.g. C-, X- and D-type asteroids),
while those with $PC_{1}\lesssim0$ correspond to featured spectra
that show a broad absorption band longwards of 7000 $\textrm{\AA}$
(e.g. S- and V-type asteroids). The value of $PC_{2}$ is related
to the overall slope of the spectrum, the larger the $PC_{2}$ the
higher the slope. For featureless spectra, $PC_{2}$ gives an idea
of how reddish is the spectrum; for featured spectra, it gives an
idea of the band depth (see \citealp{2006Icar..183..411R}).

Within these 40\,863 observations, we identified 371 observations corresponding
to 257 different Trojan asteroids listed in the database of Trojan
proper elements maintained by the PETrA Project (\citealp{2001Icar..153..391B};
\texttt{http://staff.on.br/froig/petra}). Their distribution in the
space of principal components is also shown in Fig. \ref{fig:1} (black
dots). Most of these observations of Trojan asteroids have values
of $PC_{1}\gtrsim0$ compatible with featureless spectra, and values
of $PC_{2}\gtrsim0$ indicating that they have moderate to high spectral
slopes. There are, however, some observations (circled dots in Fig.
\ref{fig:1}) that either depart significantly from the overall distribution
of other Trojan observations, or they clearly fall within the region
of featured spectra occupied by the S-type asteroids ($PC_{1}\lesssim0.05$
and $PC_{2}\lesssim0.2$). Direct inspection of the reflectance fluxes
indicates that these observations are not compatible with featureless
spectra or they show anomalous fluctuations, therefore we discarded
them as well. The final sample contains 349 observations corresponding
to 250 unique known Trojan asteroids. Hereafter we will refer to this
sample as the \textit{Sloan sample}. The Sloan sample includes 200
observations of asteroids in the L4 swarm and 149 observations of
the L5 swarm. About 40\% of these observations correspond to asteroid
family members.

It is worth mentioning that the main goal of our selection method
is that it provides a sample of good quality observations from the
SDSS-MOC3 that can be easily linked to family and to background (i.e.
non family) asteroids. Our approach is different from that introduced
by \citealp{2007MNRAS.377.1393S}, who applied a kinematic criterion
to select \textit{candidate} Trojan asteroids within the SDSS-MOC3.
These authors got a much larger sample of 1,187 observations, but
these observations cannot be separated in those corresponding to family
and to non family asteroids.

Each observation in our Sloan sample has been characterized by its
equivalent spectral slope $S$, in $\textrm{\AA}^{-1}$. The slope
was computed from a least-squares fit to a straight line passing through
the fluxes $F_{g},F_{r},F_{i}$ and $F_{z}$. This fit takes into
account the individual errors of the fluxes to estimate the slope
and its error $\Delta S$. Hereafter, we will refer to this set of
349 slopes as the \textit{Sloan slopes}. Note that the flux $F_{u}$
has not been used to compute the spectral slope. The reason for this
it twofold: (i) we know, from spectroscopic observations, that the
reflectance flux in the $u$ band usually drops off and significantly
deviates from the linear trend of the spectrum; (ii) we intend to
compare the Sloan sample to a sample of spectroscopic data, described
below, where most spectra do not cover the wavelengths $\lesssim5000$
$\textrm{\AA}$. Since $F_{u}$ does not contribute effectively to
determine the slope, there is no harm in keeping its error unconstrained
as we did.

Table \ref{tab:1} provides the list of all the know Trojan asteroids
contained in our Sloan sample. This table also gives the estimated
spectral slope, $S$, with its correspondig error, $\Delta S$,
and the number of observations, $N_{\mathrm{obs}}$,
in the sample. For $N_{\mathrm{obs}}\geq2$, the slope given in this
table is the weighted mean of the individual Sloan slopes, with the
weights defined as $1/\left(\Delta S\right)^{2}$.

\subsection{The Spectroscopic sample}

The second data set analyzed here is a collection of 91 spectra corresponding
to 74 individual Trojan asteroids published in the literature. All
the spectra are defined in the visible wavelength range and have been
obtained by different observational surveys, in particular: 3 spectra
come from the SMASS1 survey \citep{1995Icar..115....1X}, 2 spectra
from the SMASS2 survey\citep{2002Icar..158..146B}, 33 spectra from
\citet{2004Icar..168..374B}, 25 spectra from \citet{2004Icar..172..221F},
13 spectra from the S3OS2 survey \citep{2004Icar..172..179L}, and
15 spectra from \citet{2006Icar..183..420D}. Hereafter we will refer
to this data set as the \textit{Spectroscopic sample}. This sample
includes 52 spectra of asteroids in the L4 swarm and 39 spectra of
the L5 swarm. About 40\% of these spectra correspond to asteroid family
members.

To determine the spectral slope, each spectrum in the sample was first
rebinned by applying a 20 $\textrm{\AA}$ running box average. The
rebinned spectrum was then normalized to 1 at 6230 $\textrm{\AA}$
to make it comparable to the Sloan fluxes. Finally, the slope of the
normalized spectrum was computed from a least-squares fit to a straight
line in the interval 5000-9200 $\textrm{\AA}$. This wavelength interval
is similar to the one adopted to compute the Sloan slopes, and it
is well covered by most spectra in our sample except for a few cases
for which the fit had to be done over a smaller available range. Following
the same approach as \citet{2007Icar..190..622F}, the error in the
slope was estimated by adding an ``ad-hoc'' error of $\pm5\times10^{-6}\,\textrm{\AA}^{-1}$
to the standard error of the fit. The idea is to account for uncertainties
in the sample related to use of data obtained by different surveys.
It is worth noting that the slopes computed here are not compatible
with other published slopes (e.g. \citealp{1990AJ....100..933J},
\citealp{2007Icar..190..622F}) due to different normalization wavelengths
--usually 5500 $\textrm{\AA}$-- and also due to different wavelenghts
intervals used to fit the data. In fact, our slopes may be up to 20\%
smaller than those published in the literature. Hereafter, we will
refer to our set of 91 slopes as the \textit{Spectroscopic slopes},
to distinguish them from the Sloan slopes.

Table \ref{tab:2} provides the list of all the know Trojan asteroids
contained in our Spectroscopic sample. For asteroids with $N_{\mathrm{obs}}\geq2$,
the slope shown in this table has been computed as the weighted mean
of the individual Spectroscopic slopes. It is worth recalling that
in Table \ref{tab:1} an asteroid with $N_{\mathrm{obs}}\geq2$ had
all its observations made by the same survey, i.e. the SDSS, while
in Table \ref{tab:2} an asteroid with $N_{\mathrm{obs}}\geq2$ had
its observations made by different spectroscopic surveys.

\subsection{Accuracy of the samples}

\begin{figure}[t]
\centering 
\includegraphics[clip,width=1\columnwidth]{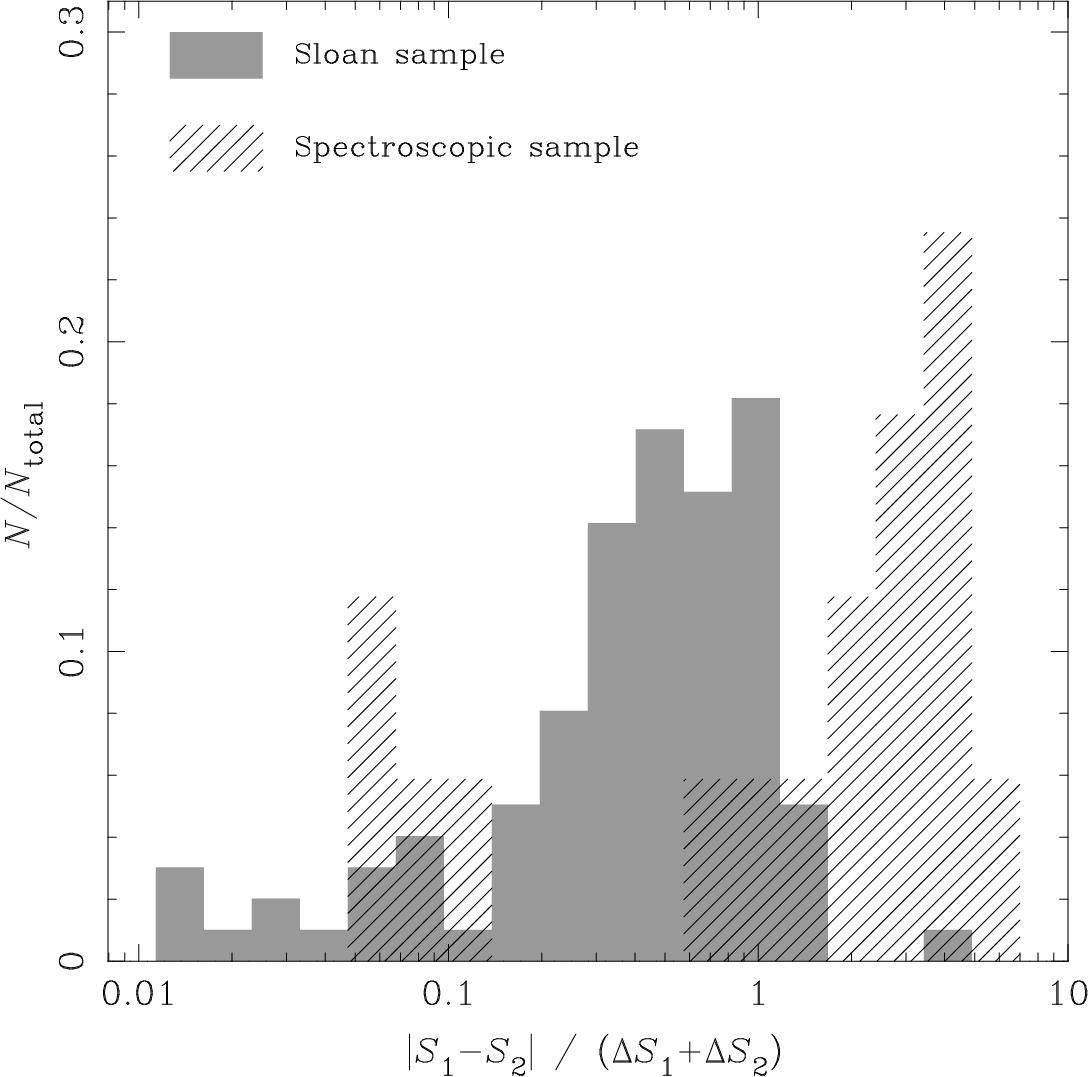} 
\caption{Distribution of the parameter $\epsilon$ (see text) for the Sloan
sample (gray histogram) and for the Spectroscopic sample (hatched
histogram). Each histogram has been normalized such that its area
is 1.}
\label{fig:2} 
\end{figure}

The Sloan sample is $\sim4$ times larger than the Spectroscopic sample,
which in terms of statistics does not appear to be a significant improvement.
However, the Sloan sample is expected to be more homogeneous than
the Spectroscopic sample because, in the former case, the observations
come from the same survey, while in the latter they came from different
surveys. Moreover, the spectroscopic surveys have been usually dedicated
either to observe family members only (e.g. \citealp{2004Icar..172..221F},
\citealp{2006Icar..183..420D}, \citealp{2007Icar..190..622F}) or
to observe background asteroids only (e.g. \citealp{2004Icar..172..179L},
\citealp{2004Icar..168..374B}). But the Sloan sample includes both
family members and background asteroids observed by the same survey.
The Sloan sample is also expected to include a significant amount
of very small Trojans that spectroscopic surveys normally do not observe.
Although the SDSS photometry is not as precise as spectroscopy, this
is not crucial in the case of the Trojan asteroids because they all
show featureless spectra that are properly characterized by the average
spectral slope.

\begin{figure*}[t]
\centering
\includegraphics[clip,width=0.47\textwidth]{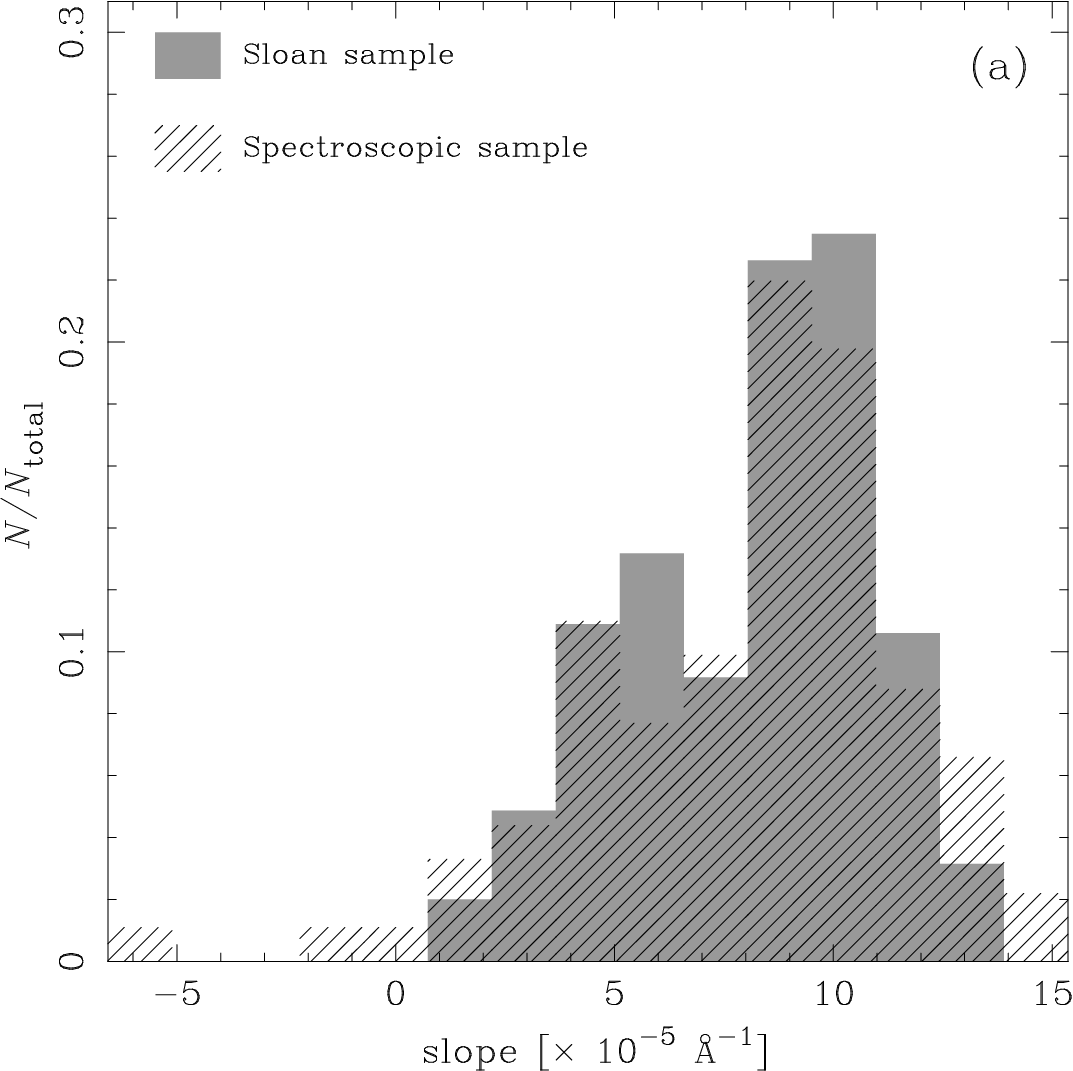}
~~~~~~~
\includegraphics[clip,width=0.47\textwidth]{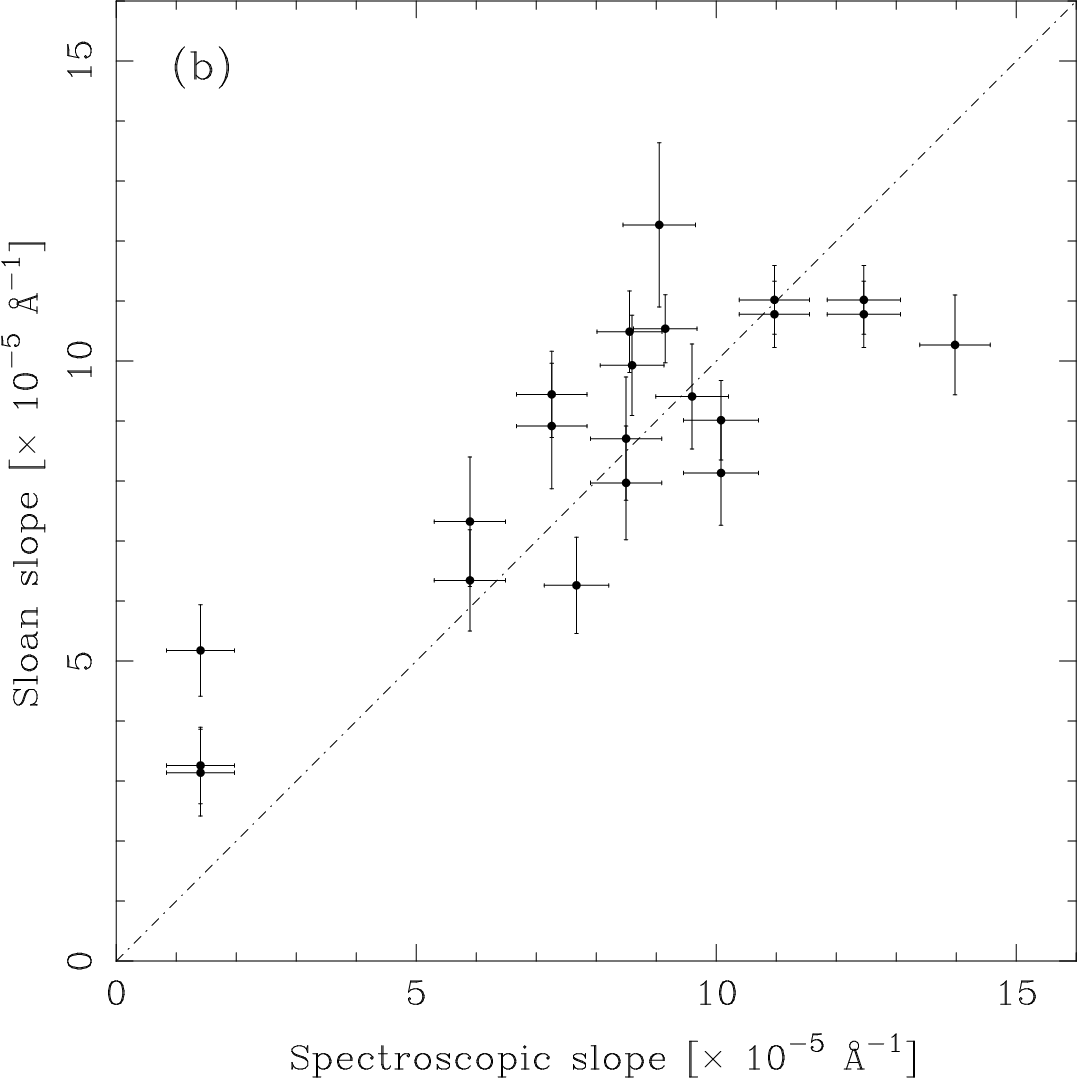} 
\caption{\textbf{(a)} Distribution of spectral slopes from the Sloan sample
(349 observations, gray histogram) and from the Spectroscopic sample
(91 observations, hatched histogram). Each histogram has been normalized
such that its area is 1. Both distributions show a bimodality related
to the presence of two taxonomic types: P-type (smaller slopes) and
D-type (larger slopes). \textbf{(b)} Comparison between the Spectroscopic
slopes and the Sloan slopes from observations of asteroids included
in both samples}
\label{fig:3} 
\end{figure*}

In order to verify the reliability of the Sloan and the Spectroscopic
samples, we performed the following test. For each asteroid with $N_{\mathrm{obs}}\geq2$
in Table \ref{tab:1} we computed the parameter\[
\epsilon=\frac{\left|S_{1}-S_{2}\right|}{\Delta S_{1}+\Delta S_{2}}\]
where $S_{i}$ are the spectral slopes of two different observations
of that asteroid. A value of $\epsilon<1$ indicates that this two
observations are self-consistent, since their differences are within
the individual errors. We can apply the same procedure to each asteroid
with $N_{\mathrm{obs}}\geq2$ in Table \ref{tab:2}, and compare the
results. In Fig. \ref{fig:2}, we show the distribution of $\epsilon$
values for the Sloan and Spectroscopic samples. The Sloan sample shows
a good self-consistency among the observations of each asteroid with
$N_{\mathrm{obs}}\geq2$. Since in most cases the individual errors
are $\sim10$\%, this result supports the idea of a quite homogeneous
sample.

On the other hand, a significant fraction of the Spectroscopic sample
shows differences among the observations of each asteroid with $N_{\mathrm{obs}}\geq2$
that are beyond their errors. This may be explained by different observational
conditions, different instrumental setup and different reduction processes
among the different surveys. The inconsistency could be minimized
by increasing the ``ad-hoc'' error introduced to estimate the slope
error, but this ``ad-hoc'' error is already of $\sim10$\%. Therefore,
the result shown in Fig. \ref{fig:2} supports the idea that the Spectroscopic
sample is less homogeneous than the Sloan sample, as expected.

\section{Global distribution of spectral slopes\label{sec:Global-distribution-of}}

In this section we analyze the distribution of spectral slopes of
the whole population of known Trojan asteroids included in our data
samples, with particular attention to the asteroid families. First,
we compare the Sloan and the Spectroscopic samples, and then we discuss
each sample separately.

\subsection{Comparisons between the samples\label{sub:Comparisons-between-the}}

In Fig. \ref{fig:3} a we show the distribution of Sloan slopes (349
observations) compared to the distribution of Spectroscopic slopes
(91 observations). Both distributions show a clear bimodality, that
is more evident in the Sloan sample. This bimodality is related to
the presence of two different taxonomic types among the Jupiter Trojans:
(i) the D-type, with spectral slopes $S\gtrsim7.5\times10^{-5}$ $\textrm{\AA}^{-1}$,
that correspond to redder surfaces, and (ii) the P-type, with slopes
$1.5\lesssim S\lesssim7.5\times10^{-5}$ $\textrm{\AA}^{-1}$, that
correspond to less reddish colors. There is also a small amount of
observations compatible with the C-type taxonomy, with slopes $S\lesssim1.5\times10^{-5}$
$\textrm{\AA}^{-1}$, that correspond to more neutral colors. The
limiting slopes between the taxonomic classes are estimated within
a $\pm0.7\times10^{-5}$ $\textrm{\AA}^{-1}$ interval of tolerance,
which is the approximate bin size in Fig. \ref{fig:3}. It is worth
stressing that these limiting slopes are \textit{totally arbitrary}.
Moreover, they are not compatible with the values adopted by the usual
taxonomies, where the separation between the P- and D-types occurs
at $S\sim5.5\times10^{-5}$ $\textrm{\AA}^{-1}$. Nevertheless, our
choice is based on the natural separation of the slopes induced by
the bimodality in their distribution and it is valid as far as no
mineralogical constraint is known to define the P and D taxonomic
types. 

\begin{figure*}[t]
\centering
\includegraphics[clip,width=0.47\textwidth]{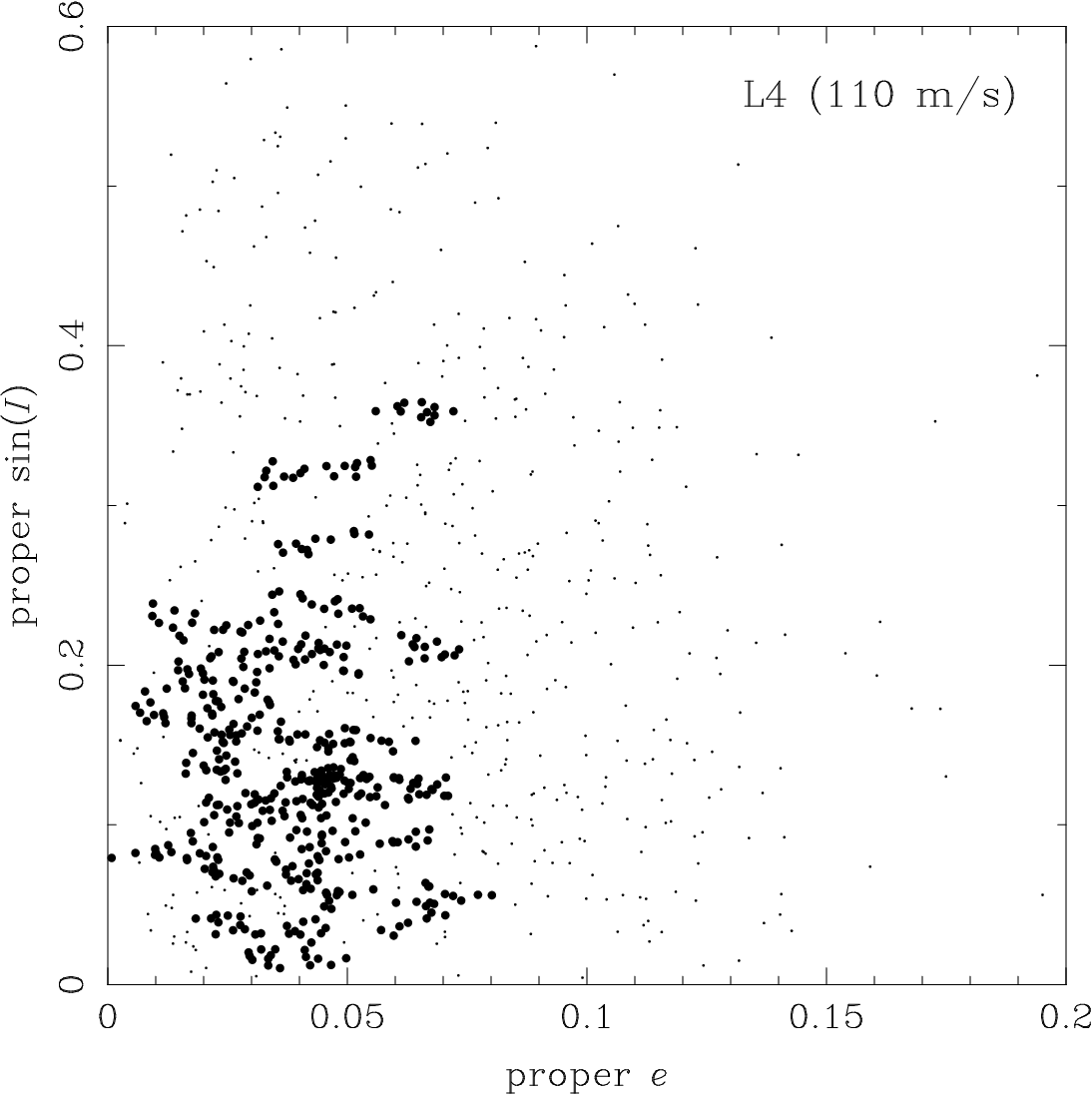}
~~~~~~~
\includegraphics[clip,width=0.47\textwidth]{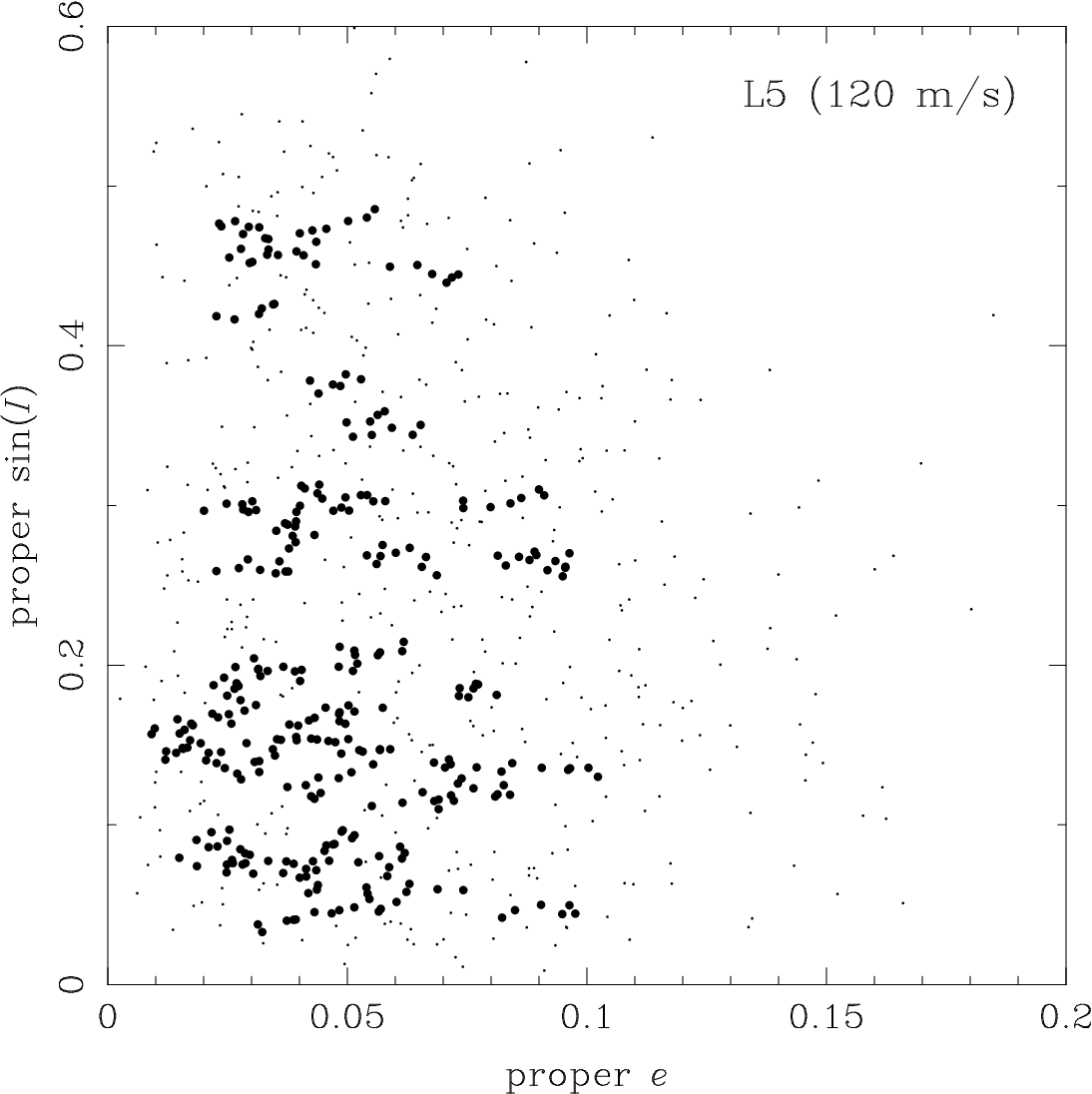} 
\caption{\textit{Left panel.} Asteroid families (big dots) identified in the
L4 swarm, projected in the space of proper eccentricity and inclination.
Background asteroids are represented by small dots. The cutoff level
is 110 $\mathrm{m\, s^{-1}}$. \textit{Right panel.} The same but
for the L5 swarm. The cutoff level is 120 $\mathrm{m\, s^{-1}}$.}
\label{fig:4} 
\end{figure*}

The D-type observations dominate over the P-type in the approximate
proportion $\frac{2}{3}:\frac{1}{3}$, but the Sloan sample shows
a larger abundance of P-type relative to D-type than the Spectroscopic
sample. The bimodality observed in Fig. \ref{fig:3}a has also been
reported by \citet{2007MNRAS.377.1393S} analyzing SDSS-MOC3 colors%
\footnote{Indeed, the principal color $t_{\mathrm{c}}^{*}$ introduced by these
authors is strongly correlated to both the second principal component
$PC_{2}$ and the spectral slope $S$.%
}. The Sloan slopes appear more tightly clustered than the Spectroscopic
slopes, which might be related to the less homogeneity of the Spectroscopic
sample. Nevertheless, the Sloan slopes appear well correlated to the
Spectroscopic slopes, as shown in Fig. \ref{fig:3}b for the few observations
corresponding to asteroids included in both samples.

To analyze the distribution of spectral slopes of the asteroid families,
we first proceeded to identify the different families in each Trojan
swarm. We used the catalog of 1702 Trojan asteroids with known resonant
proper elements maintained by the PETrA Project \citep{2001Icar..153..391B}
and applied to this catalog the Hierarchical Clustering Method (HCM,
\citet{1995Icar..116..291Z}). The mutual distance between any pair
of asteroids in the proper elements space was computed according to
the metric\[
d=\left[\frac{1}{4}\left(\frac{\delta a}{a_{0}}\right)^{2}+2\left(\delta e\right)^{2}+2\left(\delta\sin I\right)^{2}\right]^{1/2}\]
\citep{1993CeMDA..57...59M}, where $\delta a$, $\delta e$ and $\delta\sin I$
are the differences in proper semi-major axis, proper eccentricity
and proper sinus of inclination, respectively, between the given pair
of asteroids, and $a_{0}=5.2026$ AU is the average proper semi-major
axis of the Trojan population. Those bodies for which $d\leq d_{\mathrm{cut}}$
were clustered together to form the families. The cutoff value $d_{\mathrm{cut}}$
was chosen to be 110 $\mathrm{m\, s^{-1}}$ for the L4 swarm and 120
$\mathrm{m\, s^{-1}}$ for the L5 swarm, which are comparable to the
corresponding quasi-random level of each swarm%
\footnote{The quasi-random level is the maximum level of statistical significance
of the HCM.%
}. We have verified that values of $d_{\mathrm{cut}}$ within $\pm10$
$\mathrm{m\, s^{-1}}$ around the above values produce practically
the same results. Clusters with less than 8 members in the L4 swarm
and with less than 6 members in the L5 swarm were considered statistical
fluctuations and were disregarded. For a detailed explanation on the
definition of $d_{\mathrm{cut}}$ and the application of the HCM to
the Trojan case refer to \citet{2001Icar..153..391B}. The distribution
of the detected families in the space of proper eccentricity and inclination
is shown in Fig. \ref{fig:4}.

\begin{figure*}[t]
\centering
\includegraphics[clip,width=0.47\textwidth]{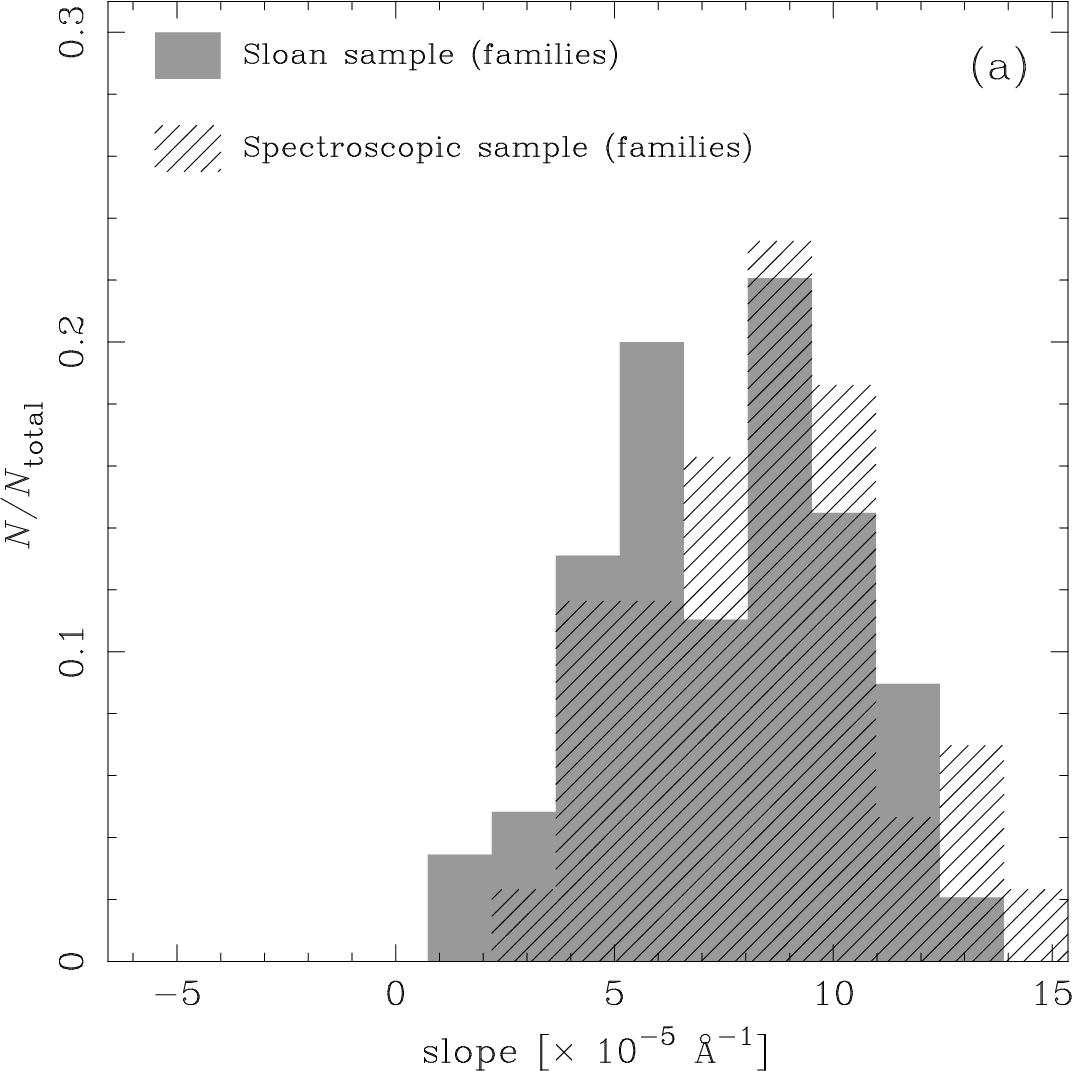}
~~~~~~~
\includegraphics[clip,width=0.47\textwidth]{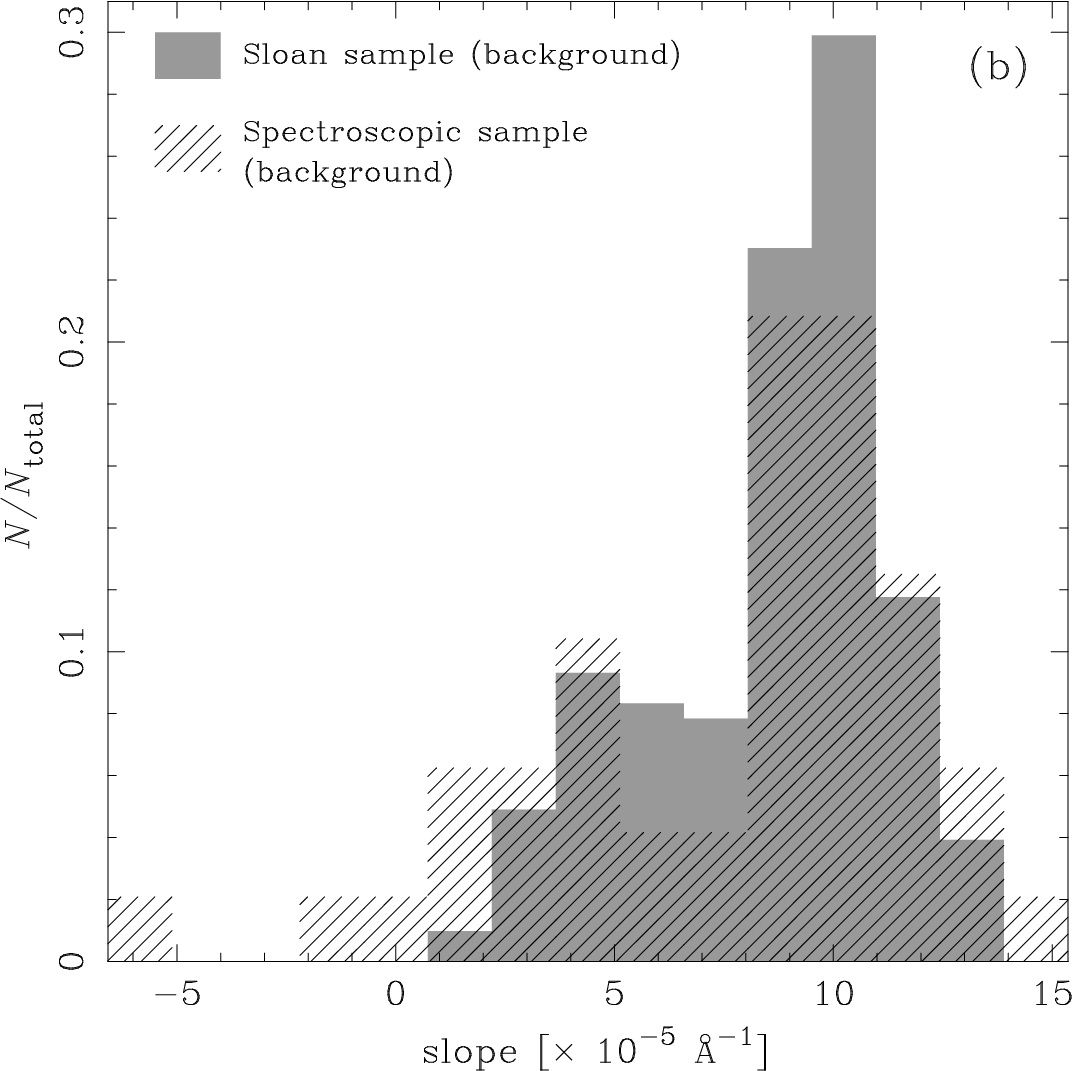} 
\caption{\textbf{(a)} Distribution of spectral slopes of observations corresponding
to family members. The gray histogram correspond to the Sloan slopes
and the hatched histogram to the Spectroscopic slopes. \textbf{(b)}
The same but for the observations corresponding to background asteroids.
Each histogram has been normalized such that its area is 1.}
\label{fig:5} 
\end{figure*}

In Fig. \ref{fig:5}, we show the slope distribution of family members
(panel a) compared to the background asteroids (panel b). The gray
histograms correspond to the Sloan sample, while the hatched histograms
correspond to the Spectroscopic sample. Except for a few background
asteroids with very small and even negative slopes (these latter observed
by \citet{2004Icar..168..374B}), the distribution of Sloan slopes
of background asteroids is comparable to the distribution of Spectroscopic
slopes, both showing a clear abundance of D-type asteroids. The situation
is quite different for the family members, since the Sloan sample
shows a clear bimodality that is not observed in the Spectroscopic
sample. It is worth noting that the Sloan sample contains about 3
times more observations of family members and goes much deeper in
absolute magnitude than the Spectroscopic sample. This result may
indicate that family members contribute with a significant amount
of the P-type asteroids found among the Trojan swarms. It also indicates
that families appear to be bluer than the background.

\subsection{Global analysis of the Sloan sample\label{sub:Global-analysis-of}}

\begin{figure*}[t]
\centering
\includegraphics[clip,width=0.47\textwidth]{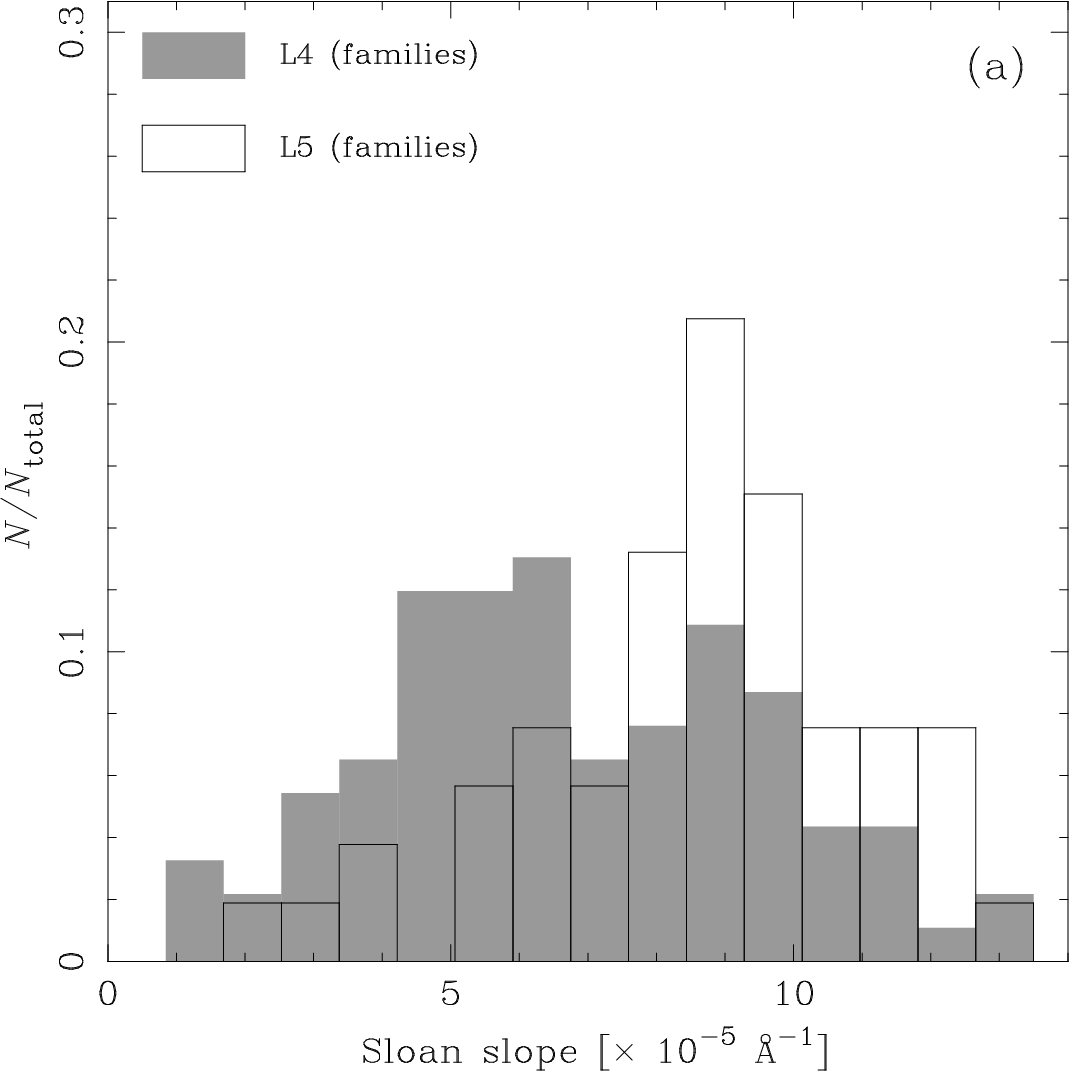}
~~~~~~~
\includegraphics[clip,width=0.47\textwidth]{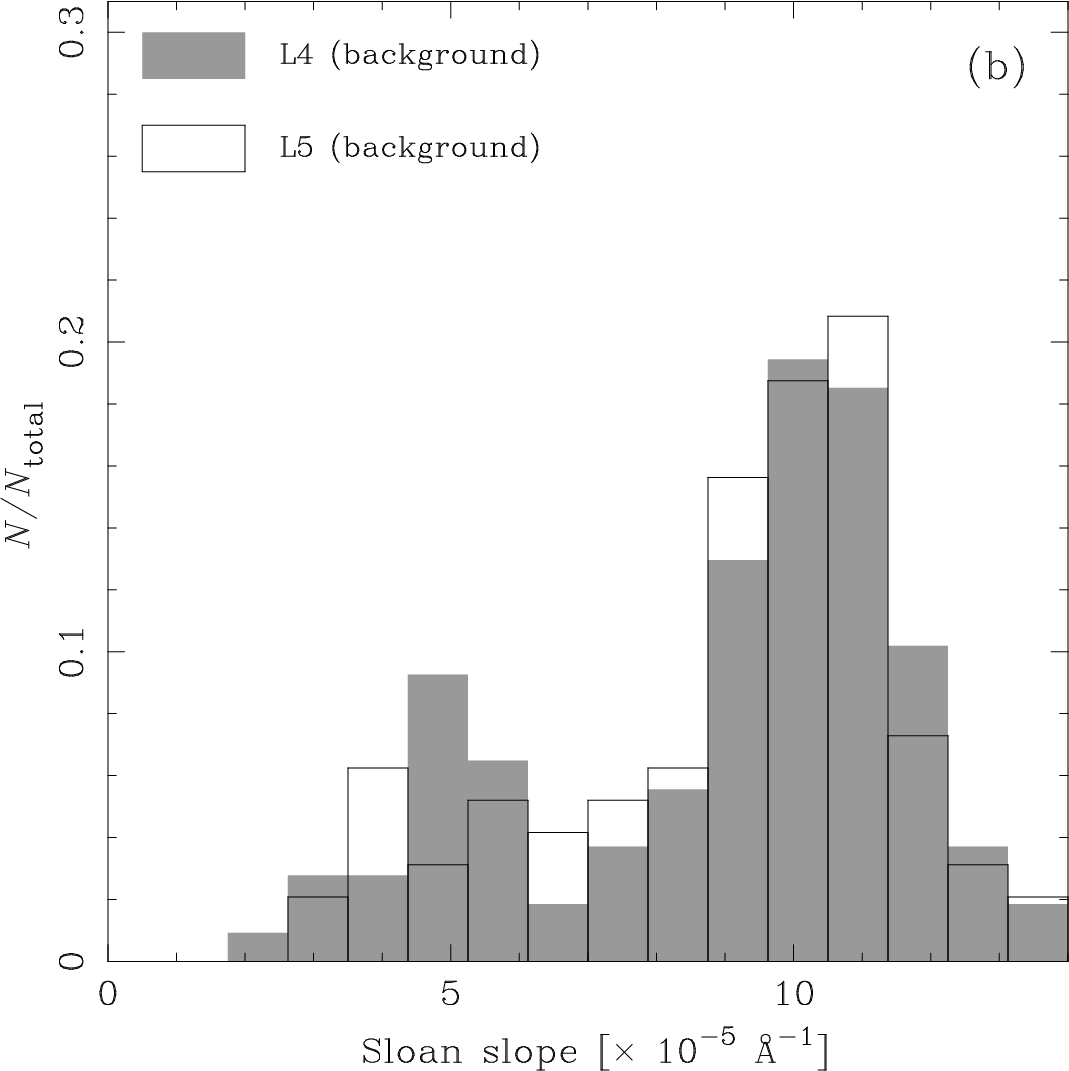}
\par\bigskip
\includegraphics[clip,width=0.47\textwidth]{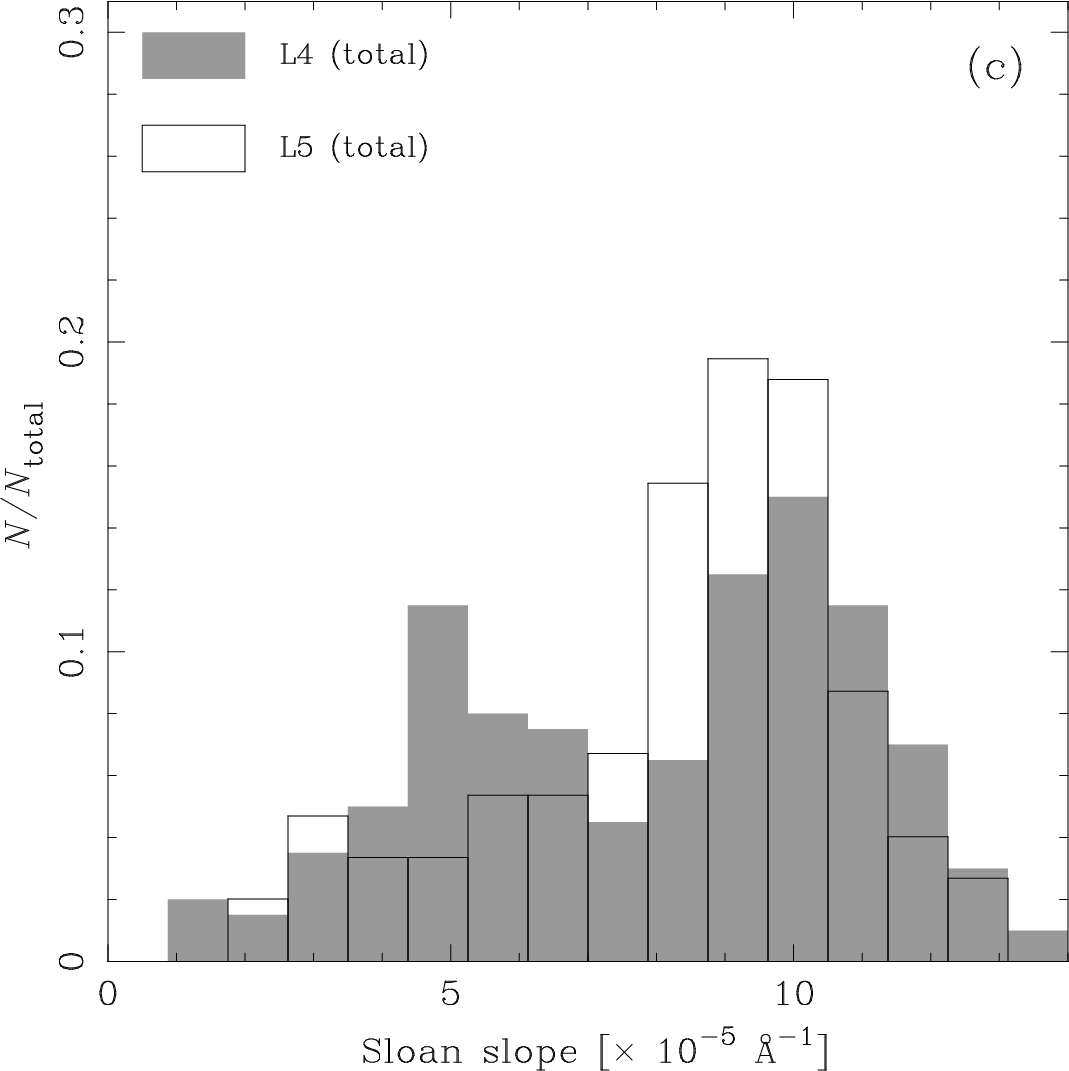}
\caption{\textbf{(a)} Distribution of the Sloan slopes of family members only.
The gray histogram correspond to the L4 swarm and the outlined histogram
to the L5 swarm. \textbf{(b)} Same as (a) but for the background asteroids
only. \textbf{(c)} Same as (a) but for both family members and background
asteroids together. Each histogram has been normalized such that its
area is 1.}
\label{fig:6} 
\end{figure*}

It is well known that asteroid families do not appear equally distributed
among the L4 and L5 swarms. While the families in L4 are more conspicuous
and tend to form large clusters, the families in L5 are smaller and
tighter. The total number of families is also larger in L4 than in
L5. Thus, it is interesting to analyze the distribution of Sloan slopes
separately in each swarm. Figure \ref{fig:6}a shows the distribution
of Sloan slopes of family members in the L4 swarm (gray histogram)
and in the L5 swarm (outlined histogram). The difference between the
swarms is notorious. While in L4 the Sloan slopes show a predominance
of P-type asteroids among the families, the L5 families appear dominated
by D-type asteroids. The families in L5 are significantly redder than
those in L4. On the other hand, the slope distribution of background
asteroids, shown in Fig. \ref{fig:6}b, is almost the same in the
two swarms, with a significant peak of D-type asteroids.

The behavior observed in Fig. \ref{fig:6}a,b may explain the different
color distributions between the L4 and L5 swarms reported by \citet{2007MNRAS.377.1393S}
from the analysis of SDSS-MOC3 colors. These authors pointed out that
the amount of redder asteroids (higher slopes), relative to the bluer
ones (smaller slopes), is much larger in the L5 swarm than in the
L4 swarm. The situation is clearly illustrated in Fig. \ref{fig:6}c.
They explained this difference on the basis of an observational selection
effect that causes to detect more asteroids with high orbital inclination,
relative to those with low orbital inclination, in the L5 swarm compared
to the L4 swarm. Since there is a clear correlation between color
and orbital inclination, such that the bluer bodies have low inclinations
while the redder ones are predominantly found at high inclinations,
and since this correlation appears to be the same in both swarms,
\citet{2007MNRAS.377.1393S} conclude that it is natural to find a
large fraction of redder bodies in the L5 swarm. The authors tried
to overcome the observational selection effect by separating their
observations in those corresponding to high inclination asteroids
($>10^{\circ}$) and those corresponding to low inclination bodies
($<10^{\circ}$), and showing that, with this separation, the differences
between L4 and L5 almost disappear.

\begin{figure*}[t]
\centering
\includegraphics[clip,width=0.47\textwidth]{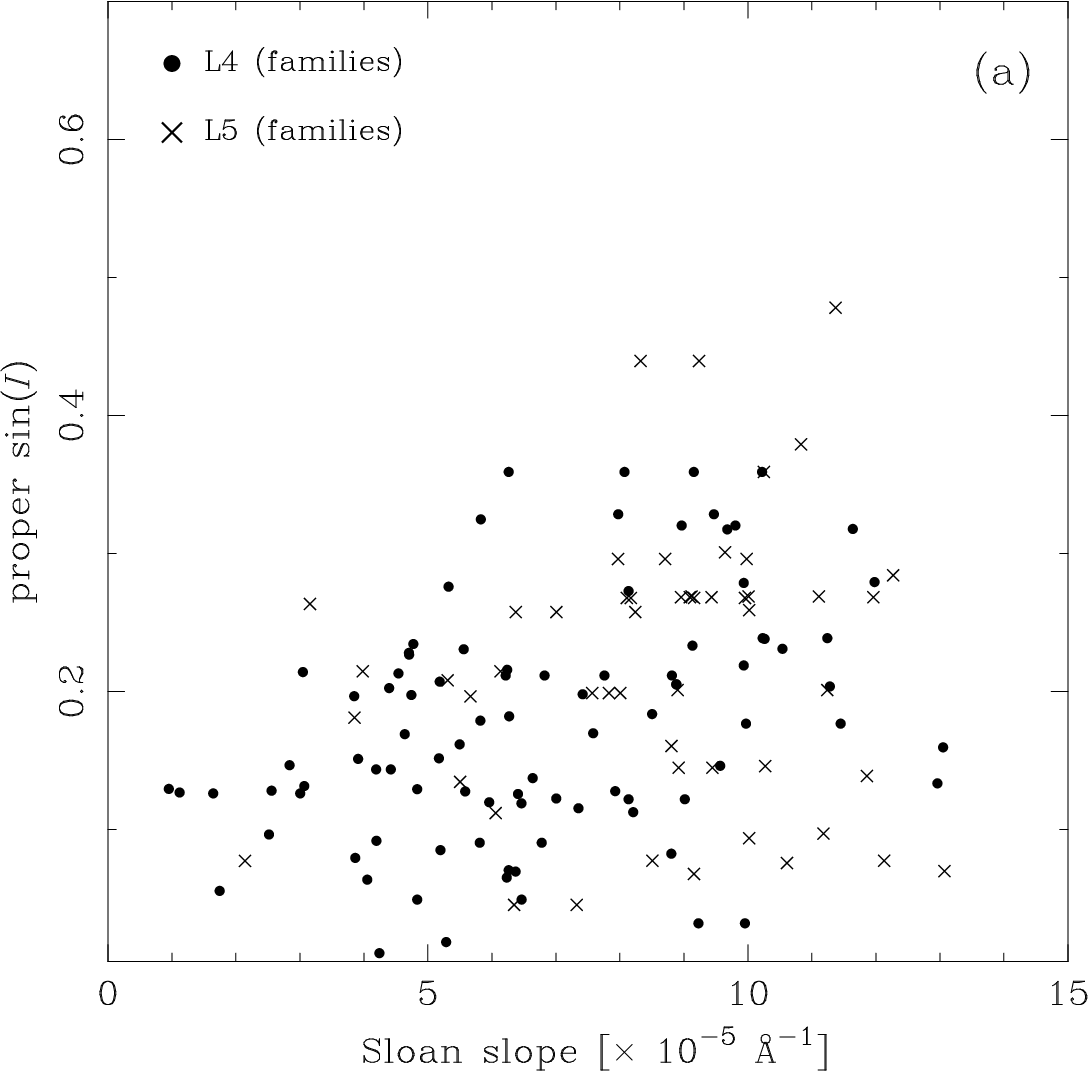}
~~~~~~~
\includegraphics[clip,width=0.47\textwidth]{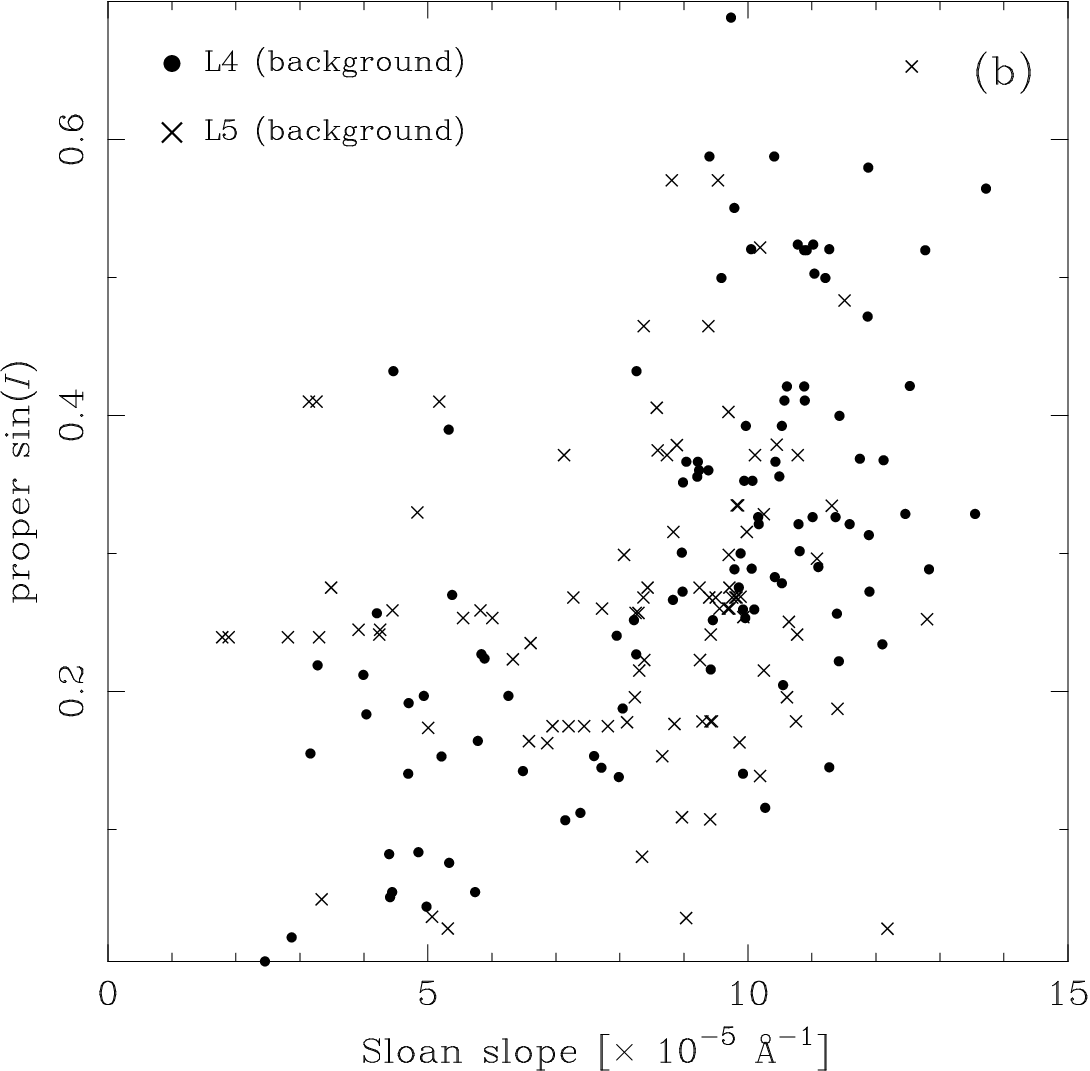} 
\caption{\textbf{(a) }Distribution of the Sloan slopes of family members as
a function of proper inclination. Dots correspond to the L4 swarm
and crosses to the L5 swarm. \textbf{(b) }Same as (a) but for the
background asteroids. Note the significant lack of high-inclination
background asteroids with small slopes (P-types).}
\label{fig:7} 
\end{figure*}

We believe, however, that a separation in terms of asteroid families
and background asteroids, instead of orbital inclinations, provides
a much better explanation, since it is clear from Fig. \ref{fig:6}
that the swarms differ in their color distributions due to the presence
of the asteroid families. The advantage of this scenario is that it
has a physical basis and does not require to invoke any strange observational
bias. It is also interesting to analyze the color-inclination correlation
in terms of asteroid families. In Fig. \ref{fig:7} we show the distribution
of asteroid family members (panel a) and background asteroids (panel
b) in the plane of spectral slope vs. orbital inclination. Dots and
crosses represent the L4 and L5 observations, respectively. The family
members do not show any apparent correlation between color and inclination,
contrary to the background that it is strongly correlated. This correlation
appears to be the same in both swarms, as \citet{2007MNRAS.377.1393S}
conjectured.

We must note that the separation in low and high inclination populations
proposed by \citet{2007MNRAS.377.1393S} partially works to explain
the different color distributions between L4 and L5 because the family
members are not uniformly distributed in terms of proper inclination.
In fact, the families in the L4 swarm are mostly concentrated at low
inclinations while the families in L5 spread over a wider range of
proper inclinations, as we can see in Fig. \ref{fig:4}. If we consider
only the high inclination asteroids ($\sin I\gtrsim0.2$), then the
L4 swarm is dominated by background asteroids (Fig. \ref{fig:4})
which are predominantly red (Fig. \ref{fig:7}b). The L5 swarm has
a larger proportion of asteroid families at high inclinations (Fig.
\ref{fig:4}), but these are also predominantly red (Fig. \ref{fig:6}a)
as the background. Thus, both swarms show the same color distribution
at large inclinations. On the other hand, if we consider the low inclination
asteroids ($\sin I\lesssim0.2$), the asteroid families significantly
contribute to the slope distribution. While the background tends to
be bluer (Fig. \ref{fig:7}b), the families cover a wider range of
colors (Fig. \ref{fig:6}a) and this tends to disguise the differences
in slope distribution between the swarms. This is precisely the result
found by \citet{2007MNRAS.377.1393S}.

Another interesting result concerns the correlation between spectral
slope and absolute magnitude (or size). Figure \ref{fig:8} is analogous
to Fig. \ref{fig:7}, but in terms of absolute magnitude instead of
orbital inclination. If we eliminate the few large bodies ($H\lesssim9$)
from the sample then the families (Fig. \ref{fig:8}a) do not show
any apparent correlation, but the background asteroids (Fig. \ref{fig:8}b)
shows a weak correlation since bodies in the range $9\lesssim H\lesssim11$
are predominantly red. Note that, if we consider together the families
and the background, the slope-size correlation is disguised and this
is probably the reason why \citet{2007MNRAS.377.1393S} did not detected
this correlation in their analysis. 

The results of Figs. \ref{fig:7}b and \ref{fig:8}b led to conclude
that large background asteroids in both Trojan swarms tend to be redder
and tend to be located at large orbital inclinations.

\subsection{Global analysis of the Spectroscopic sample\label{sub:Global-analysis-spec}}

The behavior observed in Fig. \ref{fig:6} is not reproduced among
the Spectroscopic sample. This is probably due to the fact that our
Spectroscopic sample is deficient in observations of family members
in the L4 swarm. In fact, the actual population in the L4 swarm is
at least $\sim1.6$ times larger than in the L5 swarm, but our Spectroscopic
sample includes about the same amount of family members in both swarms.

Besides, neither the correlation between slope and inclination, nor
the one between slope and size are observed in the Spectroscopic sample.
This may be due to the smaller number of objects contained in the
sample and to its less homogeneity. It is interesting to note that
the background of the Spectroscopic sample includes some large P-type
asteroids and also some high-inclination P-type asteroids that are
not observed in the Sloan sample. 

An analysis of a much larger data set of spectroscopic data has been
performed by \citet{2007Icar..190..622F}. Their sample includes the
same observations that we include in our Spectroscopic sample plus
other spectroscopic observations from \citet{1990AJ....100..933J},
\citet{1994A&A...282..634F} and from themselves, totalizing 142 different
Trojan asteroids. They found a situation very similar to the one observed
in Fig. \ref{fig:6}c, indicating that the L4 swarm has a larger fraction
of P-type asteroids, relative to D-type, compared to the L5 swarm.
\citet{2007Icar..190..622F} did not find any slope-size correlation,
although they detected that the distribution of spectral slopes is
narrower at large sizes. In fact, from figure 9 of their paper, it
is possible to infer a slight predominance of D-type asteroids among
the large asteroids ($50\lesssim D\lesssim120$ km). This situation
would be similar to the one obtained in the range ($9\lesssim H\lesssim11$)
by overlapping the two panels in Fig. \ref{fig:8}. An analysis of
the slope-size relation using the data in \citet{2007Icar..190..622F}
and separating asteroid families from background asteroids might help
to check whether the result shown in Fig. \ref{fig:8} correspond
to a real correlation or is just an artifact of our Sloan sample.

\subsection{Discussion}

The fact that only the background asteroid show correlations between
spectral slope, absolute magnitude and orbital inclination, and that
the correlations are similar in both the L4 and L5 swarms, may put
important constraints to the origin and evolution of Jupiter Trojans.
No dynamical mechanism among the Trojans is known to favor the evolution
of asteroids according to their size or to their surface physical
properties%
\footnote{The Yarkovsky effect, that depends on size and surface properties,
also depends on the Sun distance and it is negligible at 5 AU.%
}. Therefore, these correlations may have a primordial origin. Alternatively,
the correlations may be the by-product of collisional evolution. We
speculate here about two possible scenarios.

\begin{figure*}[t]
\centering
\includegraphics[clip,width=0.47\textwidth]{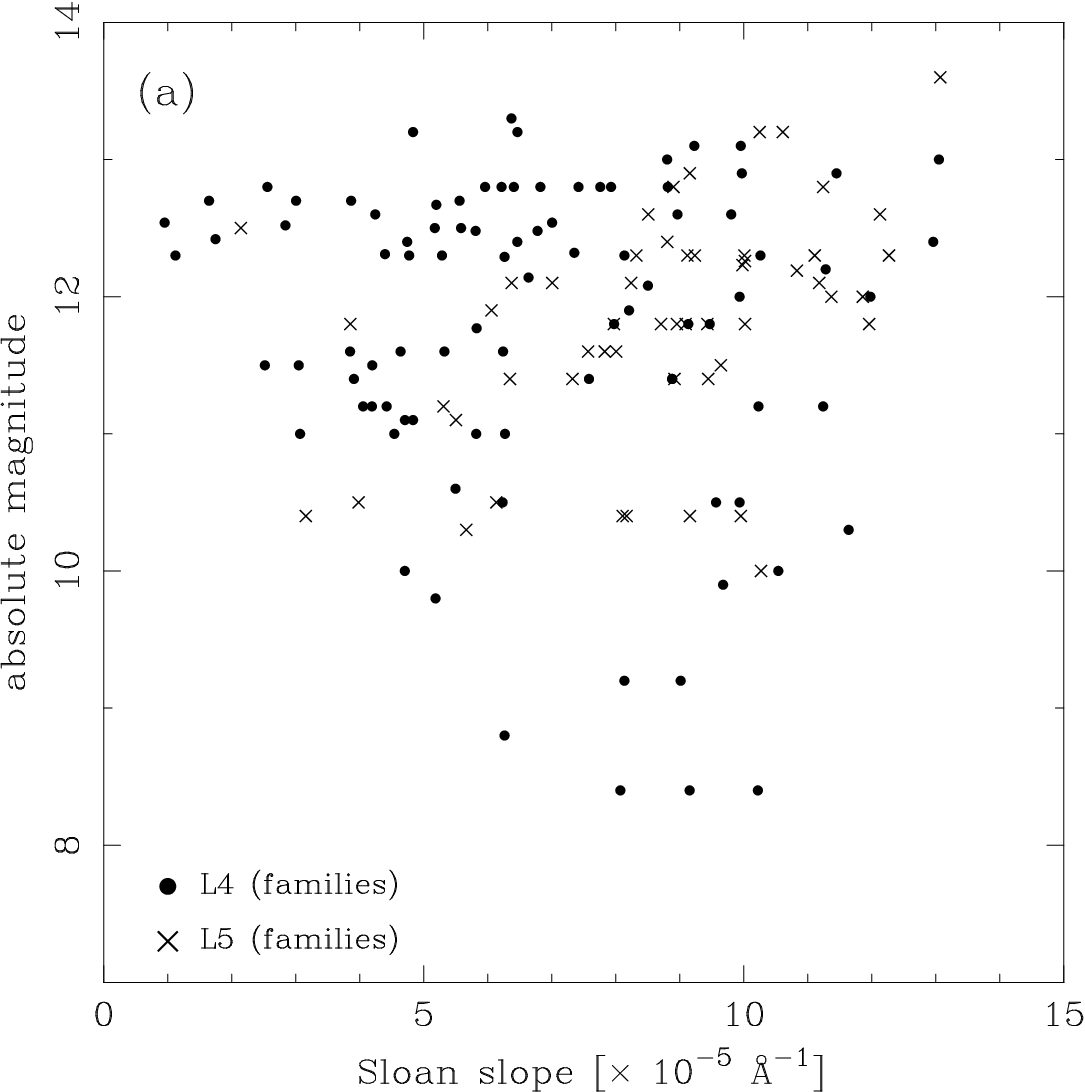}
~~~~~~~
\includegraphics[clip,width=0.47\textwidth]{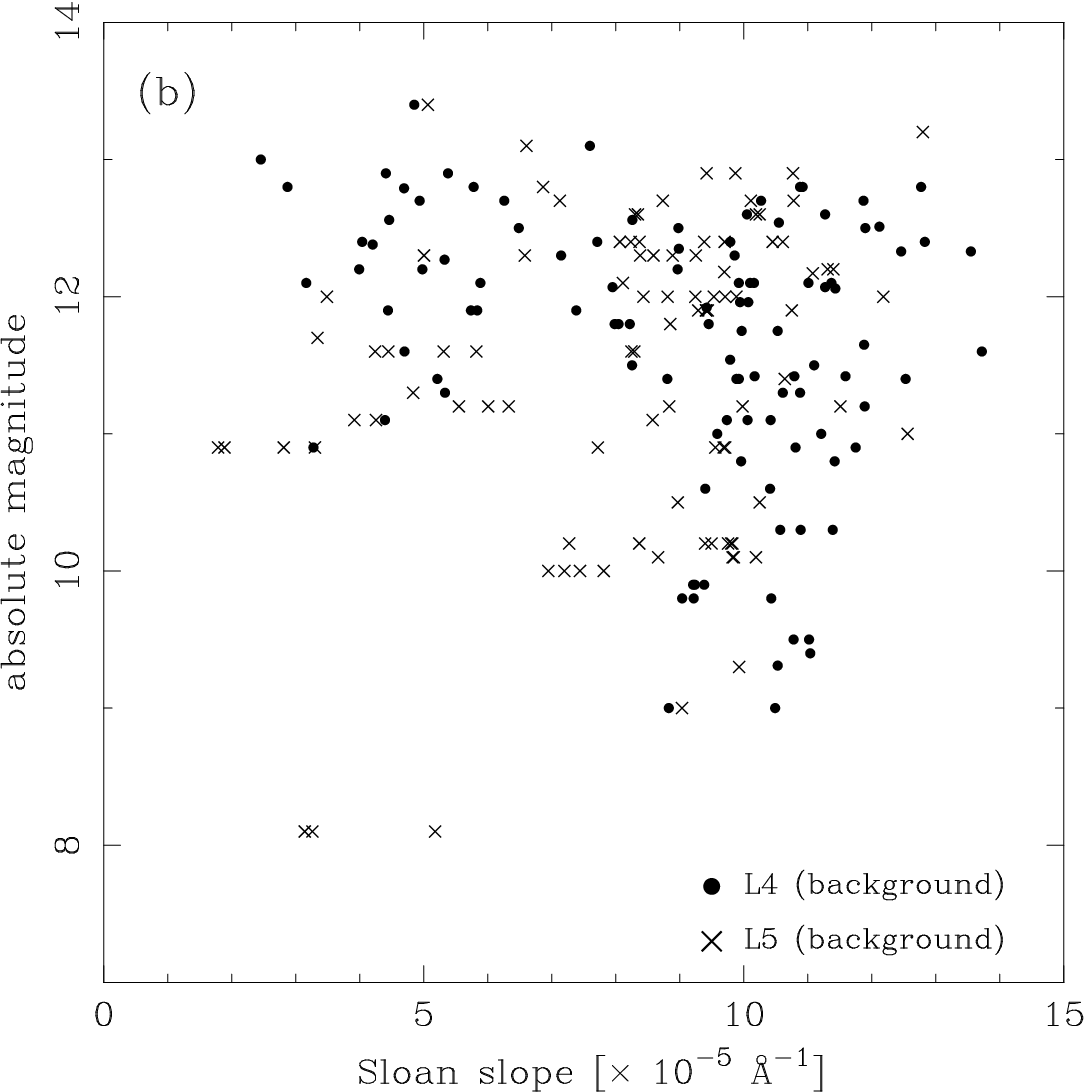} 
\caption{\textbf{(a) }Distribution of Sloan slopes of family members as a function
of absolute magnitude. Dots correspond to the L4 swarm and crosses
to the L5 swarm. \textbf{(b) }Same as (a) but for the background asteroids.
Note the significant lack of large background asteroids with small
slopes (P-types).}
\label{fig:8} 
\end{figure*}

One scenario involves the idea that the P and D classes are related
to different mineralogies and, consequently, to different material
strengths. Let us assume that P-type asteroids are easier to breakup
than D-type asteroids. Recall that this is just an assumption and
there is no evidence, neither observational nor theoretical, to support
it. Therefore, large P-type asteroids will tend to fragment in smaller
bodies while large D-type asteroids will tend to remain intact, causing
a loss of large P-type asteroids as suggested in Fig. \ref{fig:8}b.
In addition, fragments from P-type asteroids may acquire larger ejection
velocities after a breakup than fragments from D-type asteroids. Since
the islands of stability around L4 and L5 shrink at large inclinations
(e.g. \citealp{2003MNRAS.345.1091M}, \citealp{2004CeMDA..90..139S}),
many of these P-type fragments might be ejected beyond the stability
limits of the swarms causing the lack of high-inclination P-type asteroids
observed in Fig. \ref{fig:7}b. The predominance of P-type asteroids
among the L4 families is in line with this scenario but, on the other
hand, the predominance of D-type asteroids among the L5 families is
against it.

Another scenario involves the idea that the P and D classes represent
the same mineralogy but modified by some aging process, like the space
weathering. Let us assume that the space weathering produces a reddening
of the surfaces, so D-type asteroids have older surfaces than P-type
asteroids. The surfaces may be renewed either by disruptive collisions,
that expose the ``fresh'' interior of the parent body, or by resurfacing
collisions. We could expect that both collisional phenomena are more
frequent at low inclinations than at high inclinations, and more frequent
among the small bodies than among the large ones. Thus, high inclination
and large asteroids would be, on average, older (i.e. redder) than
low inclination and small ones, in agreement with Figs. \ref{fig:7}b
and \ref{fig:8}b. This scenario would also imply that families in
L5 are, on average, older than those in L4.

The above scenarios have several limitations because none of them
are well constrained. The mineralogy associated to the P- and D-types
is totally unknown. In fact, some authors claim that D-type asteroids
would be more fragile than P-types (\citealp{1997A&A...323..606D}).
The actual effect of space weathering on spectrally featureless surfaces
and the time scale to produce a significant change in the spectral
slope are also unknown. Some authors propose that the space weathering
would tend to neutralize the colors of initially red surfaces (\citealp{2004Icar..170..214M}),
so P-type asteroids would have older surfaces than D-types%
\footnote{An interesting idea is that the space weathering may have two phases:
an initial phase in which it produces a reddening of the surfaces
up to a saturation level, and a second phase in which it produces
the opposite effect, leading to more neutral color surfaces, together with
a reduction of the overall albedo. Within
this scenario, P-type asteroids could have either too young (high albedo) or too
old (low albedo) surfaces, while D-types would have mid-age surfaces. D-type asteroids
would also be, on average, more numerous because an asteroid would
spend most of its life showing a reddish surface, unless a collision
modifies it. This might be causing the overall abundance of D-type
asteroids among the Trojans. The knowledege of the albedo values for a
large amount of Jupiter Trojans might help to better contrain this scenario.
Unfortunately, up to now very few Trojans have known albedos}. 
The rate of collisional events that can produce disruption or resurfacing
depending on diameter and orbital inclination is poorly constrained.
Last, but not least, it may happen that what we observe is the product
a complex combination of  all these effects.

\section{Distribution of spectral slopes for selected asteroid families\label{sec:Distribution-of-spectral}}

In the previous section we discussed the global distribution of spectral
slopes among Trojan asteroid families and background asteroids. In
this section we analyze some particular families, selected in view
of their interest and the number of its members contained in both
the Sloan and the Spectroscopic samples. For this analysis we did
not consider all the observations available in the samples. Instead,
we used the slopes listed in Tables \ref{tab:1} and \ref{tab:2}
(i.e. for asteroids with more than one observation we consider the average
slope of the observations).

\subsection{Families in the L4 swarm\label{sub:Families-in-L4}}

\subsubsection{The Menelaus clan}

\begin{figure*}[t]
\centering
\includegraphics[clip,width=0.47\textwidth]{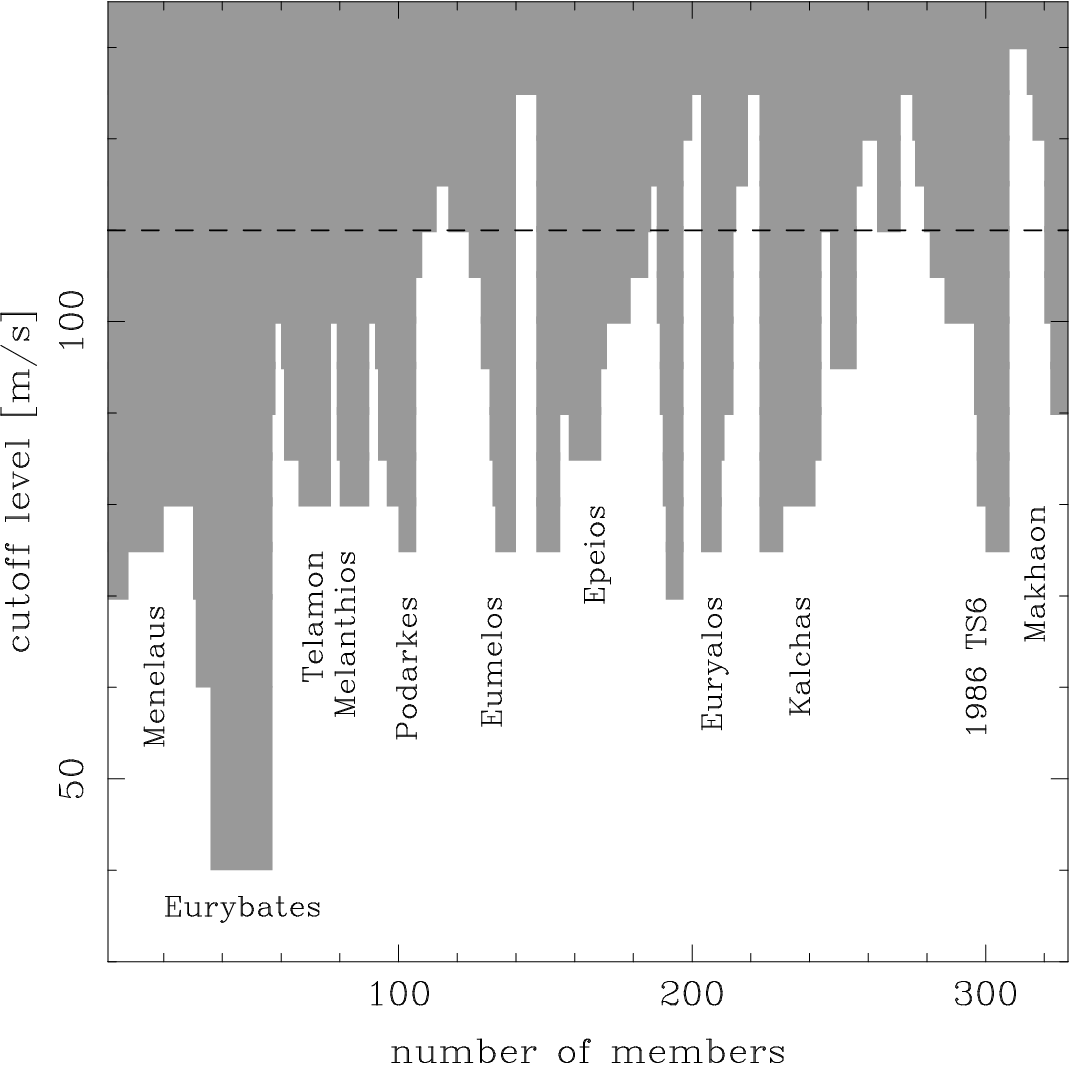}
~~~~~~~
\includegraphics[clip,width=0.47\textwidth]{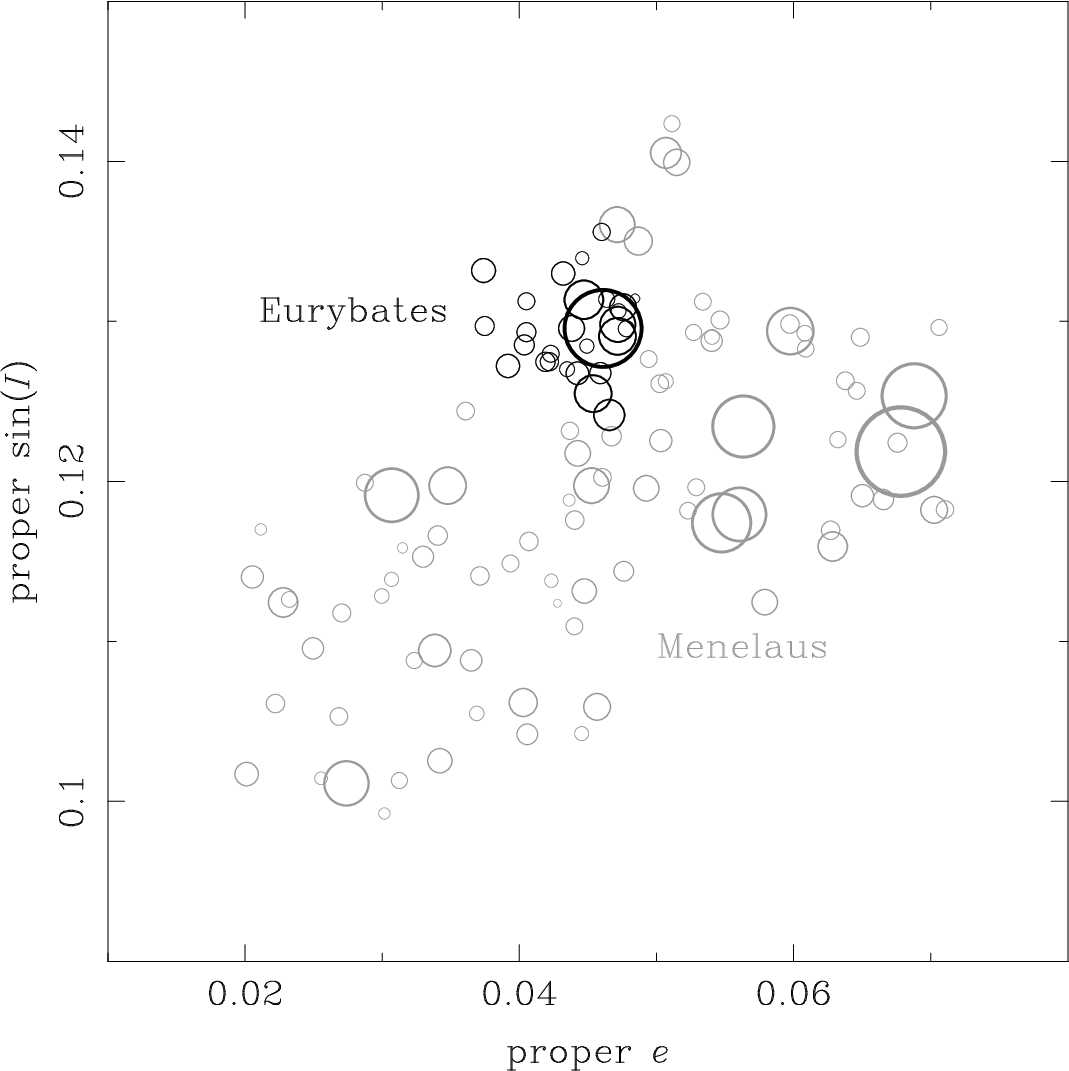} 
\caption{\textit{Left panel.} Dendogram of the Menelaus clan, indicating the main families identified.
The dashed horizontal line is the cutoff used in this study. \textit{Right panel.}
Distribution in the space of proper elements of the Menelaus family
(gray circles) as detected at $d_{\mathrm{cut}}=110$ $\mathrm{m\, s^{-1}}$,
and of the Eurybates family (black circles) as detected at $d_{\mathrm{cut}}=70$
$\mathrm{m\, s^{-1}}$. The size of each circle is proportional to
the asteroid size.}
\label{fig:10} 
\end{figure*}

\begin{figure*}[t]
\includegraphics[clip,width=0.325\textwidth]{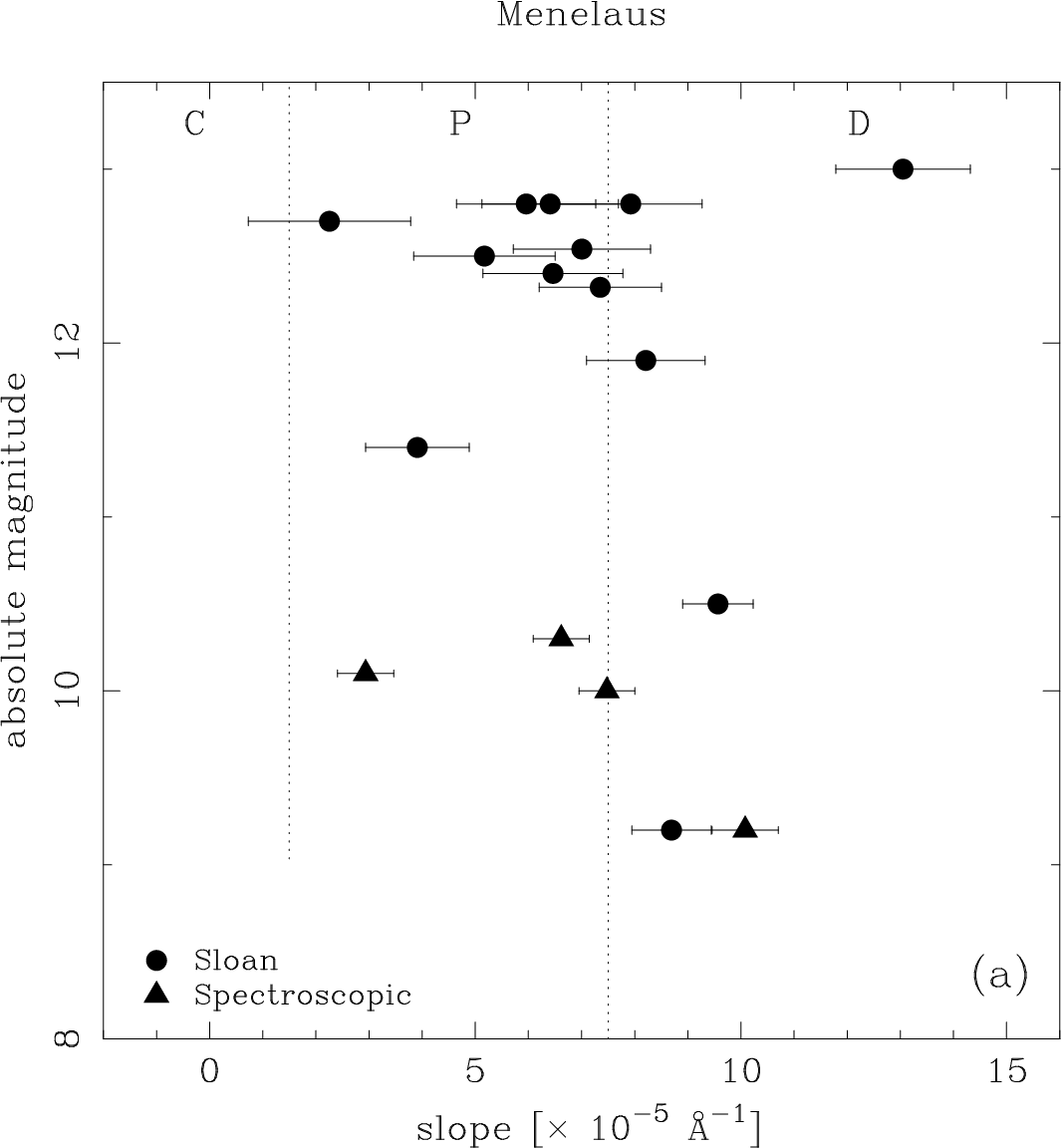}
~~~~
\includegraphics[clip,width=0.325\textwidth]{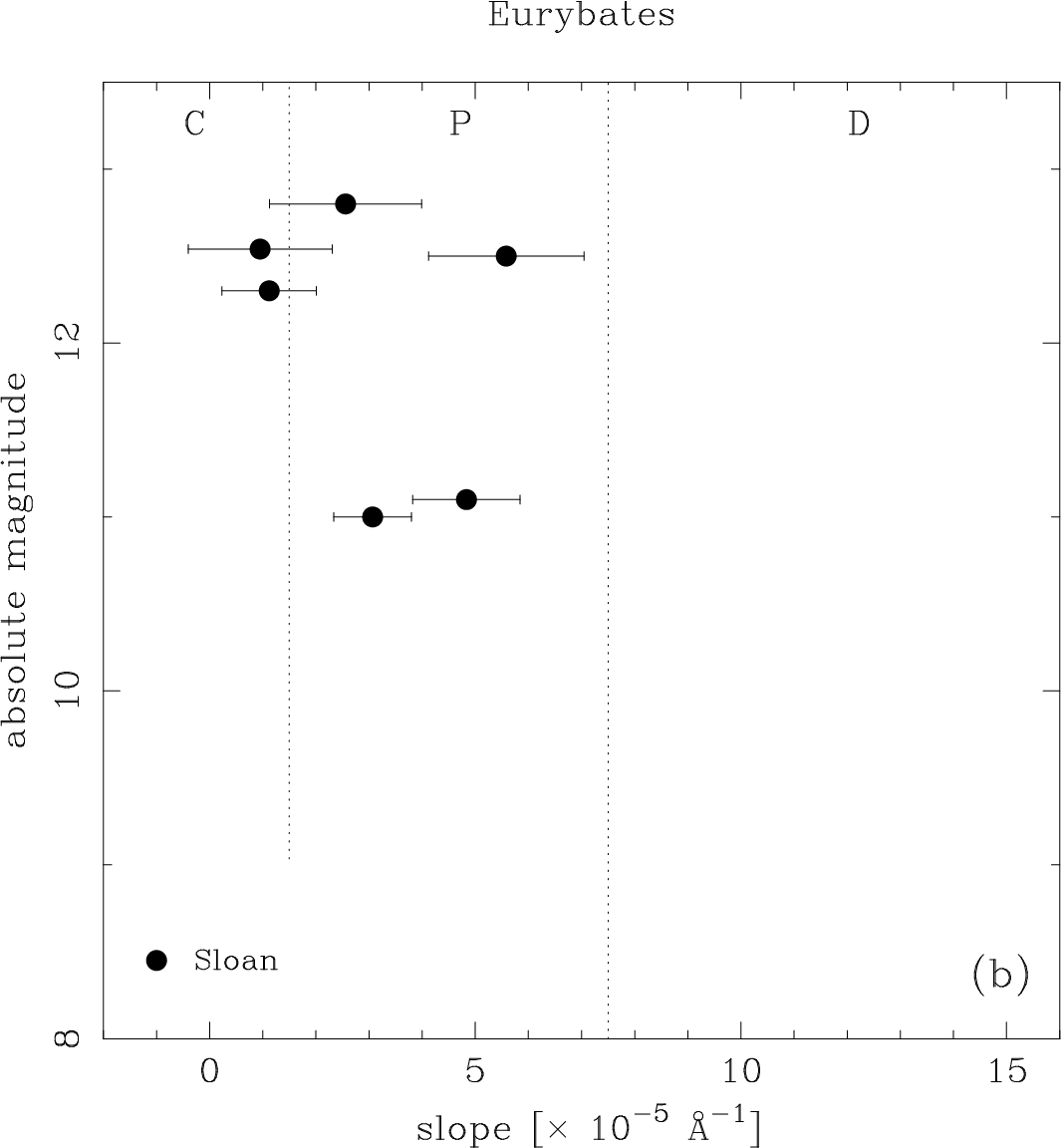}
\par\bigskip
\sidecaption
\includegraphics[clip,width=0.325\textwidth]{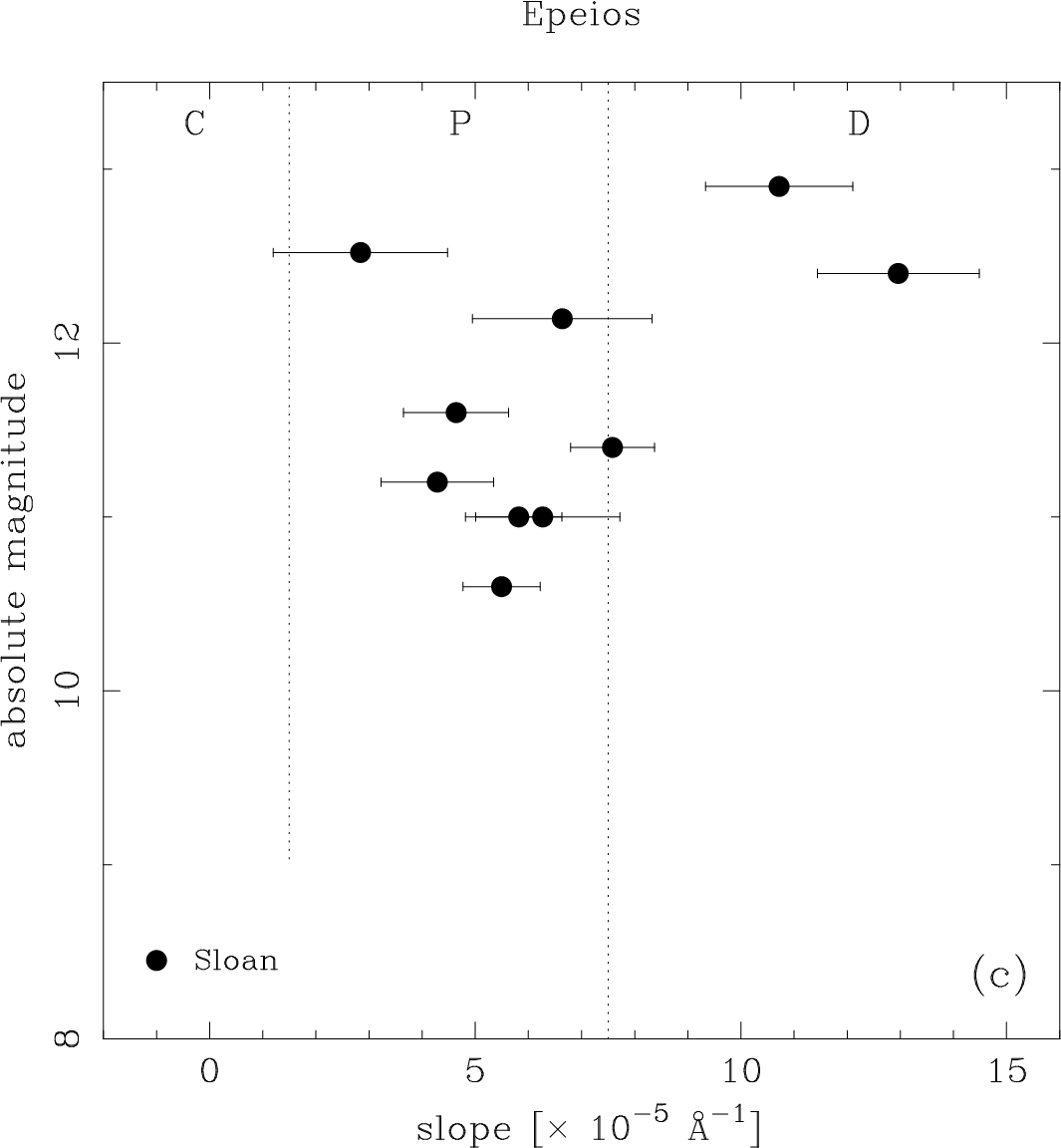}
~~~~
\includegraphics[clip,width=0.325\textwidth]{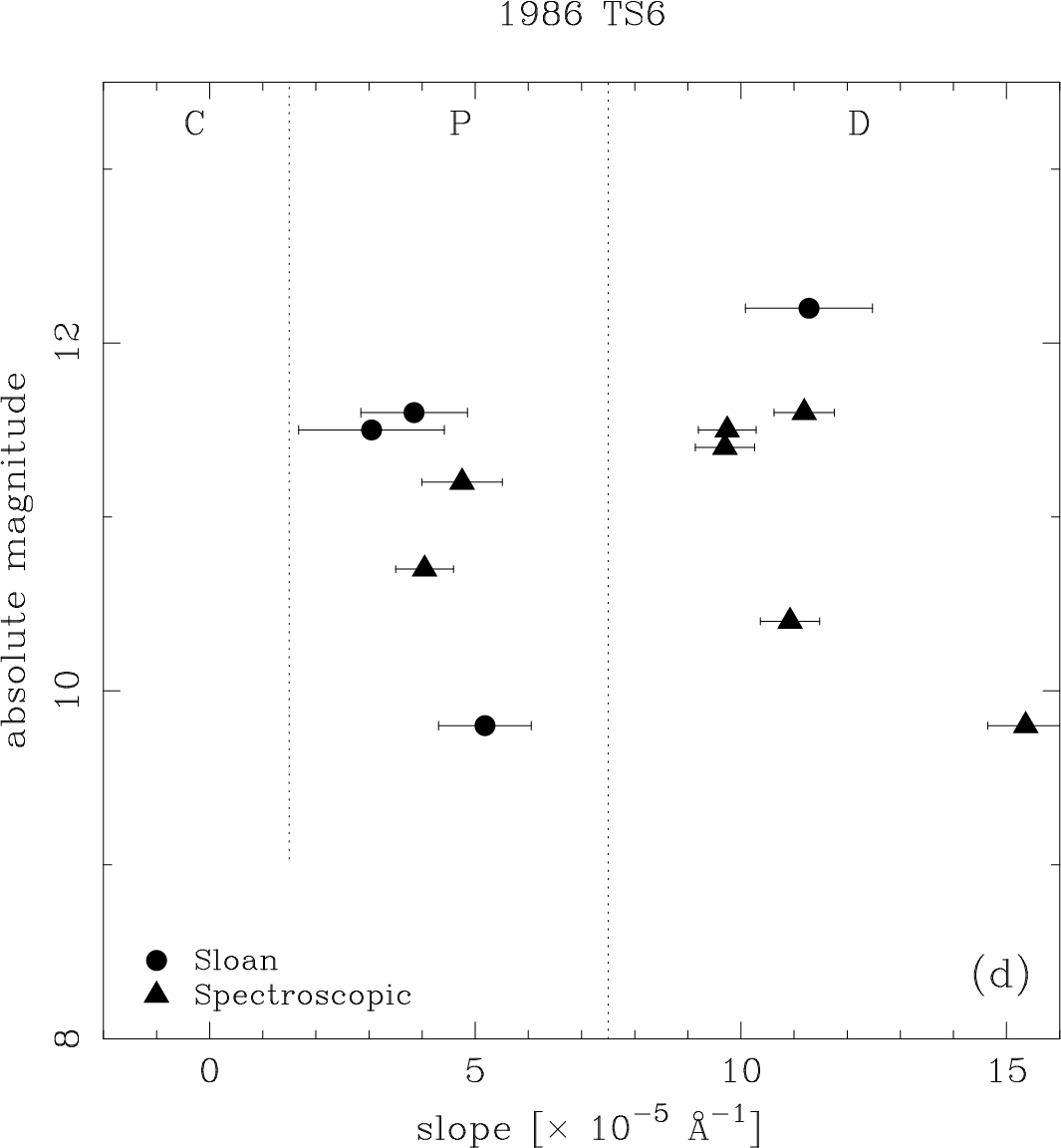} 
\caption{Distribution of Sloan slopes against absolute magnitude for four families
of the Menelaus clan. Full circles correspond to Sloan slopes. Triangles
correspond to Spectroscopic slopes. The vertical dotted lines define
the slope transition, within $\pm0.7\times10^{-5}\,\textrm{\AA}^{-1}$,
between the different taxonomic classes indicate above the plots.}
\label{fig:12} 
\end{figure*}

\begin{figure*}[t]
\sidecaption
\includegraphics[clip,width=0.325\textwidth]{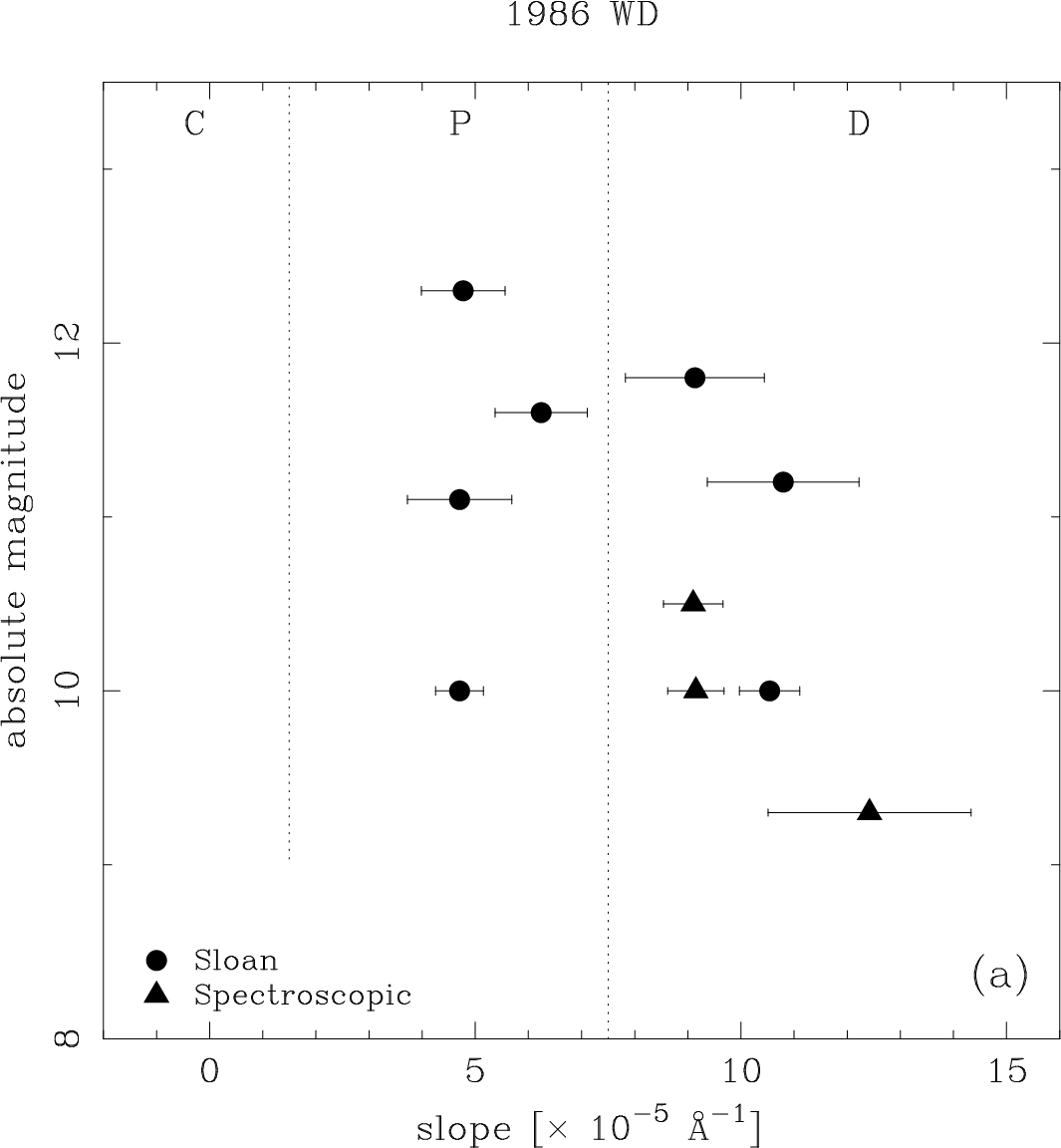}
~~~~
\includegraphics[clip,width=0.325\textwidth]{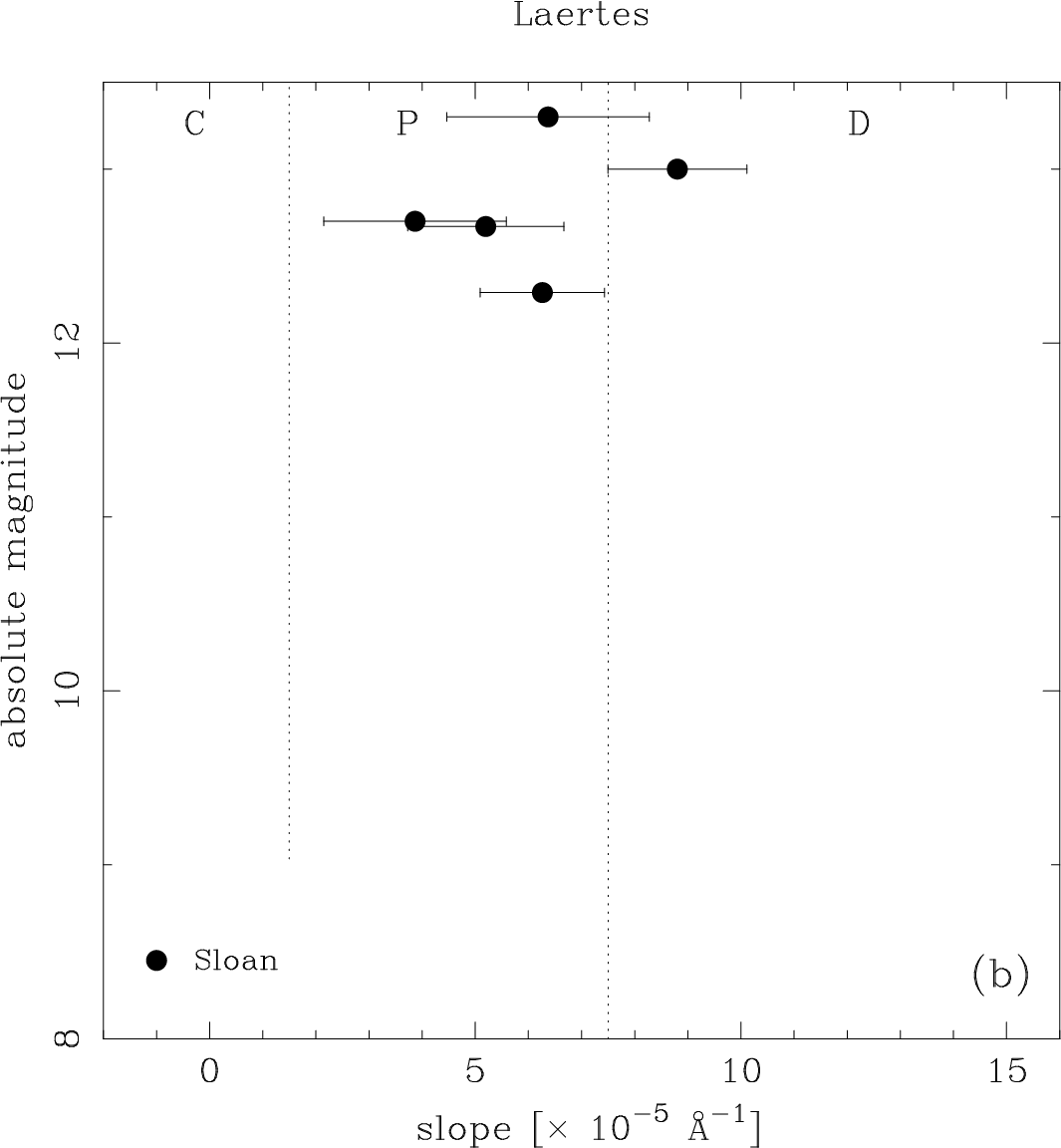}
\caption{Same as Fig. \ref{fig:12} but for two families in the L4 swarm.}
\label{fig:13} 
\end{figure*}

\begin{figure*}[t]
\centering
\includegraphics[clip,width=0.325\textwidth]{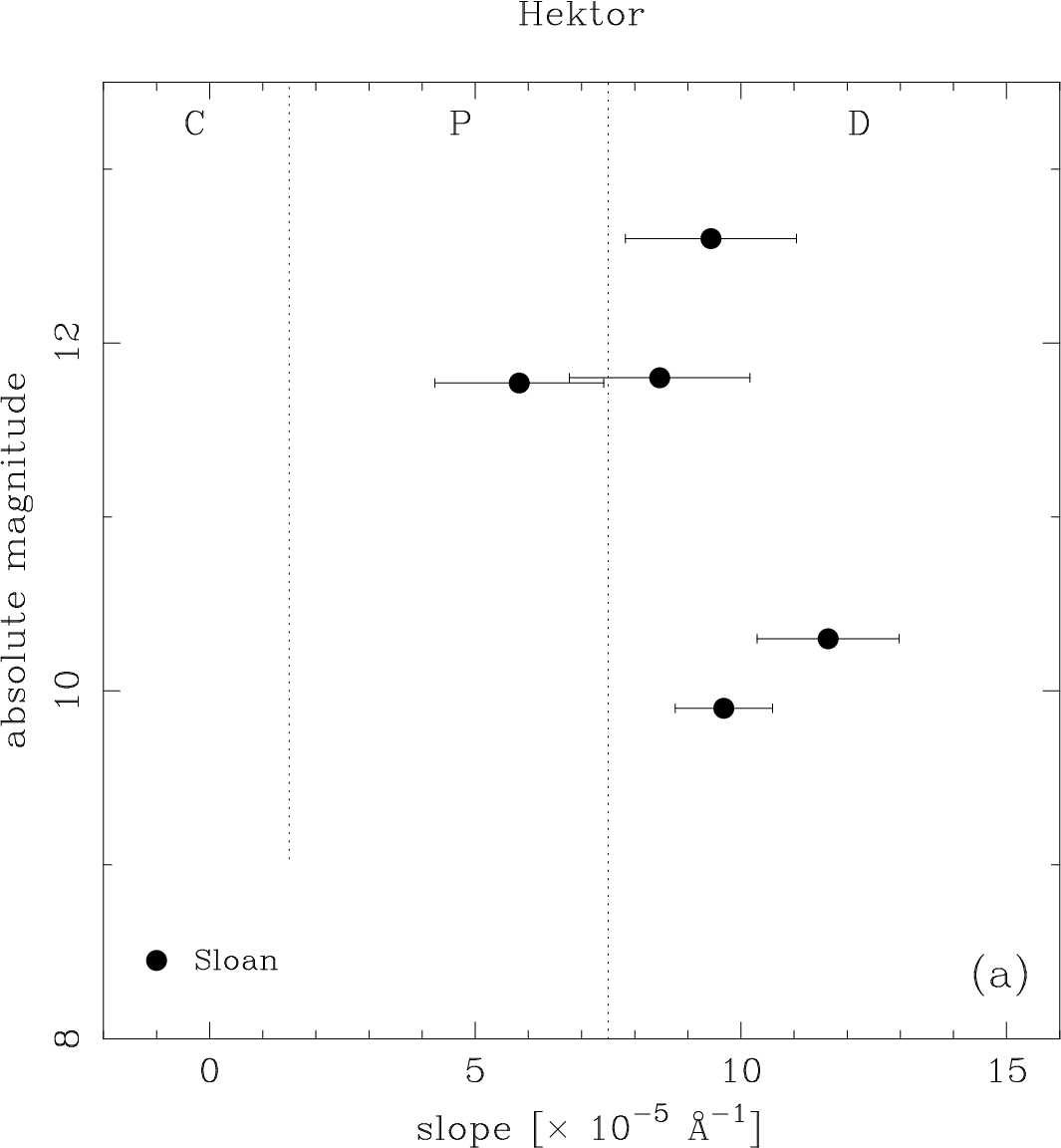}
~
\includegraphics[clip,width=0.325\textwidth]{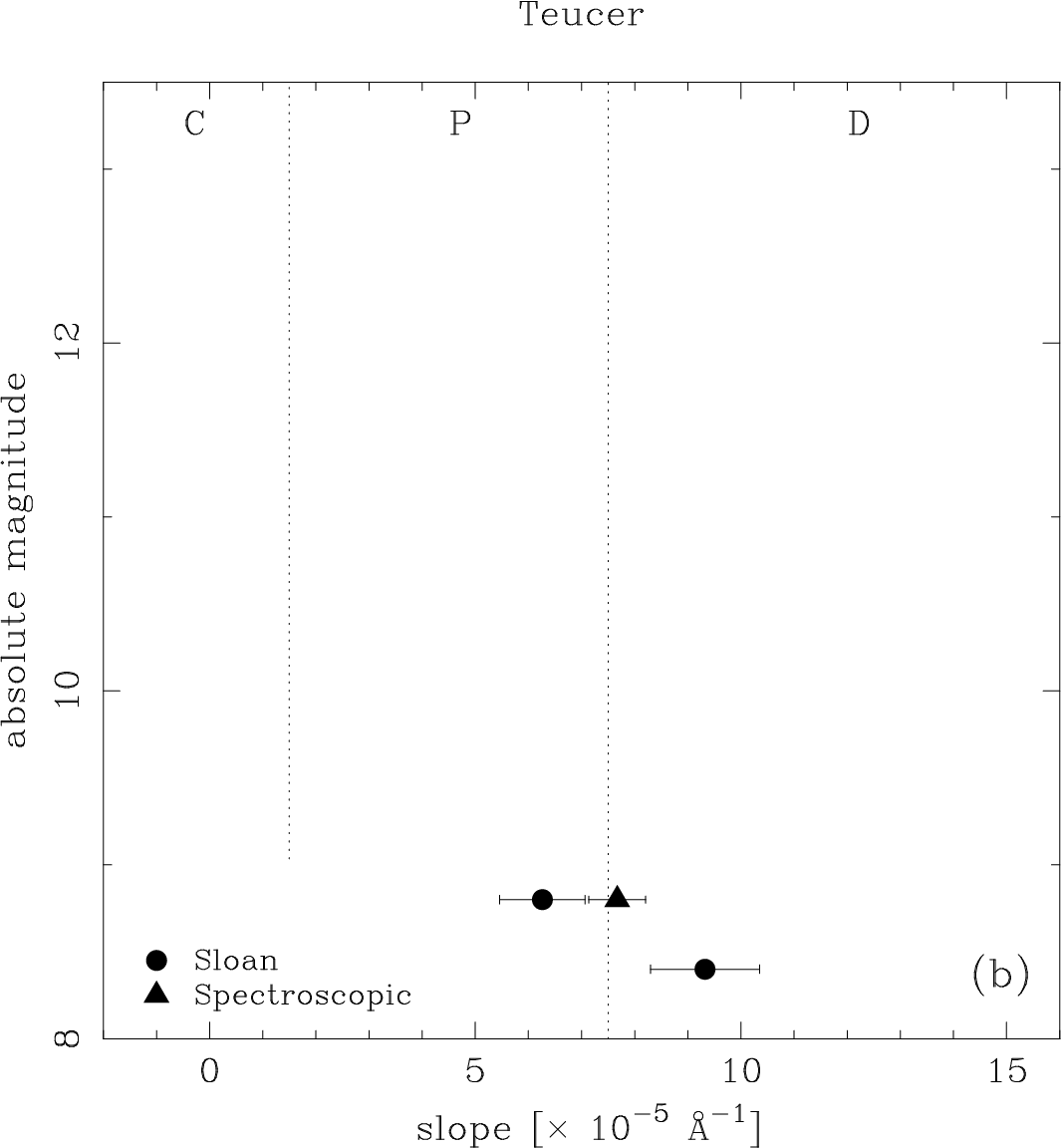}
~
\includegraphics[clip,width=0.325\textwidth]{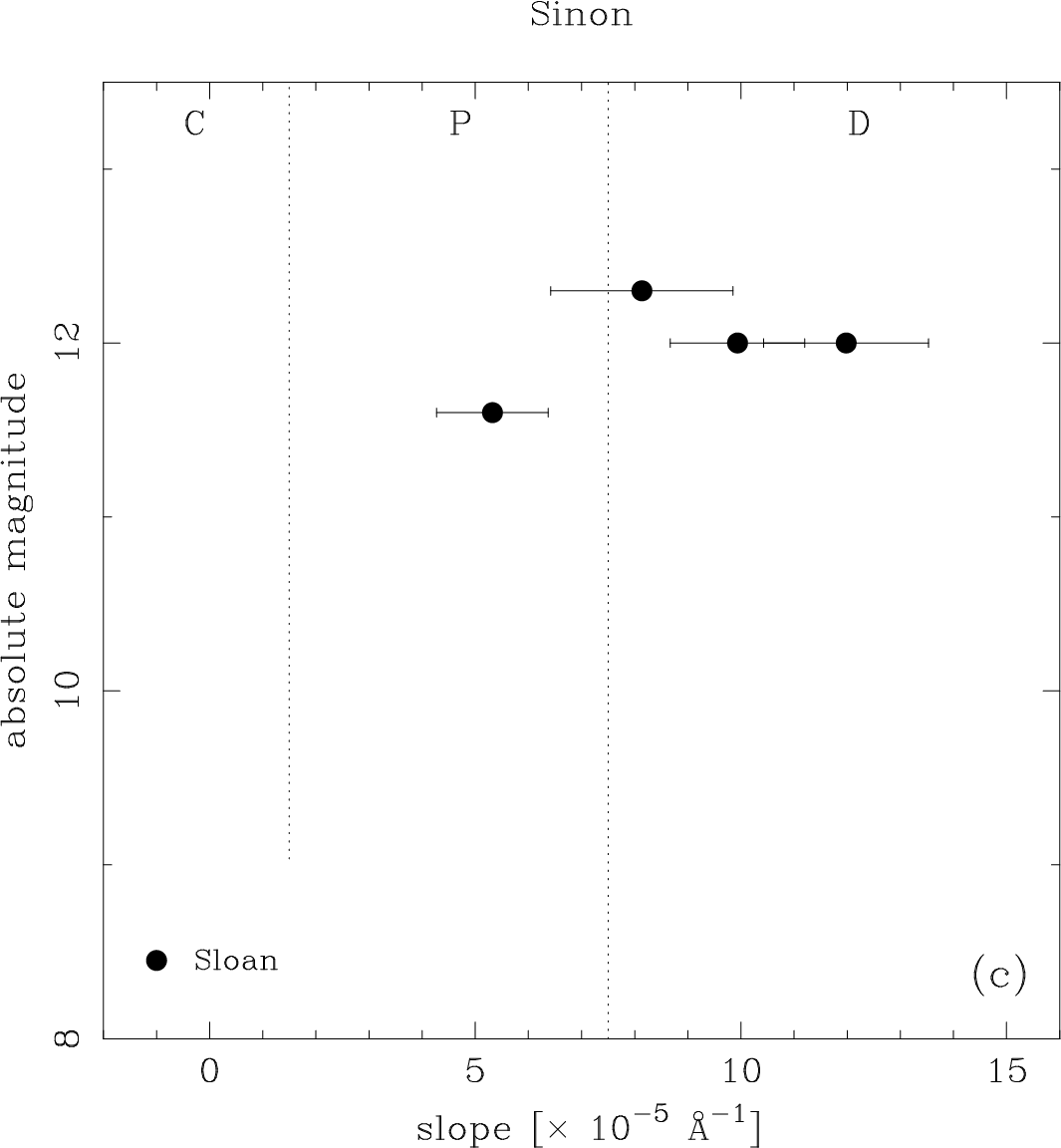}
\caption{Same as Fig. \ref{fig:12} but for three high-inclination families
in the L4 swarm.}
\label{fig:14} 
\end{figure*}

Several families in the L4 swarm merge together at high values of
the cutoff ($d_{\mathrm{cut}}>125$ $\mathrm{m\, s^{-1}}$) to form
a big clan of families, similar to the Flora clan in the inner asteroid
Main Belt. This clan gets its name after the main member, asteroid
(1647) Menelaus. The structure of the Menelaus clan is shown in Fig.
\ref{fig:10} (left) in the form of a dendogram \citep{1995Icar..116..291Z}.
Each stalactite in the dendogram represents a different family within
the clan, and it is easy to see how the families are better resolved
as we go to lower values of the cutoff. The word ``clan\char`\"{}
invokes some kind of common origin, but the fact that several families
form a clan does not necessarily imply that they all come from the
same ancestor. The taxonomic analysis of the clan members may help
to better understand this problem.

As seen in Fig. \ref{fig:10}, the more robust family of the clan
is the Menelaus family itself, which counts more than 100 members
at the quasi random level ($\sim104$ $\mathrm{m\, s^{-1}}$) and
represents the largest family in the L4 swarm. The small families
of Telamon, Melanthios and Podarkes separate from the Menelaus family
at lower cutoffs but they soon disappear. On the other hand, the Eurybates
family appears as a very robust cluster that survives down to very
small cutoffs. Indeed, the Eurybates family forms a very tight cluster
within the Menelaus family, as shown in Fig. \ref{fig:10} (right). From the
solely analysis of Fig. \ref{fig:10} it is difficult
to decide whether the Eurybates family is a sub-cluster of the Menelaus
family, i.e. a family formed by the secondary breakup of a former
Menelaus family member, or the Eurybates and Menelaus families are
two different families that simply overlap in the space of proper
elements.

The taxonomy of these two families has been analyzed by \citeauthor{2006Icar..183..420D}
(2006 - hereafter D06) and by \citeauthor{2007Icar..190..622F} (2007
- hereafter F07), who obtained spectra of 3 members of the Menelaus
family and 17 members of the Eurybates family. These authors found
that the Menelaus family is mostly a D-type family, but the Eurybates
family is dominated by C-type asteroids. A slightly different result
is obtained from the analysis of our data samples. 

Figures \ref{fig:12}a,b show the spectral slopes of the Menelaus
and Eurybates families as a function of the absolute magnitude. At
large sizes ($H<11$, that correspond to $\sim40$ km), the Menelaus
family shows a slight predominance of D-type asteroids --(1749) Telamon,
(5258) 1989 AU1 and (13362) 1998 UQ16-- compared to one P-type asteroid
--(5244) Amphilochos--, and another asteroid --(1647) Menelaus-- that
appears to be a P-type but could be classified as D-type if we account
for its error and recall that the limiting slope of $7.5\times10^{-5}\,\textrm{\AA}^{-1}$
between the P- and D-types has a $\pm0.7\times10^{-5}\,\textrm{\AA}^{-1}$
uncertainty (actually, D06 classified this asteroid as D-type). On
the other hand, at the small sizes ($H>11$) the family is clearly
dominated by P-type asteroids.

The results for the Eurybates family (Fig. \ref{fig:12}b) are more
in line with the findings of F07, although we do not detect a predominance
of C-type asteroids due to the small amount of observations in the
Sloan sample. Three asteroids --(9818) Eurymachos, (18060) 1999 XJ166
and (24426) 2000 CR12-- are P-type and one asteroid --(43212) 2000
AL113-- is C-type. These four asteroids got the same taxonomic classification
by F07. There is also one asteroid --(65225) 2002 EK44-- that appears
to be a P-type but due to its error can be classified as C-type, in
agreement with F07. The last body --2002 AE166-- is a C-type asteroid
and was not observed by F07.

Figures \ref{fig:12}c,d show the spectral slopes of other members
of the Menelaus clan: the Epeios and 1986 TS6 families. The Epeios
family has not been previously observed by any spectroscopic survey,
so the Sloan slopes shown here provide the first taxonomic information
about this family, which appears constituted mostly by P-type asteroids,
especially at the large sizes. On the other hand, the 1986 TS6 family
has been observed by D06%
\footnote{These authors refer to it as the Makhaon family due to the large cutoff
used, but it is clear from Fig. \ref{fig:10} that Makhaon is a different
family. This has been correctly addressed by F07.%
} and F07. The available data, including the Sloan slopes, indicate
that this family has two well separated components, one P-type and
one D-type, regardless of bodies size. Note, however, that these two
components cannot be resolved in terms of proper elements, i.e. the
P- and D-type members are mixed in the same cluster even for the smallest
possible cutoffs. The largest asteroid in the family, (5025) 1986
TS6, shows significantly different values of Sloan slope ($5\times10^{-5}\,\textrm{\AA}^{-1}$)
and Spectroscopic slope ($15\times10^{-5}\,\textrm{\AA}^{-1}$), but
the latter has been computed from a very noisy spectrum of \citet{2004Icar..168..374B}
so its anomalous large value should be considered with care.

The above results points to the idea that not only the Menelaus clan
as a whole, but also the individual families are quite heterogeneous
in terms of taxonomic classes, including from the reddest D-type asteroids
to the neutral-color C-type ones. Note also that the spectral slopes
do not show any particular trend with size.

\subsubsection{Other families}

Figure \ref{fig:13} shows the distributions of spectral slopes in
terms of absolute magnitude of two families that are not members of
the Menelaus clan: 1986 WD and Laertes. The 1986 WD family has been
studied by D06 and F07 who found a wide range of slopes, from the
D- to the C-type. The Sloan slopes tend to confirm these findings.
This family is small ($\sim15$-20 members) and at cutoff values slightly
smaller than the quasi random level it looses half of its members.
The family is no longer identified at cutoffs smaller than 95 $\mathrm{m\, s^{-1}}$.
Therefore, the diversity of taxonomic classes may be related to a
significant contamination of interlopers. 

The case of the Laertes family is somehow different. This family is
also small ($\sim15$-20 members) but it survives down to cutoff 80
$\mathrm{m\, s^{-1}}$ loosing only 40\% of its members. All the members
are small bodies ($H>11$), including (11252) Laertes. The few members
contained in the Sloan sample (the family has never been observed
spectroscopically) show a quite homogeneous distribution of spectral
slopes, all belonging to the P-type. Unfortunately, this family is
located at a very small inclination ($\sin I\sim0.08$), and it cannot
be distinguished from the background also dominated by P-type asteroids.
The identification of the Laertes family as a real P-type family relies
more in the accuracy of the HCM than in the distribution of its spectral
slopes.

Other interesting cases are the high inclination families in the L4
swarm. There are only three of these families, with $\sin I>0.25$:
Hektor, Teucer and Sinon. The distribution of the respective spectral
slopes are shown in Fig. \ref{fig:14}. Hektor and Sinon families
have not been observed by previous spectroscopic surveys, and the
Sloan slopes provide the first clues about their taxonomic composition.
The slopes in Fig. \ref{fig:14} point to a predominance of D-type
asteroids, making these families indistinguishable from the background.

\subsection{Families in the L5 swarm\label{sub:Families-in-L5}}

\subsubsection{The Anchises clan}

The L5 swarm has its own clan of families, although it is somehow
different from the Menelaus clan in L4. The Anchises clan, named after
asteroid (1173) Anchises, is quite tight and constituted by only five
families identified at $d_{\mathrm{cut}}=120$ $\mathrm{m\, s^{-1}}$:
Panthoos, Polydoros, Sergestus, Agelaos and 1999 RV165. All these
families merge in the clan at $d_{\mathrm{cut}}=150$ $\mathrm{m\, s^{-1}}$.
The taxonomic analysis of this clan indicates that it is populated
by both P- and D-type asteroids covering a wide range of spectral
slopes. But at variance with the Menelaus clan, the individual families
of the Anchises clan appear to be more homogeneous in terms of taxonomy.

\begin{figure}[t]
\centering 
\includegraphics[clip,width=1\columnwidth]{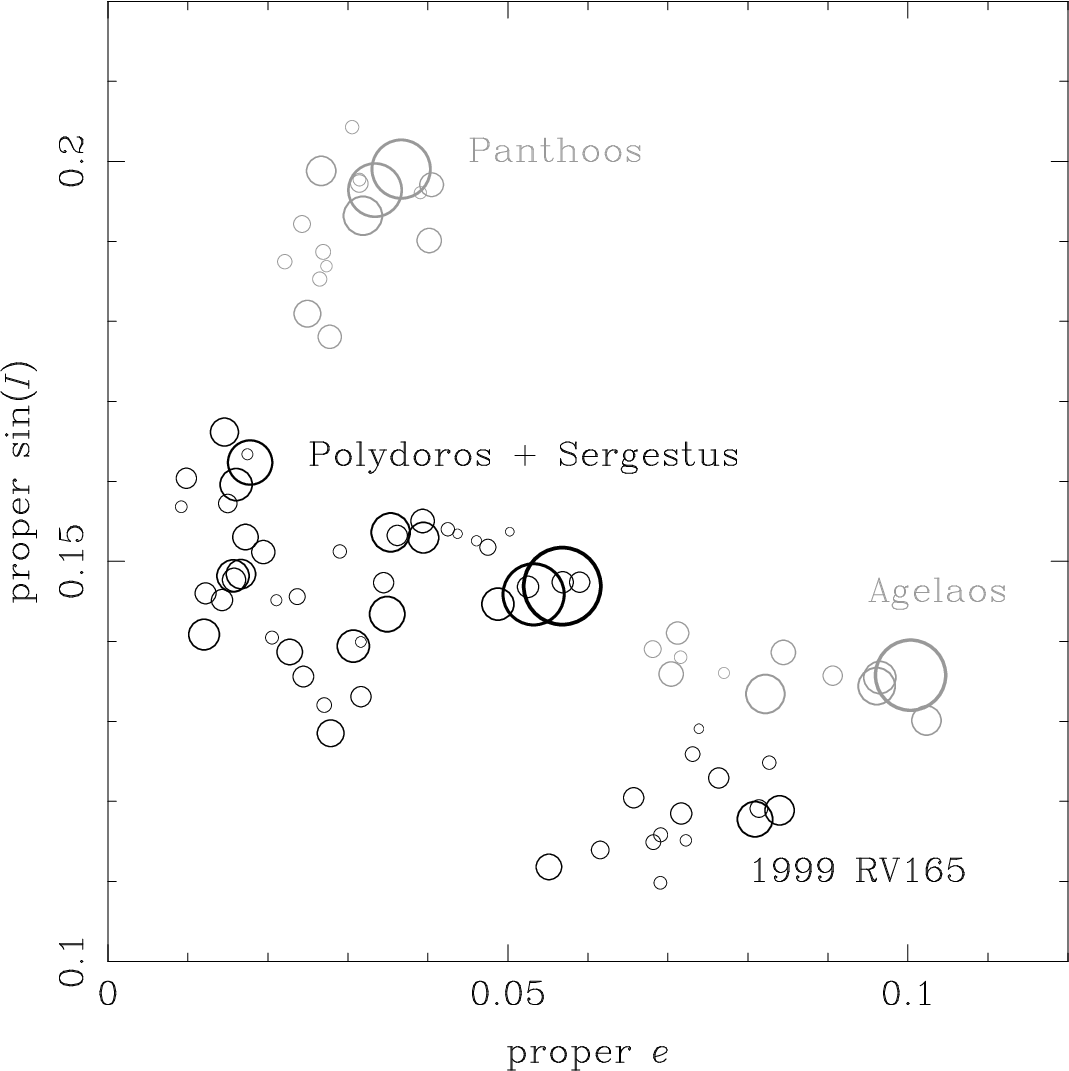}
\caption{Distribution in the space of proper elements of the Anchises clan
as detected at $d_{\mathrm{cut}}=130$ $\mathrm{m\, s^{-1}}$. It
is constituted by five families: Panthoos, Polydoros, Sergestus, Agelaos
and 1999 RV165. (1173) Anchises is incorporated to the Agelaos family
at $d_{\mathrm{cut}}>135$ $\mathrm{m\, s^{-1}}$. The size of each
circle is proportional to the asteroid size.}
\label{fig:15} 
\end{figure}

\begin{figure*}[t]
\sidecaption
\includegraphics[clip,width=0.325\textwidth]{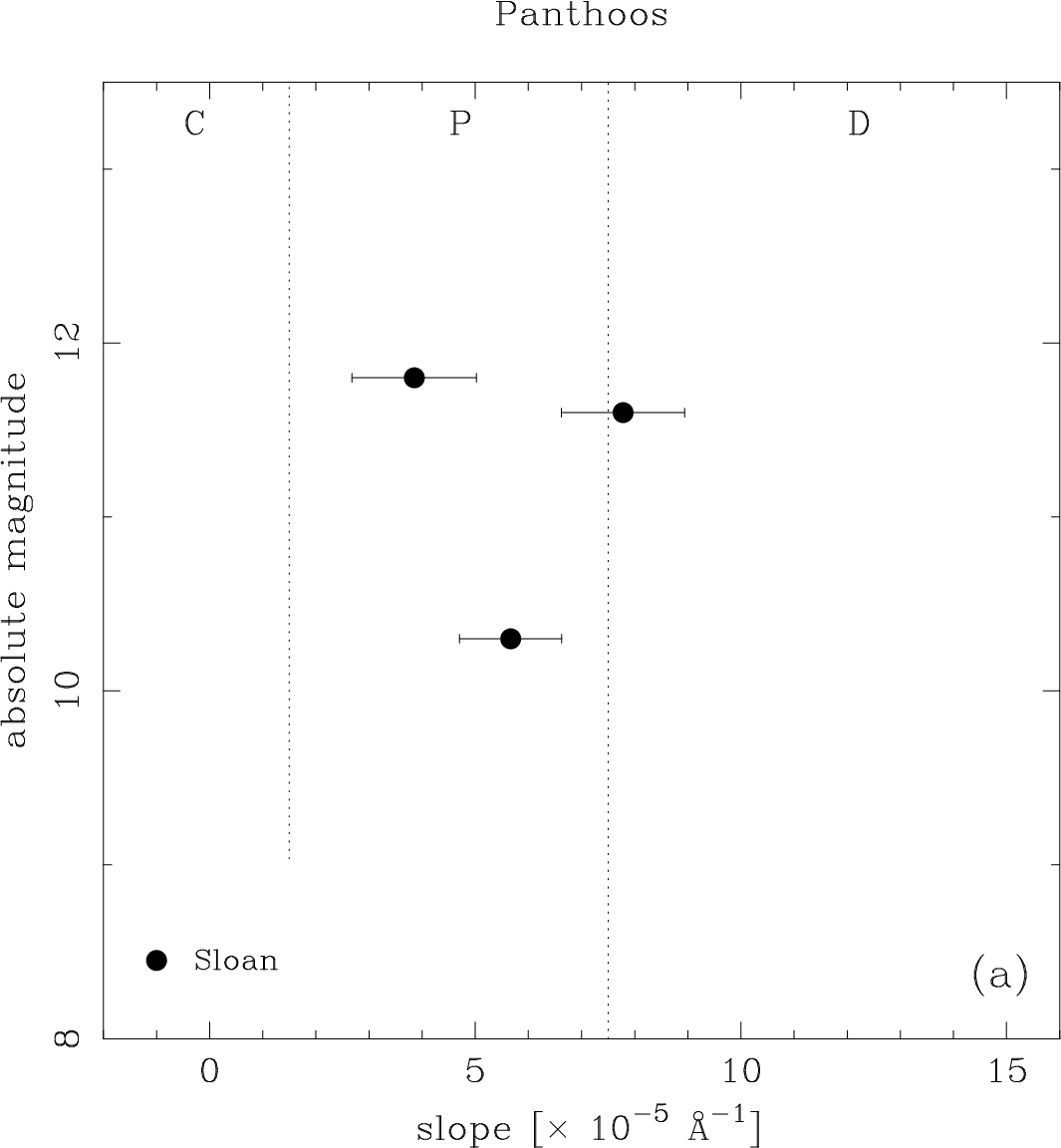}
~~~~
\includegraphics[clip,width=0.325\textwidth]{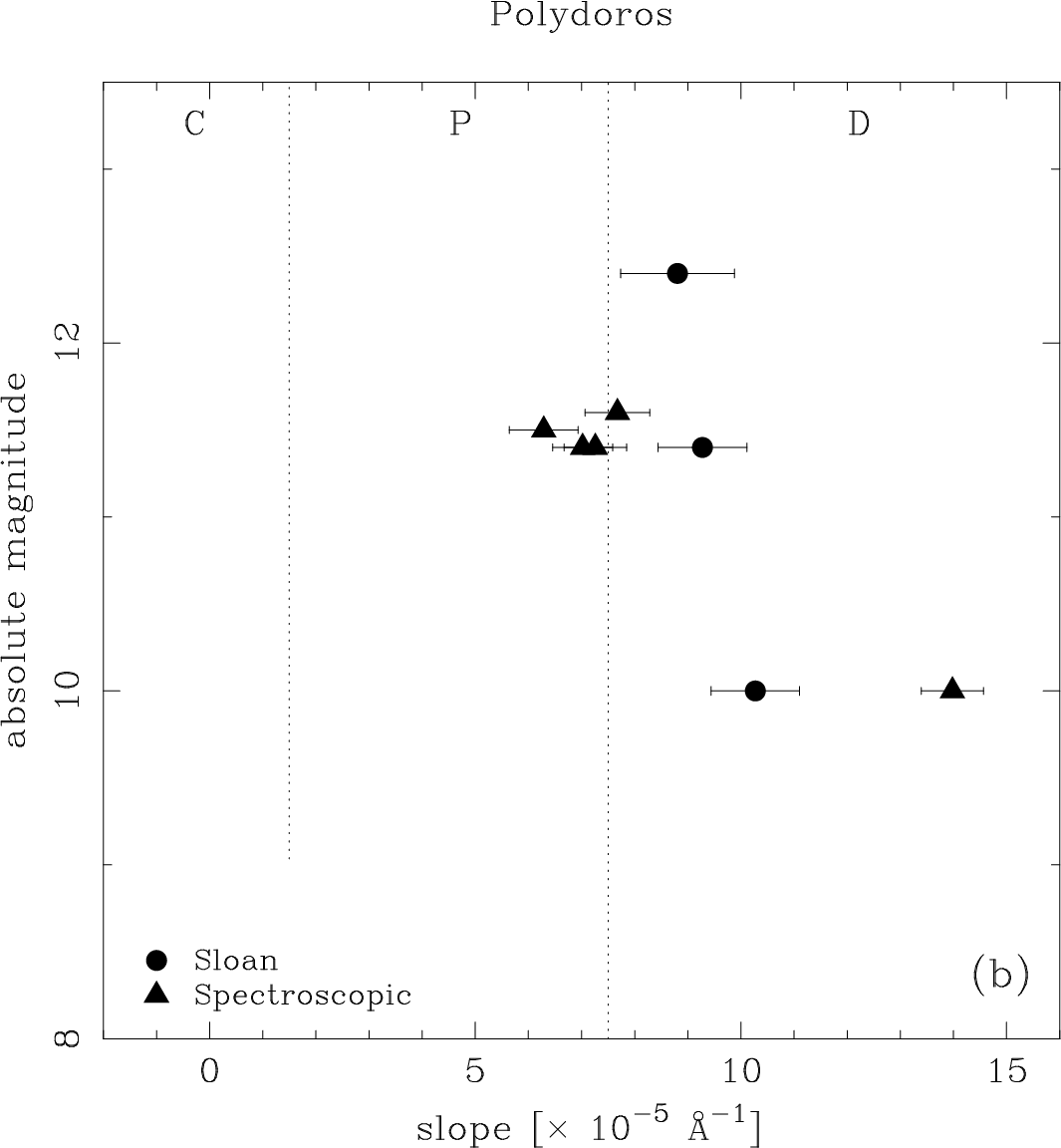}
\caption{Distribution of Sloan slopes against absolute magnitude for three
families of the Anchises clan: \textbf{(a)} Panthoos, \textbf{(b)}
Polydoros and Sergestus. Full circles correspond to Sloan slopes.
Triangles correspond to Spectroscopic slopes. The vertical dotted
lines define the slope transition, within $\pm0.7\times10^{-5}\,\textrm{\AA}^{-1}$,
between the different taxonomic classes indicate above the plots.}
\label{fig:16} 
\end{figure*}

The Panthoos family appear to be a P-type family (Fig. \ref{fig:16}a)
and it is well distinguishable from the background, dominated by D-type
asteroids. This is a quite robust family that remains isolated over
a wide range of cutoff values, from 90 to 140 $\mathrm{m\, s^{-1}}$.
Its distribution in the space of proper elements for $d_{\mathrm{cut}}=130$
$\mathrm{m\, s^{-1}}$ is shown in Fig. \ref{fig:15}. It is worth
noting that F07 studied this family and found that it is a D-type
family. However, due to an incorrect choice of the cutoff level, all
the 8 asteroids that they used to perform their classification are
not actual members of the Panthoos family but of the Sergestus family.
These two families merge together at $d_{\mathrm{cut}}>140$ $\mathrm{m\, s^{-1}}$.

The Polydoros family is another example of a quite robust family that
is detected down to cutoff 90 $\mathrm{m\, s^{-1}}$. This family
merges with the Sergestus family for $d_{\mathrm{cut}}\geq130$ $\mathrm{m\, s^{-1}}$
to form the single Polydoros family shown in Fig. \ref{fig:15}. The
distribution of spectral slopes of the Polydoros (+Sergestus) family,
shown in Fig. \ref{fig:16}b, indicates that this is a quite homogeneous
D-type family. Interestingly, the spectral slope of (4829) Sergestus
measured by F07 indicates that this asteroid is likely to be a P-type,
so it may be an interloper. Recall, however, that the Polydoros and
Sergestus families are taxonomically indistinguishable from the background
and this makes difficult the discussion about interlopers.

The Agelaos and 1999 RV165 families are somehow different from the
other families of the Anchises clan. None of them survive down to
small cutoffs and they appear less homogeneous in terms of taxonomy.
Their distribution in proper elements is shown in Fig. \ref{fig:15}.
At $d_{\mathrm{cut}}>135$ $\mathrm{m\, s^{-1}}$, the Agelaos family
incorporates asteroid (1173) Anchises and becomes the Anchises family.
At the same cutoff, the 1999 RV165 family becomes the Antenor family
after incorporating asteroid (2207) Antenor. These two families merge
together at $d_{\mathrm{cut}}>145$ $\mathrm{m\, s^{-1}}$. 

The 1999 RV165 family has only one member in the Sloan sample classified
as P-type, so we cannot say too much about it. The Agelaos family
has two members observed in the Sloan sample, one P- and one D-type,
but this family has also been observed by F07 who identified it as
the Anchises family. Analyzing the slopes provided by these authors
together with the Sloan slopes, we conclude that 3 members are P-type
--(1173), (23549), (24452)-- and 3 members are D-type --(47967), (52511),
2001 SB173--. It is worth noting that the Sloan slope of (24452) is
compatible with the slope published by F07.

\subsubsection{Other families}

\begin{figure*}[t]
\includegraphics[clip,width=0.325\textwidth]{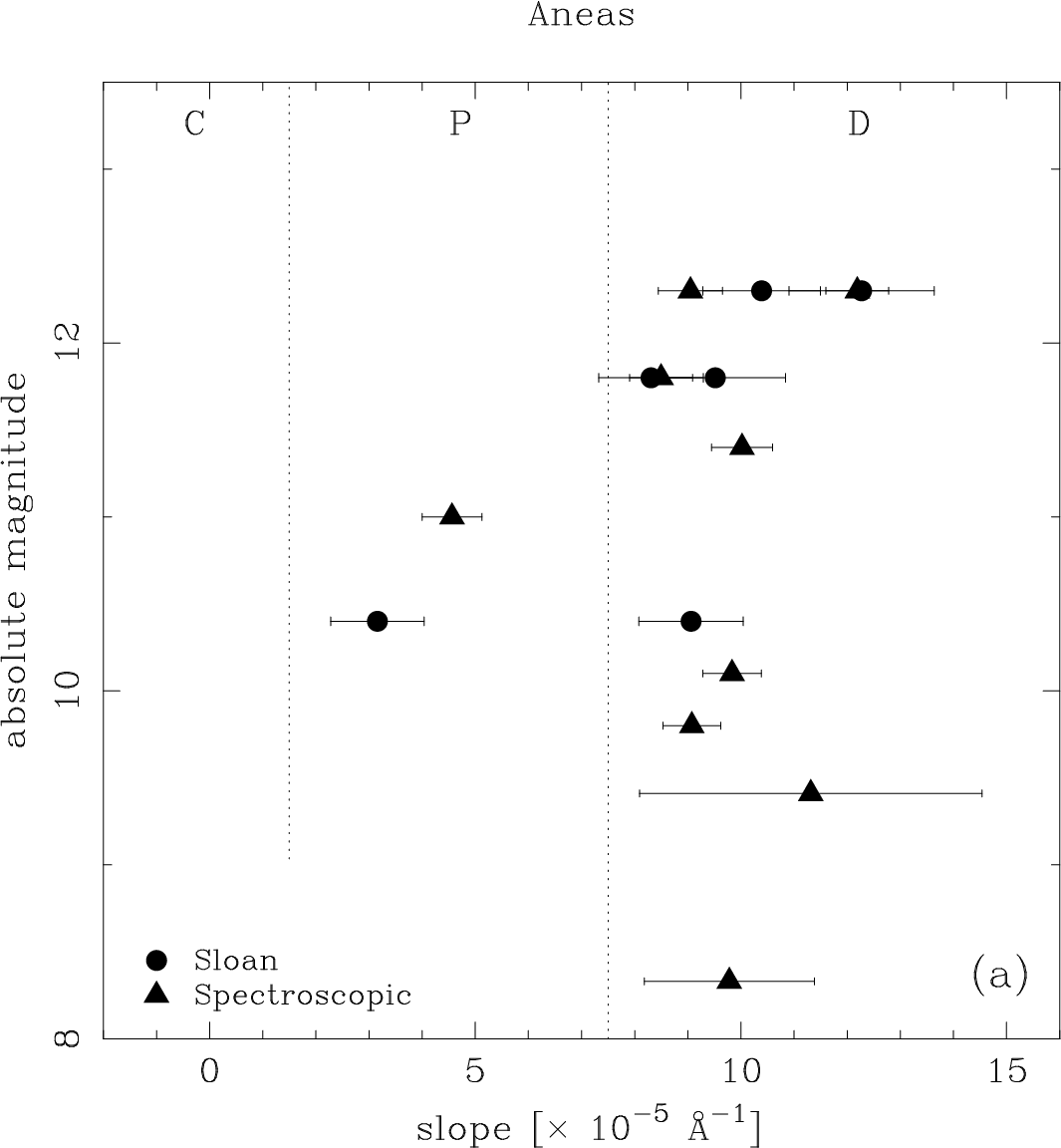}
~~~~
\includegraphics[clip,width=0.325\textwidth]{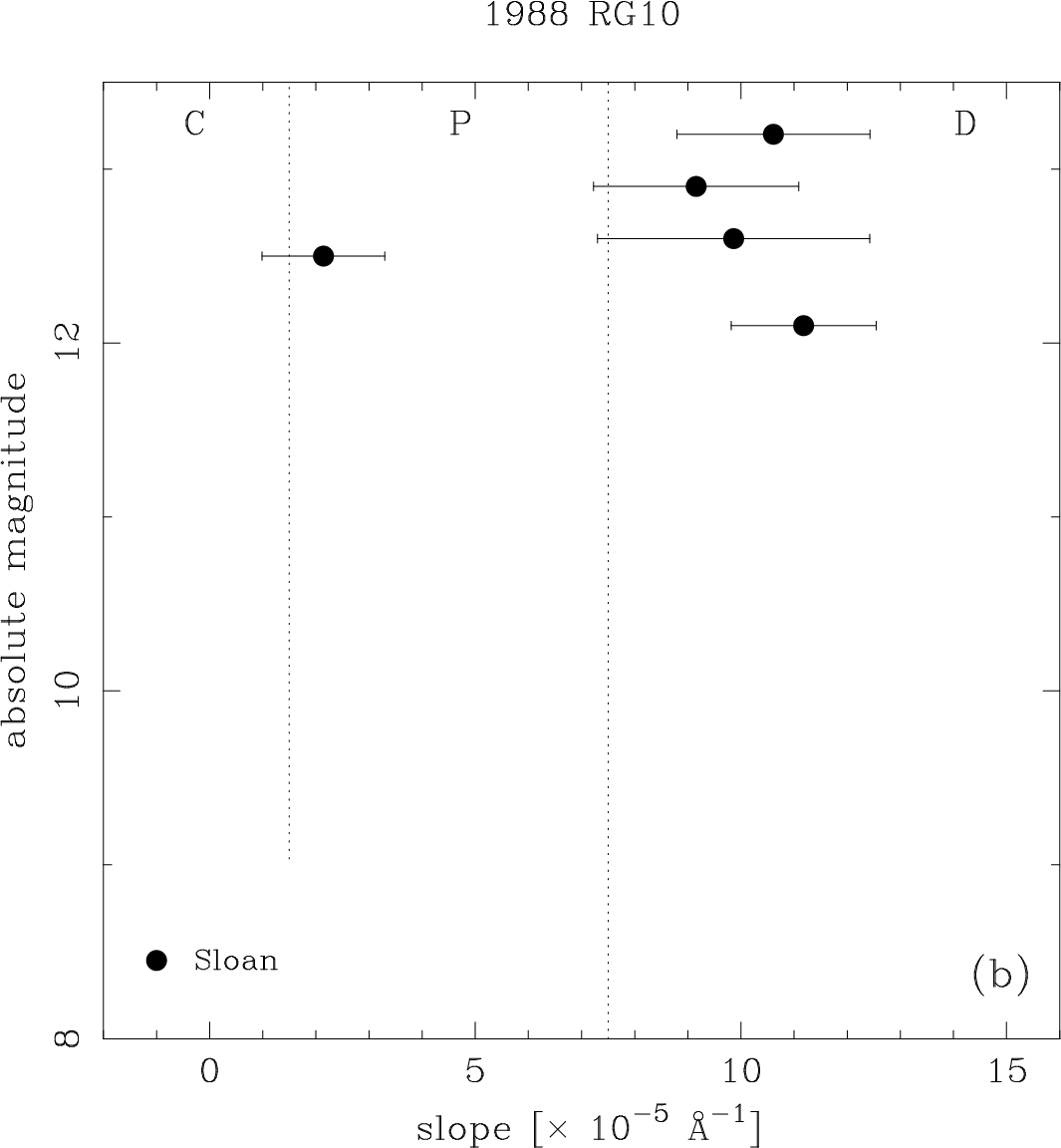}
\par\bigskip
\sidecaption
\includegraphics[clip,width=0.325\textwidth]{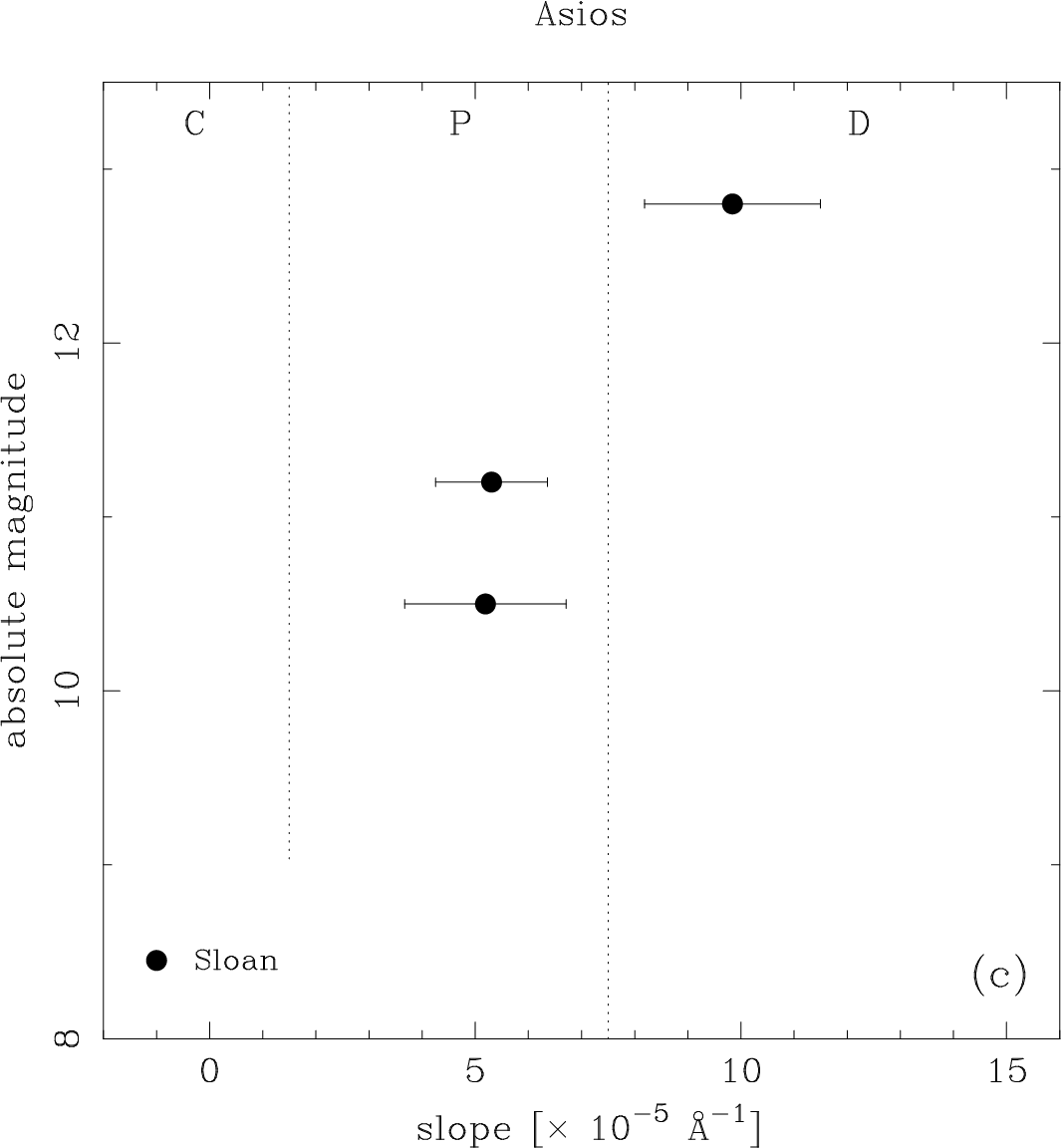}
~~~~
\includegraphics[clip,width=0.325\textwidth]{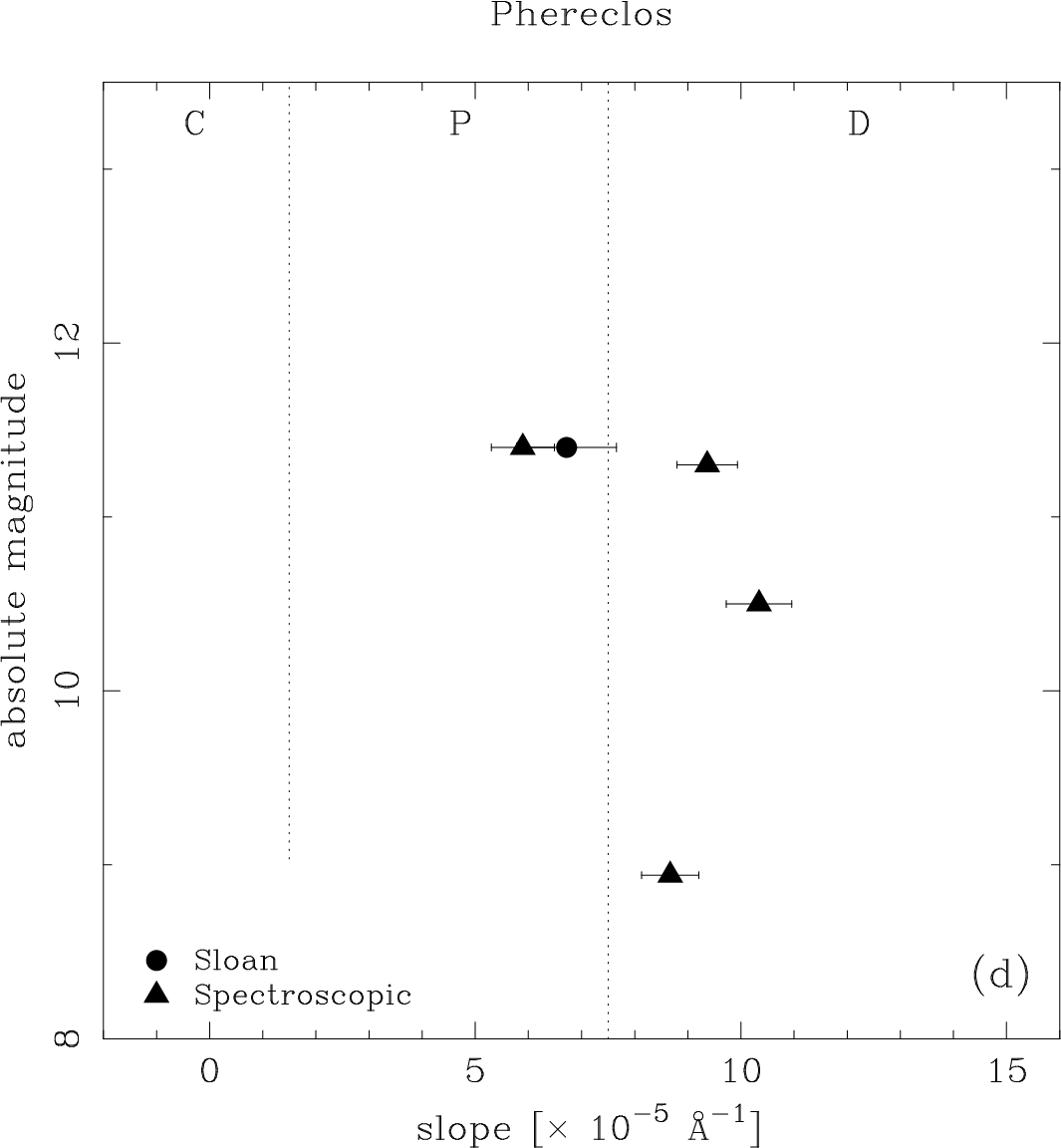} 
\caption{Same as Fig. \ref{fig:16} but for four different families in the
L5 swarm.}
\label{fig:17} 
\end{figure*}

Figure \ref{fig:17} shows the distributions of spectral slopes in
terms of absolute magnitude of four L5 families: Aneas, Phereclos,
1988 RG10 and Asios. 

The Aneas family has been studied by (\citeauthor{2004Icar..172..221F}
2004 - hereafter F04) and by F07, who treated it as the Sarpedon family.
This family is actually formed from the merging of two families: Sarpedon
and 1988 RN10. The Sarpedon family is resolved at $d_{\mathrm{cut}}<130$
$\mathrm{m\, s^{-1}}$, and the 1988 RN10 family is resolved at $d_{\mathrm{cut}}<140$
$\mathrm{m\, s^{-1}}$. Both families are identified down to $d_{\mathrm{cut}}=90$
$\mathrm{m\, s^{-1}}$. In Fig. \ref{fig:17}a we show the spectral
slopes of the whole Aneas family (i.e. Aneas + Sarpedon + 1988 RN10).
The values indicate that this is a quite homogeneous D-type family,
in agreement with F07%
\footnote{ Two members of the Sarpedon family observed by F07 --(48252) and
(84709)--, that are not included in our Spectroscopic sample, are
also classified as D-type.%
}. The only two P-type members shown in Fig. \ref{fig:17}a abandon
the family at $d_{\mathrm{cut}}<115$ $\mathrm{m\, s^{-1}}$, so they
are probably interlopers. So far, this family is one of the most homogeneous
families in terms of taxonomy already detected, together with the
Eurybates family in L4.

The 1988 RG10 and the Asios families, shown in Figs. \ref{fig:17}b,c,
have not been observed before by spectroscopic surveys. The distribution
of Sloan slopes indicates that the 1988 RG10 family would be a quite
homogeneous D-type family. For the Asios family the results are inconclusive.
The last family to be discussed here is the Phereclos family shown
in Fig. \ref{fig:17}d. It has been analyzed by F04 and F07 and their
results point to a quite homogeneous D-type family. The only observation
contained in the Sloan sample correspond to asteroid (18940), already
observed by F04, and its Sloan slope is compatible with its Spectroscopic
slope.

\subsection{Discussion}

While the individual families in L5 appear to be taxonomically homogeneous,
the individual families in L4 show a wide range of spectral slopes
and a mixture of the C-, P- and D-types. There are, at least,
two possibilities to explain the presence of different taxonomic classes
within a single family: 

\begin{itemize}
\item The family contains several interlopers, i.e. background asteroids
that overlap with the family in proper elements. The amount of interlopers
is significant because the presently observed family members are so
sparse that we need to use large cutoff values to detect the family.
Therefore, only the families detected at small cutoffs, like the Eurybates
family, can be considered, for the time being, as the less contaminated
by interlopers. Actually, the analysis of the Eurybates family made
by F07 indicates that it would be the most homogeneous family of the
Menelaus clan in terms of taxonomy. Note that the contamination by
background interlopers would not introduce significant inhomogeneities
in the taxonomy of L5 families, even if these families were not well
defined, because most of these families are taxonomically indistinguishable
from the background (in other words, the L5 families are dominated
by D-type asteroids and the L5 background too).
\item We may invoke some aging process on the surfaces of the asteroids,
like the space weathering, that originates the wide range of slopes
observed within a single family. In this case we must presume that
the surfaces of the members of the family do not have the same age.
In fact, many small members in the family may have been formed by
secondary collisions, thus showing younger surfaces. Also, the small
members, having a larger collision probability, may be more frequently
affected by collisional resurfacing processes. Unfortunately, the
space weathering is not well constrained in the case of the Trojan
asteroids, and we cannot say whether it produces a reddening of the
spectra with age, or viceversa \citep{2004Icar..170..214M}, or both.
The lack of a clear correlation between spectral slope and size among
the Trojan families, together with the still small amount of asteroids
with known spectral slope and the poorly constrained collisional evolution
of Trojan families, prevents to perform a reliable analysis of any
aging process. Moreover, the apparent taxonomical homogeneity of the
L5 families seems to play against the surface aging scenario.
\end{itemize}
Nevertheless, the case of the Eurybates/Menelaus families constitutes
an interesting paradigm of the possible effect of space weathering
on the surfaces of Trojan asteroids. The compactness of the Eurybates
family, compared to the Menelaus family, may be interpreted as a rough
measure of youthfulness. There are many examples in the Main Belt
that support this idea: the Karin family inside the Koronis family
\citep{2002Natur.417..720N}, the Baptistina family inside the Flora
clan \citep{2007DPS....39.5002B}, and the Veritas family \citep{2003ApJ...591..486N}.
Within this hypothesis, the color distribution of the Eurybates/Menelaus
families may be explained if we assume that the space weathering causes
a reddening of the surfaces with age. Then, we may speculate that
the members of the Eurybates family are the fresh fragments from the
interior of a former member of the Menelaus family. The remaining
members of the Menelaus family would have much older surfaces thus
being much redder. An analysis of the family ages based on purely
dynamical/collisional arguments is mandatory to better address this
issue.

\section{Conclusions\label{sec:Conclusions}}

We have analyzed the distribution of spectral slopes and colors of
Trojan asteroids using a sample of data from the SDSS-MOC3 together
with a collection of spectra obtained from several surveys. Our analysis
has been focused on the Trojan asteroid families. We have studied
the global properties of the sample as well as the properties of some
individual families. Our results can be summarized as follows:

\begin{itemize}
\item The analysis of photometric data from the SDSS-MOC3 produces reliable
results that are comparable with those obtained from the analysis
of spectroscopic data.
\item The distribution of spectral slopes among the Trojan asteroids shows
a clear bimodality. About 2/3 of the Trojan population is constituted
by reddish objects that may be classified as D-type asteroids. The
remaining bodies show less reddish colors compatible with the P-type,
and only a small fraction (less than 10\%) is constituted of bodies
with neutral colors compatible with the C-type.
\item The members of asteroid families show a bimodal distribution with
a very slight predominance of D-type asteroids. The background, on
the contrary, is significantly dominated by D-type asteroids.
\item The L4 and L5 swarms show significantly different distributions of
spectral slopes. The distribution in L4 is bimodal with a slight predominance
of D-type asteroids. The distribution in L5 is unimodal with a clear
peak of D-type asteroids. These differences can be attributed to the
presence of asteroid families.
\item The background asteroids show the same spectral slope distributions
in both swarms, with a significant fraction ($\sim80$\%) of D-type
asteroids. The families in L4 are dominated by P- and C-type asteroids,
while the families in L5 are dominated by D-type asteroids.
\item The background asteroids show correlations between spectral slope
and orbital inclination and between spectral slope and size. D-type
asteroids dominate among the high inclination bodies and also among
the large bodies. Low inclination bodies are slightly dominated by
P-type asteroids. These correlations are most probably the result
of the background collisional evolution, either by fragmentation or
by collisional resurfacing. Similar correlations are not observed
among the family members.
\item Individual families in the L5 swarm are taxonomically homogeneous,
but in the L4 swarm show a mixture of taxonomic types. This may be
attributed to the presence of interlopers or to a surface aging effect.
\item Any taxonomic analysis of individual families must be accompanied
by a detailed analysis of the families structure as a function of
the cutoff level of detection. An estimation of the family ages is
also mandatory to complement these analysis.
\end{itemize}

%
%%%%%%%%%%%%%%%%%%%%%%%%%%%%%% Acknowledgements.
\begin{acknowledgements}
We wish to thank Sonia Fornasier, Elisabetta Dotto, Phillipe Bendjoya
and Alberto Cellino who kindly allowed us to use their spectroscopic
data. Fruitful discussions with Jorge Carvano are also highly appreciated. 
This work has been supported by CNPq (Brazil) and SECYT (Argentina). 
\end{acknowledgements}

%
%%%%%%%%%%%%%%%%%%%%%%%%%%%%%% Bilbiography.
\bibliographystyle{aa}
\bibliography{trojan}

%
%%%%%%%%%%%%%%%%%%%%%%%%%%%%%% Long tables.
%
\longtabL{1}{%
\begin{landscape}
\begin{longtable}{rlrcl|rlrcl}
\caption{\label{tab:1}%
Known Trojan asteroids included in our Sloan sample. The spectral
slope $S$ was computed by a linear fit to the $g,r,i,z$ reflectance
fluxes, normalized to 1 at the $r$ band. For asteroids with two or
more observations ($N_{\mathrm{obs}}>1$), the table gives the average
weighted slope of the observations. Family membership is indicated
in the last column. Families were defined at a cutoff level of 110
$\mathrm{m\, s^{-1}}$ for the L4 swarm and 120 $\mathrm{m\, s^{-1}}$
for the L5 swarm.} \tabularnewline
\hline\hline
\multicolumn{5}{l|}{L4 swarm} & \multicolumn{5}{l}{L5 swarm} \tabularnewline
\hline 
 No. & Name & $S\:[\times10^{-5}\textrm{\AA}^{-1}]$ & $N_{\mathrm{obs}}$ & Family & No. & Name & $S\:[\times10^{-5}\textrm{\AA}^{-1}]$ & $N_{\mathrm{obs}}$ & Family \tabularnewline
\hline
\endfirsthead
\caption{continued.} \tabularnewline
\hline\hline
\multicolumn{5}{l|}{L4 swarm} & \multicolumn{5}{l}{L5 swarm} \tabularnewline
\hline 
 No. & Name & $S\:[\times10^{-5}\textrm{\AA}^{-1}]$ & $N_{\mathrm{obs}}$ & Family & No. & Name & $S\:[\times10^{-5}\textrm{\AA}^{-1}]$ & $N_{\mathrm{obs}}$ & Family \tabularnewline
\hline
\endhead
\hline
\endfoot 
  1749 & Telamon            & $  8.69\pm 0.75$ &  2 & Menelaus       &   1870 & Glaukos            & $  8.97\pm 0.92$ &  1 &                \tabularnewline
  2260 & Neoptolemus        & $ 10.53\pm 0.93$ &  1 &                &   1872 & Helenos            & $  5.71\pm 1.09$ &  2 &                \tabularnewline
  2759 & Idomeneus          & $  9.82\pm 0.82$ &  3 &                &   2674 & Pandarus           & $  9.03\pm 0.98$ &  1 &                \tabularnewline
  2797 & Teucer             & $  9.32\pm 1.03$ &  3 & Teucer         &   3451 & Mentor             & $  3.76\pm 1.09$ &  3 &                \tabularnewline
  3564 & Talthybius         & $  8.83\pm 0.66$ &  1 &                &   3708 & 1974 FV1           & $  9.93\pm 0.84$ &  1 &                \tabularnewline
  3793 & Leonteus           & $  6.26\pm 0.80$ &  1 & Teucer         &   4792 & Lykaon             & $ 10.27\pm 0.83$ &  1 & Polydoros      \tabularnewline
  4489 & 1988 AK            & $ 10.49\pm 0.68$ &  1 &                &   4827 & Dares              & $  8.66\pm 0.68$ &  1 &                \tabularnewline
  4946 & Askalaphus         & $  9.31\pm 0.88$ &  2 &                &   5257 & 1988 RS10          & $ 12.18\pm 1.68$ &  1 &                \tabularnewline
  5012 & Eurymedon          & $  4.39\pm 0.78$ &  1 &                &   5638 & Deikoon            & $  7.32\pm 0.95$ &  4 &                \tabularnewline
  5023 & Agapenor           & $  4.70\pm 0.45$ &  1 & 1986 WD        &   6997 & Laomedon           & $ 10.25\pm 0.56$ &  1 &                \tabularnewline
  5025 & 1986 TS6           & $  5.18\pm 0.87$ &  1 & 1986 TS6       &  11273 & 1988 RN11          & $  5.31\pm 1.55$ &  1 &                \tabularnewline
  5027 & Androgeos          & $ 11.04\pm 1.04$ &  1 &                &  11554 & Asios              & $  5.19\pm 1.52$ &  2 & Asios          \tabularnewline
  5028 & Halaesus           & $  9.21\pm 0.66$ &  1 &                &  11869 & 1989 TS2           & $  9.81\pm 1.03$ &  4 &                \tabularnewline
  5259 & Epeigeus           & $ 11.39\pm 0.74$ &  1 &                &  16560 & 1991 VZ5           & $  2.38\pm 0.77$ &  4 &                \tabularnewline
  5264 & Telephus           & $ 10.90\pm 0.56$ &  2 &                &  16667 & 1993 XM1           & $  9.26\pm 0.94$ &  4 &                \tabularnewline
  5284 & Orsilocus          & $  9.68\pm 0.92$ &  1 & Hektor         &  17414 & 1988 RN10          & $ 10.39\pm 1.11$ &  3 & 1988 RN10      \tabularnewline
  6545 & 1986 TR6           & $ 10.54\pm 0.57$ &  1 & 1986 WD        &  17416 & 1988 RR10          & $ 12.27\pm 1.37$ &  1 & Sarpedon       \tabularnewline
  7214 & Antielus           & $  4.77\pm 0.79$ &  1 & 1986 WD        &  17419 & 1988 RH13          & $  3.16\pm 0.88$ &  1 & 1988 RN10      \tabularnewline
  8241 & Agrius             & $  5.33\pm 1.14$ &  1 &                &  17420 & 1988 RL13          & $  7.11\pm 1.05$ &  3 & 1988 RL13      \tabularnewline
  9590 & 1991 DK1           & $  7.38\pm 0.87$ &  1 &                &  18046 & 1999 RN116         & $ 10.19\pm 0.72$ &  1 &                \tabularnewline
  9818 & Eurymachos         & $  3.07\pm 0.73$ &  1 & Eurybates      &  18137 & 2000 OU30          & $  9.28\pm 0.84$ &  2 & Polydoros      \tabularnewline
 10247 & Amphiaraos         & $  6.23\pm 1.38$ &  1 & Kalchas        &  18940 & 2000 QV49          & $  6.72\pm 0.94$ &  2 & Phereclos      \tabularnewline
 11396 & 1998 XZ77          & $  9.94\pm 0.66$ &  1 & 1998 XZ77      &  19020 & 2000 SC6           & $  9.06\pm 0.98$ &  4 & 1988 RN10      \tabularnewline
 13331 & 1998 SU52          & $  4.05\pm 1.22$ &  1 & Demophon       &  19844 & 2000 ST317         & $ 12.56\pm 0.97$ &  1 &                \tabularnewline
 13362 & 1998 UQ16          & $  9.57\pm 0.66$ &  1 & Menelaus       &  24018 & 1999 RU134         & $  5.19\pm 1.00$ &  2 &                \tabularnewline
 13387 & Irus               & $  7.98\pm 1.71$ &  1 &                &  24452 & 2000 QU167         & $  5.50\pm 0.91$ &  1 & Agelaos        \tabularnewline
 13475 & Orestes            & $  3.91\pm 0.97$ &  1 & Menelaus       &  24454 & 2000 QF198         & $  3.34\pm 1.06$ &  1 &                \tabularnewline
 14235 & 1999 XA187         & $  9.92\pm 1.11$ &  1 &                &  25347 & 1999 RQ116         & $  8.31\pm 0.98$ &  2 & Sarpedon       \tabularnewline
 15536 & 2000 AG191         & $  9.96\pm 0.83$ &  1 &                &  29314 & 1994 CR18          & $  4.16\pm 1.13$ &  2 &                \tabularnewline
 15663 & Periphas           & $  9.82\pm 1.04$ &  2 &                &  30499 & 2000 QE169         & $ 10.02\pm 0.98$ &  1 & 2000 QE169     \tabularnewline
 16152 & 1999 YN12          & $  4.98\pm 0.99$ &  1 &                &  30505 & 2000 RW82          & $ 11.51\pm 1.01$ &  1 &                \tabularnewline
 18060 & 1999 XJ156         & $  4.83\pm 1.01$ &  1 & Eurybates      &  30705 & Idaios             & $  9.84\pm 0.80$ &  2 &                \tabularnewline
 18062 & 1999 XY187         & $ 10.06\pm 0.72$ &  1 &                &  30708 & Echepolos          & $  8.58\pm 1.11$ &  1 &                \tabularnewline
 19725 & 1999 WT4           & $  5.50\pm 0.73$ &  1 & Epeios         &  31814 & 1999 RW70          & $ 11.18\pm 1.37$ &  1 & 1988 RG10      \tabularnewline
 20424 & 1998 VF30          & $ 10.76\pm 0.72$ &  2 &                &  32339 & 2000 QA88          & $  9.32\pm 0.89$ &  2 &                \tabularnewline
 20995 & 1985 VY            & $ 10.79\pm 1.01$ &  3 &                &  32482 & 2000 ST354         & $  5.67\pm 0.96$ &  1 & Panthoos       \tabularnewline
 21370 & 1997 TB28          & $  8.21\pm 1.11$ &  1 & Menelaus       &  32499 & 2000 YS11          & $  9.08\pm 0.99$ &  7 &                \tabularnewline
 21599 & 1998 WA15          & $  8.88\pm 0.96$ &  2 & 1998 XZ77      &  32811 & Apisaon            & $  4.83\pm 1.10$ &  1 &                \tabularnewline
 22049 & 1999 XW257         & $  5.15\pm 1.14$ &  2 &                &  34298 & 2000 QH159         & $  8.11\pm 1.33$ &  1 &                \tabularnewline
 22052 & 2000 AQ14          & $  6.27\pm 1.45$ &  1 & Epeios         &  38257 & 1999 RC13          & $  4.24\pm 0.91$ &  1 &                \tabularnewline
 22404 & 1995 ME4           & $  4.04\pm 0.98$ &  1 &                &  47969 & 2000 TG64          & $  9.52\pm 1.32$ &  4 & 1988 RN10      \tabularnewline
 23075 & 1999 XV83          & $ 11.42\pm 1.31$ &  1 &                &  48249 & 2001 SY345         & $  9.41\pm 0.87$ &  1 &                \tabularnewline
 23123 & 2000 AU57          & $  7.58\pm 0.79$ &  1 & Epeios         &  51345 & 2000 QH137         & $  9.89\pm 0.74$ &  1 &                \tabularnewline
 23144 & 2000 AY182         & $  5.82\pm 0.81$ &  1 & Epeios         &  51346 & 2000 QX158         & $  8.85\pm 1.11$ &  1 &                \tabularnewline
 23285 & 2000 YH119         & $ 10.81\pm 0.62$ &  1 &                &  51364 & 2000 SU333         & $  6.33\pm 1.07$ &  1 &                \tabularnewline
 23382 & Epistrophos        & $  8.61\pm 1.14$ &  2 &                &  51935 & 2001 QK134         & $ 10.64\pm 1.22$ &  1 &                \tabularnewline
 23706 & 1997 SY32          & $  6.46\pm 1.32$ &  1 & Menelaus       &  51994 & 2001 TJ58          & $  9.38\pm 1.41$ &  2 &                \tabularnewline
 23939 & 1998 TV33          & $  5.21\pm 0.53$ &  1 &                &  52273 & 1988 RQ10          & $ 11.40\pm 1.13$ &  1 &                \tabularnewline
 23963 & 1998 WY8           & $  2.52\pm 1.08$ &  1 & Kalchas        &  52511 & 1996 GH12          & $ 11.86\pm 1.10$ &  1 & Agelaos        \tabularnewline
 24225 & 1999 XV80          & $  4.64\pm 0.99$ &  1 & Epeios         &  52767 & 1998 MW41          & $  7.78\pm 1.16$ &  3 & Panthoos       \tabularnewline
 24233 & 1999 XD94          & $  4.71\pm 0.98$ &  1 & 1986 WD        &  54596 & 2000 QD225         & $  2.14\pm 1.16$ &  1 & 1988 RG10      \tabularnewline
 24403 & 2000 AX193         & $ 11.75\pm 1.02$ &  1 &                &  55457 & 2001 TH133         & $  8.26\pm 1.02$ &  2 &                \tabularnewline
 24426 & 2000 CR12          & $  5.58\pm 1.46$ &  1 & Eurybates      &  55460 & 2001 TW148         & $  6.06\pm 1.48$ &  1 & 1999 RV165     \tabularnewline
 24485 & 2000 YL102         & $  4.54\pm 1.16$ &  1 & 1998 XZ77      &  55678 & Lampos             & $ 10.19\pm 1.29$ &  1 &                \tabularnewline
 24498 & 2001 AC25          & $  3.05\pm 1.37$ &  1 & 1986 TS6       &  56976 & 2000 SS161         & $  5.31\pm 1.05$ &  1 & Asios          \tabularnewline
 24505 & 2001 BZ            & $ 11.89\pm 1.20$ &  1 &                &  57013 & 2000 TD39          & $  9.64\pm 1.22$ &  1 & 2000 SA191     \tabularnewline
 24508 & 2001 BL26          & $  5.29\pm 1.25$ &  1 & 1999 XM78      &  57626 & 2001 TE165         & $  3.85\pm 1.17$ &  1 & Panthoos       \tabularnewline
 24539 & 2001 DP5           & $  8.04\pm 1.15$ &  1 &                &  58008 & 2002 TW240         & $  7.53\pm 2.93$ &  4 &                \tabularnewline
 24882 & 1996 RK30          & $  4.29\pm 1.06$ &  2 & Epeios         &  58084 & Hiketaon           & $  9.87\pm 1.27$ &  1 &                \tabularnewline
 31835 & 2000 BK16          & $ 10.80\pm 1.43$ &  2 & 1986 WD        &  62201 & 2000 SW54          & $  8.35\pm 1.32$ &  1 &                \tabularnewline
 32498 & 2000 XX37          & $ 10.70\pm 0.88$ &  2 &                &  63955 & 2001 SP65          & $ 11.31\pm 1.23$ &  1 &                \tabularnewline
 33822 & 2000 AA231         & $  9.86\pm 0.99$ &  1 &                &  64270 & 2001 TA197         & $  9.84\pm 1.65$ &  2 & Asios          \tabularnewline
 35272 & 1996 RH10          & $ 11.03\pm 2.15$ &  2 &                &  65590 & Archeptolemos      & $  6.58\pm 1.29$ &  1 &                \tabularnewline
 36259 & 1999 XM74          & $  9.89\pm 1.48$ &  1 &                &  73795 & 1995 FH8           & $  8.59\pm 1.29$ &  1 &                \tabularnewline
 36279 & 2000 BQ5           & $ 10.65\pm 1.20$ &  2 &                &  76820 & 2000 RW105         & $  8.70\pm 1.16$ &  2 &                \tabularnewline
 38052 & 1998 XA7           & $  2.87\pm 1.39$ &  1 &                &  76824 & 2000 SA89          & $  6.87\pm 1.49$ &  1 &                \tabularnewline
 38606 & 1999 YC13          & $ 11.10\pm 1.29$ &  1 &                &  76837 & 2000 SL316         & $ 10.45\pm 1.17$ &  1 &                \tabularnewline
 38614 & 2000 AA113         & $  5.32\pm 1.05$ &  1 & Sinon          &  77891 & 2001 SM232         & $  6.60\pm 1.54$ &  1 &                \tabularnewline
 38617 & 2000 AY161         & $  3.85\pm 1.00$ &  1 & 1986 TS6       &        & 1988 SJ2           & $ 12.80\pm 1.43$ &  1 &                \tabularnewline
 38619 & 2000 AW183         & $  8.25\pm 1.31$ &  1 &                &        & 2000 QZ75          & $ 10.37\pm 1.48$ &  2 &                \tabularnewline
 38621 & 2000 AG201         & $  5.89\pm 1.21$ &  1 &                &        & 2000 RE29          & $ 11.37\pm 1.44$ &  1 & Deiphobus      \tabularnewline
 39264 & 2000 YQ139         & $ 11.64\pm 1.34$ &  1 & Hektor         &        & 2000 SG187         & $  5.01\pm 1.16$ &  1 &                \tabularnewline
 39287 & 2001 CD14          & $  3.87\pm 1.72$ &  1 & Laertes        &        & 2000 SK47          & $  9.06\pm 1.56$ &  4 &                \tabularnewline
 39293 & 2001 DQ10          & $  9.13\pm 1.31$ &  1 & 1986 WD        &        & 2000 SM250         & $ 11.08\pm 1.45$ &  1 &                \tabularnewline
 41268 & 1999 XO64          & $ 10.00\pm 1.19$ &  2 &                &        & 2000 SP92          & $ 10.25\pm 1.44$ &  1 & 2000 RO85      \tabularnewline
 42168 & 2001 CT13          & $  3.28\pm 0.98$ &  1 &                &        & 2000 SR79          & $  8.81\pm 1.07$ &  1 & Sergestus      \tabularnewline
 42179 & 2001 CP25          & $  4.19\pm 1.26$ &  1 & Kalchas        &        & 2000 SZ135         & $  9.12\pm 0.93$ &  2 &                \tabularnewline
 42403 & Andraimon          & $  6.49\pm 1.18$ &  1 &                &        & 2000 TU44          & $  8.92\pm 1.17$ &  2 &                \tabularnewline
 43212 & 2000 AL113         & $  1.12\pm 0.89$ &  1 & Eurybates      &        & 2001 QM257         & $  8.89\pm 1.10$ &  1 &                \tabularnewline
 43706 & Iphiklos           & $  5.56\pm 1.17$ &  1 & Makhaon        &        & 2001 RN122         & $ 10.02\pm 1.94$ &  1 & 1988 RL13      \tabularnewline
 51378 & 2001 AT33          & $  9.74\pm 0.97$ &  1 &                &        & 2001 SA220         & $ 13.07\pm 1.88$ &  1 & 1988 RG10      \tabularnewline
 53477 & 2000 AA54          & $  6.24\pm 0.87$ &  1 & 1986 WD        &        & 2001 SC101         & $  9.86\pm 2.56$ &  2 & 1988 RG10      \tabularnewline
 55568 & 2002 CU15          & $ 12.53\pm 1.29$ &  1 &                &        & 2001 SC137         & $  9.16\pm 1.93$ &  1 & 1988 RG10      \tabularnewline
 55571 & 2002 CP82          & $  4.70\pm 0.98$ &  1 &                &        & 2001 SD30          & $  9.98\pm 1.49$ &  1 & 2000 SA191     \tabularnewline
 57920 & 2002 EL153         & $  8.80\pm 1.31$ &  1 & Laertes        &        & 2001 TK131         & $  8.98\pm 1.68$ &  2 &                \tabularnewline
 58473 & 1996 RN7           & $  4.85\pm 1.32$ &  1 &                &        & 2001 TO108         & $  9.70\pm 1.12$ &  1 &                \tabularnewline
 58479 & 1996 RJ29          & $  6.41\pm 1.28$ &  1 & Menelaus       &        & 2001 VB52          & $ 10.83\pm 1.56$ &  1 & 2000 SY317     \tabularnewline
 60383 & 2000 AR184         & $ 10.42\pm 0.80$ &  1 &                &        & 2001 WX20          & $  5.07\pm 1.66$ &  1 &                \tabularnewline
 63202 & 2000 YR131         & $  4.24\pm 0.98$ &  1 & 1999 XM78      &        & 2001 XV105         & $  8.74\pm 1.67$ &  2 &                \tabularnewline
 63210 & 2001 AH13          & $ 11.28\pm 1.20$ &  1 & 1986 TS6       &        & 2002 VH107         & $  8.83\pm 1.24$ &  2 & Bitias         \tabularnewline
 63257 & 2001 BJ79          & $  4.74\pm 1.32$ &  1 & Euryalos       &        & 2003 WQ25          & $ 10.61\pm 1.82$ &  1 & 1988 RG10      \tabularnewline
 63259 & 2001 BS81          & $  5.44\pm 1.74$ &  2 & Demophon       & & & & & \tabularnewline
 63265 & 2001 CP12          & $ 10.27\pm 1.31$ &  1 &                & & & & & \tabularnewline
 63272 & 2001 CC49          & $  7.71\pm 1.52$ &  1 &                & & & & & \tabularnewline
 63286 & 2001 DZ68          & $  7.15\pm 1.32$ &  1 &                & & & & & \tabularnewline
 63291 & 2001 DU87          & $  5.96\pm 1.31$ &  1 & Menelaus       & & & & & \tabularnewline
 63292 & 2001 DQ89          & $  3.99\pm 1.29$ &  1 &                & & & & & \tabularnewline
 63294 & 2001 DQ90          & $ 11.57\pm 1.55$ &  3 &                & & & & & \tabularnewline
 65000 & 2002 AV63          & $  8.47\pm 1.70$ &  2 & Hektor         & & & & & \tabularnewline
 65134 & 2002 CH96          & $  2.45\pm 1.07$ &  1 &                & & & & & \tabularnewline
 65194 & 2002 CV264         & $ 10.45\pm 2.06$ &  2 &                & & & & & \tabularnewline
 65209 & 2002 DB17          & $  9.94\pm 1.26$ &  1 & Sinon          & & & & & \tabularnewline
 65224 & 2002 EJ44          & $  5.38\pm 2.01$ &  1 &                & & & & & \tabularnewline
 65225 & 2002 EK44          & $  2.56\pm 1.43$ &  1 & Eurybates      & & & & & \tabularnewline
 65583 & Theoklymenos       & $  7.93\pm 1.34$ &  1 & Menelaus       & & & & & \tabularnewline
 79444 & 1997 UM26          & $  5.77\pm 1.31$ &  2 &                & & & & & \tabularnewline
 80302 & 1999 XC64          & $  8.14\pm 1.71$ &  1 & Sinon          & & & & & \tabularnewline
 83975 & 2002 AD184         & $  5.78\pm 1.25$ &  1 &                & & & & & \tabularnewline
 83977 & 2002 CE89          & $  5.17\pm 1.33$ &  1 & Menelaus       & & & & & \tabularnewline
 83983 & 2002 GE39          & $  8.81\pm 1.57$ &  1 &                & & & & & \tabularnewline
 88225 & 2001 BN27          & $ 13.72\pm 1.35$ &  1 &                & & & & & \tabularnewline
 89829 & 2002 BQ29          & $  5.84\pm 1.07$ &  1 &                & & & & & \tabularnewline
 89871 & 2002 CU143         & $ 10.72\pm 1.39$ &  2 & Epeios         & & & & & \tabularnewline
 89924 & 2002 ED51          & $  9.47\pm 1.37$ &  2 & 1999 XM78      & & & & & \tabularnewline
       & 1995 QC6           & $  4.20\pm 1.14$ &  1 &                & & & & & \tabularnewline
       & 1996 TA58          & $ 10.55\pm 1.55$ &  1 &                & & & & & \tabularnewline
       & 1997 WA12          & $  8.97\pm 1.42$ &  1 &                & & & & & \tabularnewline
       & 1999 XJ55          & $ 11.88\pm 1.74$ &  1 &                & & & & & \tabularnewline
       & 2000 AG90          & $ 11.98\pm 1.55$ &  1 & Sinon          & & & & & \tabularnewline
       & 2000 AJ114         & $ 10.26\pm 1.18$ &  1 & Makhaon        & & & & & \tabularnewline
       & 2000 AL8           & $ 11.27\pm 1.75$ &  1 &                & & & & & \tabularnewline
       & 2000 BV1           & $  4.41\pm 1.38$ &  1 &                & & & & & \tabularnewline
       & 2000 YB131         & $ 10.92\pm 1.31$ &  3 &                & & & & & \tabularnewline
       & 2000 YC112         & $  7.30\pm 1.35$ &  4 & 1998 XZ77      & & & & & \tabularnewline
       & 2000 YS109         & $  4.39\pm 1.14$ &  1 & 1998 XZ77      & & & & & \tabularnewline
       & 2001 AG51          & $  5.20\pm 1.47$ &  1 & Laertes        & & & & & \tabularnewline
       & 2001 BD49          & $ 11.43\pm 1.23$ &  1 &                & & & & & \tabularnewline
       & 2001 BS16          & $ 10.01\pm 1.73$ &  2 &                & & & & & \tabularnewline
       & 2001 DL10          & $  9.42\pm 1.31$ &  1 &                & & & & & \tabularnewline
       & 2001 DO93          & $ 12.87\pm 1.86$ &  2 &                & & & & & \tabularnewline
       & 2001 FV58          & $ 12.96\pm 1.52$ &  1 & Epeios         & & & & & \tabularnewline
       & 2002 AE166         & $  0.95\pm 1.36$ &  1 & Eurybates      & & & & & \tabularnewline
       & 2002 CH109         & $  6.64\pm 1.69$ &  1 & Epeios         & & & & & \tabularnewline
       & 2002 CL109         & $  6.49\pm 1.51$ &  2 & 1998 US24      & & & & & \tabularnewline
       & 2002 CL130         & $  6.26\pm 1.17$ &  1 & Laertes        & & & & & \tabularnewline
       & 2002 CN130         & $  9.79\pm 1.10$ &  1 &                & & & & & \tabularnewline
       & 2002 CQ186         & $  6.37\pm 1.91$ &  1 & Laertes        & & & & & \tabularnewline
       & 2002 CZ256         & $  2.26\pm 1.53$ &  2 & Menelaus       & & & & & \tabularnewline
       & 2002 DD1           & $ 10.77\pm 1.67$ &  2 &                & & & & & \tabularnewline
       & 2002 DW15          & $  9.43\pm 1.61$ &  2 & Hektor         & & & & & \tabularnewline
       & 2002 DX12          & $  8.50\pm 1.25$ &  1 & Euryalos       & & & & & \tabularnewline
       & 2002 EK51          & $ 12.10\pm 1.77$ &  1 &                & & & & & \tabularnewline
       & 2002 EP106         & $  3.16\pm 1.35$ &  1 &                & & & & & \tabularnewline
       & 2002 ES83          & $ 10.29\pm 1.66$ &  2 &                & & & & & \tabularnewline
       & 2002 ET136         & $  8.99\pm 1.54$ &  1 &                & & & & & \tabularnewline
       & 2002 EU14          & $  6.52\pm 2.69$ &  2 &                & & & & & \tabularnewline
       & 2002 EX5           & $  2.84\pm 1.64$ &  1 & Epeios         & & & & & \tabularnewline
       & 2002 FL37          & $  5.32\pm 1.70$ &  1 &                & & & & & \tabularnewline
       & 2002 FM7           & $ 12.12\pm 2.42$ &  1 &                & & & & & \tabularnewline
       & 2002 GG33          & $ 13.05\pm 1.27$ &  1 & Menelaus       & & & & & \tabularnewline
       & 2002 GO150         & $  1.75\pm 1.68$ &  1 & Demophon       & & & & & \tabularnewline
       & 2003 FJ64          & $  7.01\pm 1.29$ &  1 & Menelaus       & & & & & \tabularnewline
       & 2003 FR72          & $  5.83\pm 1.59$ &  1 & Hektor         & & & & & \tabularnewline
       & 2003 GU35          & $  7.95\pm 1.54$ &  1 &                & & & & & \tabularnewline
       & 2003 GX7           & $ 11.87\pm 1.41$ &  1 &                & & & & & \tabularnewline
       & 2004 HS1           & $  7.60\pm 1.45$ &  1 &                & & & & & \tabularnewline
       & 2004 JO43          & $  4.69\pm 1.76$ &  1 &                & & & & & \tabularnewline
       & 2004 KJ4           & $  7.35\pm 1.15$ &  1 & Menelaus       & & & & & \tabularnewline
       & 5214 T-2           & $  7.42\pm 1.11$ &  1 & Euryalos       & & & & & \tabularnewline
\end{longtable}%
\end{landscape}
}

%%%%%%%%%%%%
%
\longtabL{2}{%
\begin{landscape}
\begin{longtable}{rlrcl|rlrcl}
\caption{\label{tab:2}%
Trojan asteroids included in our Spectroscopic sample. The spectral
slope $S$ was computed by a linear fit, in the interval 5000--9200 $\textrm{\AA}$,
of the rebinned spectra normalized to 1 at 6240 $\textrm{\AA}$. For asteroids with more
than one observation ($N_{\mathrm{obs}}>1$), the table gives the average
weighted slope of the observations. We recall that different observations of the same
asteroid come from different surveys. Family membership is indicated
in the last column. Families were defined at a cutoff level of 110
$\mathrm{m\, s^{-1}}$ for the L4 swarm and 120 $\mathrm{m\, s^{-1}}$
for the L5 swarm.} \tabularnewline
\hline\hline
\multicolumn{5}{l|}{L4 swarm} & \multicolumn{5}{l}{L5 swarm} \tabularnewline
\hline 
 No. & Name & $S\:[\times10^{-5}\textrm{\AA}^{-1}]$ & $N_{\mathrm{obs}}$ & Family & No. & Name & $S\:[\times10^{-5}\textrm{\AA}^{-1}]$ & $N_{\mathrm{obs}}$ & Family \tabularnewline
\hline
\endfirsthead
\caption{continued.} \tabularnewline
\hline\hline
\multicolumn{5}{l|}{L4 swarm} & \multicolumn{5}{l}{L5 swarm} \tabularnewline
\hline 
 No. & Name & $S\:[\times10^{-5}\textrm{\AA}^{-1}]$ & $N_{\mathrm{obs}}$ & Family & No. & Name & $S\:[\times10^{-5}\textrm{\AA}^{-1}]$ & $N_{\mathrm{obs}}$ & Family \tabularnewline
\hline
\endhead
\hline
\endfoot 
   588 & Achilles           & $  2.09\pm 0.64$ &  1 &                &   1172 & Aneas              & $  9.78\pm 1.60$ &  2 & Aneas          \tabularnewline
   911 & Agamemnon          & $  9.12\pm 4.14$ &  2 &                &   1871 & Astyanax           & $  5.45\pm 0.56$ &  1 &                \tabularnewline
  1143 & Odysseus           & $ 11.07\pm 0.63$ &  1 &                &   2223 & Sarpedon           & $ 11.31\pm 3.22$ &  2 & Sarpedon       \tabularnewline
  1647 & Menelaus           & $  6.62\pm 0.53$ &  1 & Menelaus       &   2357 & Phereclos          & $  8.67\pm 0.54$ &  1 & Phereclos      \tabularnewline
  1749 & Telamon            & $ 10.08\pm 0.62$ &  1 & Menelaus       &   2895 & Memnon             & $ -1.59\pm 0.72$ &  1 &                \tabularnewline
  1868 & Thersites          & $  8.78\pm 0.60$ &  1 &                &   3317 & Paris              & $  7.94\pm 3.70$ &  2 &                \tabularnewline
  2920 & Automedon          & $ 10.34\pm 0.59$ &  1 &                &   3451 & Mentor             & $  1.40\pm 0.57$ &  1 &                \tabularnewline
  3063 & Makhaon            & $  8.33\pm 0.52$ &  1 & Makhaon        &   3708 & 1974 FV1           & $  8.60\pm 0.53$ &  1 &                \tabularnewline
  3709 & Polypoites         & $ 11.19\pm 0.62$ &  2 &                &   4348 & Poulydamas         & $  4.12\pm 0.53$ &  1 &                \tabularnewline
  3793 & Leonteus           & $  7.67\pm 0.54$ &  1 & Teucer         &   4715 & 1989 TS1           & $ 14.61\pm 0.59$ &  1 &                \tabularnewline
  4035 & 1986 WD            & $ 12.42\pm 1.91$ &  2 & 1986 WD        &   4792 & Lykaon             & $ 13.98\pm 0.59$ &  1 & Polydoros      \tabularnewline
  4060 & Deipylos           & $  1.75\pm 2.46$ &  2 &                &   5130 & Ilioneus           & $  9.08\pm 0.54$ &  1 & Sarpedon       \tabularnewline
  4063 & Euforbo            & $  8.28\pm 0.55$ &  2 &                &   5511 & Cloanthus          & $ 13.27\pm 0.57$ &  1 & Cloanthus      \tabularnewline
  4068 & Menestheus         & $ 10.77\pm 1.61$ &  2 &                &   5648 & 1990 VU1           & $ 10.46\pm 2.35$ &  2 &                \tabularnewline
  4138 & Kalchas            & $  6.53\pm 0.65$ &  1 & Kalchas        &   6998 & Tithonus           & $  9.36\pm 0.57$ &  1 & Phereclos      \tabularnewline
  4489 & 1988 AK            & $  8.55\pm 0.54$ &  1 &                &   7352 & 1994 CO            & $  6.46\pm 3.52$ &  2 & 1994 CO        \tabularnewline
  4833 & Meges              & $ 10.90\pm 0.54$ &  2 &                &   9430 & Erichthonios       & $ 10.34\pm 0.62$ &  1 & Phereclos      \tabularnewline
  4834 & Thoas              & $ 11.01\pm 0.54$ &  1 &                &  11089 & 1994 CS8           & $  3.96\pm 0.55$ &  1 & Anchises       \tabularnewline
  4835 & 1989 BQ            & $  5.89\pm 3.39$ &  2 &                &  15502 & 1999 NV27          & $  9.83\pm 0.55$ &  1 & Aneas          \tabularnewline
  4836 & Medon              & $  8.80\pm 0.55$ &  1 &                &  15977 & 1998 MA11          & $  7.50\pm 0.57$ &  1 & 1998 MA11      \tabularnewline
  4902 & Thessandrus        & $  8.30\pm 0.54$ &  1 &                &  17416 & 1988 RR10          & $  9.05\pm 0.60$ &  1 & Sarpedon       \tabularnewline
  5025 & 1986 TS6           & $ 15.36\pm 0.71$ &  1 & 1986 TS6       &  18137 & 2000 OU30          & $  7.26\pm 0.59$ &  1 & Polydoros      \tabularnewline
  5126 & Achaemenides       & $  0.76\pm 0.53$ &  1 &                &  18268 & Dardanos           & $ 12.19\pm 0.59$ &  1 & Sarpedon       \tabularnewline
  5244 & Amphilochos        & $  2.94\pm 0.53$ &  1 & Menelaus       &  18493 & 1996 HV9           & $  4.56\pm 0.56$ &  2 & Sarpedon       \tabularnewline
  5254 & Ulysses            & $  9.99\pm 0.52$ &  1 &                &  18940 & 2000 QV49          & $  5.89\pm 0.59$ &  1 & Phereclos      \tabularnewline
  5258 & 1989 AU1           & $  7.48\pm 0.53$ &  1 & Menelaus       &  23694 & 1997 KZ3           & $  7.02\pm 0.56$ &  1 & Sergestus      \tabularnewline
  5264 & Telephus           & $ 11.68\pm 1.05$ &  2 &                &  24467 & 2000 SS165         & $ 10.02\pm 0.57$ &  1 & Sarpedon       \tabularnewline
  5283 & Pyrrhus            & $ -6.58\pm 0.56$ &  1 &                &  25347 & 1999 RQ116         & $  8.50\pm 0.59$ &  1 & Sarpedon       \tabularnewline
  5285 & Krethon            & $  6.53\pm 0.59$ &  1 &                &  30698 & Hippokoon          & $  7.68\pm 0.61$ &  1 & Sergestus      \tabularnewline
  6090 & 1989 DJ            & $ 12.54\pm 0.64$ &  1 &                &  32430 & 2000 RQ83          & $  6.29\pm 0.65$ &  1 & Sergestus      \tabularnewline
  6545 & 1986 TR6           & $  9.15\pm 0.53$ &  1 & 1986 WD        &  32615 & 2001 QU277         & $  8.34\pm 0.57$ &  1 &                \tabularnewline
  7152 & Euneus             & $  5.61\pm 0.61$ &  1 & Euneus         &  34785 & 2001 RG87          & $  2.58\pm 0.59$ &  1 &                \tabularnewline
  7641 & 1986 TT6           & $  4.66\pm 0.66$ &  1 &                &  48249 & 2001 SY345         & $  9.60\pm 0.60$ &  1 &                \tabularnewline
 11351 & 1997 TS25          & $  9.10\pm 0.56$ &  1 & 1986 WD        & & & & & \tabularnewline
 12917 & 1998 TG16          & $ 11.19\pm 0.57$ &  1 & 1986 TS6       & & & & & \tabularnewline
 12921 & 1998 WZ5           & $  4.05\pm 0.54$ &  1 & 1986 TS6       & & & & & \tabularnewline
 13463 & Antiphos           & $  4.75\pm 0.76$ &  2 & 1986 TS6       & & & & & \tabularnewline
 15094 & 1999 WB2           & $  2.59\pm 0.55$ &  1 &                & & & & & \tabularnewline
 15535 & 2000 AT177         & $ 10.92\pm 0.56$ &  2 & 1986 TS6       & & & & & \tabularnewline
 20738 & 1999 XG191         & $  9.70\pm 0.56$ &  1 & 1986 TS6       & & & & & \tabularnewline
 24390 & 2000 AD177         & $  9.74\pm 0.54$ &  1 & 1986 TS6       & & & & & \tabularnewline
\end{longtable}%
\end{landscape}
}

\end{document}